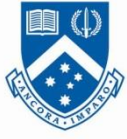
MONASH University

# Data over dialogue: Why artificial intelligence is unlikely to humanise medicine

## Joshua James Hatherley

*BA(Hons), MBioethics*

*A thesis submitted for the degree of Doctor of Philosophy at Monash University in 2023*

Department of Philosophy

School of Philosophical, Historical, and International Studies

Faculty of Arts, Monash University

## COPYRIGHT NOTICE

# DEDICATION

To my late grandfather, Max Williamson, who died just before I was accepted into the PhD program in late 2018, and who – I am assured – would have been enormously proud were he alive today.

# TABLE OF CONTENTS

# ABSTRACT

Recently, a growing number of experts in artificial intelligence (AI) and medicine have begun to suggest that AI systems, particularly machine learning (ML) systems, are likely to revolutionise the practice of medicine and the delivery of healthcare services. In particular, experts anticipate that medical ML systems will improve patient health and safety by improving the quality of clinicians' judgements and reducing the frequency of medical errors. Experts also anticipate that medical ML systems will advance health equity by improving the quality of care available to under-resourced populations, and improve efficiency in the performance of clinical tasks and the use of scarce healthcare resources. Recently, however, Eric Topol – who is arguably the foremost advocate of AI in medicine – has argued that the most substantial effects of medical ML systems will not come from their benefits to patient health, health equity, or efficiency in medicine. Rather, the most substantial effects of these systems will come from their benefits to clinician-patient relationships. As Topol himself expresses, the "greatest opportunity offered by AI is not reducing errors or workloads, or even curing cancer: it is the opportunity to restore the previous and time-honored connection and trust – the human touch – between patients and doctors" (Topol 2019a: 18).

In this thesis, however, I argue that this vision for the future of clinician-patient relationships in the coming age of AI in medicine is fundamentally misguided, since medical ML systems are more likely to negatively impact these relationships than to improve them. In particular, I argue that the use of medical ML systems is likely to comprise the quality of trust, care, empathy, understanding, and communication between clinicians and patients. I suggest that, to protect and preserve clinician-patient relationships in the coming age of AI in medicine, stakeholders must resist being carried away by excitement over the anticipated benefits of these systems. Stakeholders must also more carefully consider the risks of medical ML systems to avoid overinvesting in these systems on the basis of exaggerated assessments of their likely benefits. I conclude that minimising the risks associated with these systems depends not only on improving their accuracy and performance, but also on attending to how these systems impact relationships between human beings, and how human beings relate to these systems themselves.

# DECLARATION

This thesis contains no material which has been accepted for the award of any other degree or diploma at any university or equivalent institution and that, to the best of my knowledge and belief, this thesis contains no material previously published or written by another person, except where due reference is made in the text of the thesis.

**Signature:**     [redacted]

**Print Name:**   Joshua Hatherley

**Date:**     30/9/2023

## PUBLICATIONS DURING ENROLMENT

(*) Publications included in thesis.

## THESIS INCLUDING PUBLISHED WORKS DECLARATION

I hereby declare that this thesis contains no material which has been accepted for the award of any other degree or diploma at any university or equivalent institution and that, to the best of my knowledge and belief, this thesis contains no material previously published or written by another person, except where due reference is made in the text of the thesis.

This thesis includes material from three (3) original papers published in peer reviewed journals. The core theme of the thesis is ethical issues associated with the use of medical artificial intelligence systems. The ideas, development and writing up of all the papers in the thesis were the principal responsibility of myself, the student, working within the School of Philosophical, Historical, and International Studies under the supervision of Professor Robert Sparrow.

The inclusion of co-authors reflects the fact that the work came from active collaboration between researchers and acknowledges input into team-based research.

In the case of chapters two (2), four (4), and five (5) my contribution to the work involved the following:

| Chapter | Title | Status | Nature and % of student contribution | Co-author name(s) Nature and % of Co-author's contribution* | Co-author(s), Monash student Y/N* |
|---|---|---|---|---|---|
| 2 | Limits of trust in medical AI | Published | 100% | N/A | N |
| 4 | The virtues of interpretable medical artificial intelligence | Published | 60% Conceptualisation, drafting, revisions, editing. | Robert Sparrow (30%): Conceptualisation, revisions, editing Mark Howard (10%): Revisions, editing | N |

| 5 | Diachronic and synchronic variation in the performance of adaptive machine learning systems: the ethical challenges | Published | 60% Conceptualisation, drafting, revisions, editing | Robert Sparrow (40%): Conceptualisation, revisions, editing | N |

I have not renumbered sections of submitted or published papers in order to generate a consistent presentation within the thesis.

**Student signature:**  [redacted]  **Date:** 30/9/2023

The undersigned hereby certify that the above declaration correctly reflects the nature and extent of the student's and co-authors' contributions to this work. In instances where I am not the responsible author I have consulted with the responsible author to agree on the respective contributions of the authors.

**Main supervisor signature:**  [redacted]  **Date:** 27/9/2023

# ACKNOWLEDGEMENTS[1]

Completing this project has been substantially harder than I anticipated. Were it not for the guidance, friendship, and support of a great many people, I would have certainly abandoned the idea of finishing it long ago.

First of all, I am deeply thankful for the supervision and mentorship provided by Robert Sparrow and Justin Oakley during the course of this project. Their substantial contributions have been instrumental in shaping and improving the quality of my research. I am especially appreciative of Rob's mentorship, which has profoundly influenced my approach to research in applied ethics. His consistently insightful and constructive feedback have been crucial to the success of this work and my academic growth. I also extend my gratitude to Justin for his valuable comments on many parts of this thesis, and for his unwavering support during the many of the more challenging phases of this degree.

During the course of my PhD, I was also given the opportunity to work with Mark Howard and Julian Koplin on several projects relating to the ethics of artificial intelligence systems in various domains. I am grateful for their mentorship and collaboration. I am also grateful to Tim Bayne, Isabelle Ford, Rosa Martorana, Courtney McMahon, Tommy Ness, Graham Oppy, Nina Roxburgh, and Steph Slack who each provided valuable feedback on parts of this thesis.

I spent almost every working day while writing this thesis in office S603 of the Menzies building at Monash Clayton. While working in this office, I was fortunate enough to benefit from the companionship of a group of excellent researchers, understanding colleagues, and caring friends. For their solidarity, collaboration, and support, I am grateful to Courtney Baker, Lawrence Bradford, Andrew Corcoran, James Croxford, Alba Cuenca, Zach Daus, Gene Flenady, Isabelle Ford, Tessa Holzman, Eliana Horn, Mark Howard, Molly Johnston, Manu Kirberg, Julian Koplin, Rachel Lansell, Adam Manoogian, Marvin Martin, Rosa Martorana, Pat McConville, James McGuire, Andy McKilliam, Courtney McMahon, Niccolò Negri, Ricardo Rojas, Kelsey Perrykkad, Andrew Raivers, Kate Rivington, Nina Roxburgh, Becks Shin, Maks Sipowicz, Craig Stanbury, Laura Teerijoki, Isis Urgell Menéndez, Jasmine Walter, and James Wells. I am specially indebted to Samiksha Goyal, Tommy Ness, Steph Slack, and Chanelle Warton for their invaluable friendship, and for their unwavering support during the darker moments of this process.

---

[1] This work was supported by an Australian Government Research Training Program (RTP) scholarship.

I would also not have been able to complete this project without the friendship and support of many people beyond Monash University, including Niamh Clemens, Giulia Di Sipio, Ella Healy, Gab Howard, Jordy Kornfeld, Robin Kornfeld, Peter Kornfeld, Marisha Mazumdar, Aiden McNamara, Adrian Pacione, Sam Slade, Eve Slade, and Eleanor Yewers. I am especially grateful to Jordy Kornfeld, Sam Slade, and Guilia Di Sipio, who made themselves available to care for me at a moment's notice during an extended period of acute sickness.

Finally, I would not have even attempted this undertaking in the first place were it not for my parents, Peter Hatherley and Susanne Williamson, and my stepfather, Stuart Borg, who have each invested an enormous amount of their time, energy, and emotional well-being toward celebrating my victories and supporting me through my many failures. To each of them I owe a debt that, despite my best efforts, I am sure I could not possibly ever repay.

INTRODUCTION

Public interest and research funding for artificial intelligence (AI) are currently at an all-time high due to recent technical advances in the AI subfield of machine learning (ML). ML in particular has recently come to the attention of AI researchers, policy makers, and the general public by facilitating a series of enormous improvements in a rapidly growing list of revolutionary technologies, including driverless cars, voice assistants (e.g. Alexa, Siri, and Google Home), content recommender systems (e.g. Netflix, YouTube, and TikTok), and image, voice, and text-generation algorithms (e.g. DALL-E, Stable Diffusion, and ChatGPT). Experts anticipate that these developments in ML will lead these systems to have an enormous impact across the entire spectrum of industries and professions (Brynjolfsson and McAfee 2016; Susskind and Susskind 2015; Tegmark 2017).

Arguably, however, ML systems are expected to have their most positive and substantial impact in the domains of medicine and healthcare. One reason for this is that developers of ML systems have benefited substantially from the digitisation of medical record-keeping over the past two decades. In particular, the collection of vast repositories of health data in electronic health records (EHRs), wearable sensors (e.g. FitBits and smartphones), and publicly available online datasets (e.g. the International Skin Imaging Collaboration) has enabled researchers and AI developers to train a dazzling array of ML systems designed to assist clinicians across the entire spectrum of medical specialisations. As a result, the number of medical ML systems that have been approved for sale and use by government agencies such as the United States (US) Food and Drug Administration (FDA) has recently exploded. Indeed, 86% of all medical AI systems currently available on the US healthcare market have received FDA approval in the last 5 years alone (Lyell et al. 2021).

This recent surge of activity in the development of medical ML systems has generated soaring expectations for the coming age of AI in medicine. In particular, experts anticipate that medical ML systems will improve patient health and safety by improving the quality of clinicians' judgements and reducing the frequency of medical errors (Rajkomar, Dean, and Kohane 2019; Rajpurkar et al. 2022; Topol 2019b). Experts also anticipate that medical ML systems will advance health equity by improving the quality of care available to under-resourced populations (Guo and Li 2018; Jha and Topol 2023; Nittas et al. 2023; Vaitla et al. 2020), and improve efficiency in the performance of clinical tasks and the use of scarce healthcare resources (Nittas et al. 2023; Topol 2019a; Verghese, Shah, and Harrington 2018).

Despite historical resistance to the adoption of new health information technologies (Susskind and Susskind 2015b; Topol 2019a; Wachter 2015a), healthcare organisations and professional associations are becoming increasingly open to the adoption and use of medical ML systems. For instance, professional medical associations are releasing position statements and formalised policies on the use of ML systems in clinical practice at a rapidly accelerating pace (Academy of Medical Royal Colleges 2019; Australian Medical Association 2023; Matheny et al. 2019; Medical Radiation Practice Board of Australia 2022; Royal Australian College of General Practitioners 2021; Royal Australian and New Zealand College of Radiologists 2019; Solomonides et al. 2022). Indeed, the American Medical Association (2018) recently released its first ever set of policy guidelines on clinicians' use of what they refer to as 'augmented intelligence' in medicine (to emphasise that these systems must be designed to assist, rather than substitute or replace, human clinicians). Moreover, several influential medical journals, including the *New England Journal of Medicine*,[1] have recently launched (or announced the upcoming launch of) journals devoted to research in medical AI (El Emam and Malin 2022).

However, despite substantial enthusiasm about the potential benefits of medical ML systems, the use of these systems also presents a broad range of risks and threats. For instance, medical ML systems threaten to expand current disparities in health and healthcare due to their vulnerability to adopting the biases of their designers and the societies in which they are developed and embedded (Aquino et al. 2023; Hoffman 2021; Nadeem, Marjanovic, and Abedin 2022; Panch, Mattie, and Atun 2019; Price 2019). Medical ML systems could also contribute to intensified government surveillance of socio-politically marginalised groups, and expand the scope of power and influence of private technology corporations in medical decision-making and the delivery of healthcare services, particularly in light of developments associated with the COVID-19 pandemic (Greene, Hoffman, and Stark 2019; Mello and Wang 2020; Zhao et al. 2021). The use of medical ML systems also risks compromising patient privacy and interfering with the capacity for individuals to be held accountable for patient harm that results from their use (Bleher and Braun 2020; Gerke, Minssen, and Cohen 2020; Price and Cohen 2019). The list goes on and on (see Arnold 2021; Braun et al. 2020; Grote and Berens 2020; Keskinbora 2019; Schönberger 2019; Svensson and Jotterand 2022).

---

[1] See https://ai.nejm.org/.

While sustained attention has thus far been directed toward the impact of medical ML systems on population health and health systems at large (Panch, Mattie, and Atun 2019; Price 2019; Rajkomar et al. 2018; Starke, De Clercq, and Elger 2021), more attention to the threats these systems present to clinician-patient relationships on the frontlines of medicine is needed. The clinician-patient relationship refers to:

> a fiduciary relationship in which, by entering into the relationship, the physician agrees to respect the patient's autonomy, maintain confidentiality, explain treatment options, obtain informed consent, provide the highest standard of care, and commit not to abandon the patient without giving him or her adequate time to find a new doctor (Chipidza et al. 2015: 1).

In addition, as John Kelley and co-authors observe, the clinician-patient relationship involves both emotional and cognitive care: "Emotional care includes mutual trust, empathy, respect, genuineness, and warmth. Cognitive care includes information gathering, sharing medical information, patient education, and expectation management" (Kelley et al. 2014: 1).

More attention to the threats medical ML systems present to the clinician-patient relationship is needed because the nature and quality of these relationships have a substantial impact on several important areas of medicine, including health equity (Gupta and Carr 2008; Ferguson and Candib 2002; Lambrou et al. 2020), patient autonomy (Chin 2002; Entwistle et al. 2010), and patient health and safety (Chipidza, Wallwork, and Stern 2015; Entwistle and Quick 2006). As Teresa Hellín has expressed, the "importance of an intimate relationship between patient and physician can never be overstated because in most cases an accurate diagnosis, as well as an effective treatment, relies directly on the quality of this relationship" (Hellín 2002: 452). These relationships also hold intrinsic ethical significance due to the vulnerability that patients often experience throughout the course of their medical care, and the resulting power imbalance that emerges between them and their clinicians (Brody 1992; Frederiksen, Kragstrup, and Dehlholm-Lambertson 2010; Goodyear-Smith and Buetow 2001; Plomp and Ballast 2010). Despite this, critical engagement with the impact of medical ML systems on clinician-patient relationships has thus far been limited, which is surprising given the breadth of concerns that have been raised about the impact of new technologies on clinician-patient relationships in the past (see Anderson, Rainey, and Eysenbach 2003; Bauer 2004; Cassell 2002; Lo and Parham 2010; Norman, Aikins, and Binka 2011; Reiser 2009).

Most importantly, however, more critical attention toward the impact of medical ML systems on clinician-patient relationships is needed since a vital selling point of these systems is that they are anticipated to, as Bertalan Meskó and co-authors express, "bring forward a renaissance era in the doctor-patient relationship" (Meskó, Hetényi, and Győrffy 2018: 3). According to Abraham Verghese and co-authors, for instance:

> Human intelligence working with artificial intelligence – a well-informed, empathetic clinician armed with good predictive tools and unburdened from clerical drudgery – can bring physician closer to fulfilling Peabody's maxim that the secret of care is in 'caring for the patient' (Verghese et al. 2018: 20).

Recently, moreover, Eric Topol – who is arguably the foremost advocate of AI in medicine – has argued that the most substantial effects of medical ML systems will not come from their benefits to patient health, health equity, or efficiency in medicine. Rather, the most substantial effects of these systems will come from their benefits to clinician-patient relationships. As Topol himself expresses, the "greatest opportunity offered by AI is not reducing errors or workloads, or even curing cancer: it is the opportunity to restore the previous and time-honored connection and trust – the human touch – between patients and doctors" (Topol 2019a: 18). He anticipates that medical ML systems will benefit clinician-patient relationships by relieving a host of administrative and psychological pressures that currently preclude clinicians from developing caring, trusting, understanding, and empathetic relationships with their patients:

> Not only would we have more time to come together, enabling far deeper communication and compassion, but also we would be able to revamp how we select and train doctors. We have prized 'brilliant' doctors for decades, but the rise of machines will heighten the diagnostic skills and the fund of medical knowledge available to all clinicians. Eventually, doctors will adopt AI and algorithms as their work partners. This leveling of the medical knowledge landscape will ultimately lead to a new premium: to find and train doctors who have the highest level of emotional intelligence (Topol 2019a: 18).

In this thesis, however, I argue that Topol's vision for the future of clinician-patient relationships in the coming age of AI in medicine is fundamentally misguided. This is because, rather than ushering in a 'renaissance era' in clinician-patient relationships, medical ML systems are more likely to have an overall negative impact on the quality of these relationships. In particular, I argue that the use of medical ML systems is likely to comprise the quality of trust, care, empathy, understanding, and communication in these relationships. While medical ML systems may deliver modest improvements in narrow domains of medicine, AI developers, healthcare organisations, and policy makers need to more carefully consider the effects of these systems on the relationships between clinicians and their patients to accurately assess the costs and benefits of these technologies prior to their implementation.

Before presenting my arguments for this central thesis, however, it is first necessary to outline some background material that will underpin these later arguments. The remainder of this introduction proceeds as follows. In section one, I provide a brief technical overview of ML. In section two, I outline the current state of ML systems in medicine. In section three, I outline the anticipated benefits of medical ML systems for patient health and safety, health equity, and efficiency in medicine. Finally, in section four, I provide a chapter outline for the remainder of the thesis.

## 1. Overview of artificial intelligence and machine learning

Broadly speaking, AI refers to "a field of science and engineering concerned with computational understanding of what is commonly called intelligent behavior, and with the creation of artifacts that exhibit such behavior" (Shapiro 1992: 54; see also Nilsson 2009; Russell and Norvig 2010). However, philosophers often distinguish between 'weak' and 'strong' AI. As John Searle has expressed, for instance:

> According to weak AI, the principal value of the computer in the study of the mind is that it gives us a very powerful tool. […] But according to strong AI, the computer is not merely a tool in the study of the mind; rather, the appropriately programmed computer really *is* a mind, in the sense that computers given the right programs can be literally said to *understand* and have other cognitive states (Searle 1980: 417).

Philosophers often also distinguish between 'artificial narrow intelligence' and 'artificial general intelligence'. As Scott McLean and co-authors observe, the capabilities of artificial *narrow* intelligence systems are "task specific (or narrow) and cannot transfer to other domains with unknown and uncertain environments in which they have not been trained" (McLean et al. 2023: 649). In contrast, as Andreas Kaplan and Michael Haenlein note, systems exhibiting artificial *general* intelligence are "able to reason, plan, and solve problems autonomously for tasks they were never even designed for" (Kaplan and Haenlein 2019: 16). Despite the substantial history of recent achievements in AI, along with persistent efforts amongst AI researchers to develop strong AI and artificial general intelligence, current systems remain limited to weak AI and artificial narrow intelligence (Shevlin et al. 2019). In this thesis, therefore,

I make no attempt to engage with current debates about the ethical implications of strong AI or artificial general intelligence systems, or their potential implications in medicine.

The formalised discipline of AI was first established during the 1956 Dartmouth Summer Research Project on Artificial Intelligence. Chaired by John McCarthy, the aim of this workshop was "to find how to make machines use language, form abstractions and concepts, solve kinds of problems now reserved for humans, and improve themselves" (McCarthy et al. 2006: 12). It was also during this workshop that the term 'artificial intelligence' itself was coined. Since then, research in AI has progressed through a turbulent series of what are colloquially known as 'AI springs' and 'AI winters'. AI springs refer to periods of flourishing achievements, extensive research funding, and substantial attention and engagement from the general public. In contrast, AI winters refer to periods of disillusionment and apathy amongst funders, the general public, and AI researchers themselves towards the discipline. The longest AI winter lasted over a decade beginning in the mid-1980s, following widespread disappointment associated with expert systems (discussed in the following chapter) (see Nilsson 2009). Since the early 2010s, however, we have been in the midst of yet another AI spring that can largely be attributed to recent advances in ML.

ML refers to a subdiscipline of AI research concerned with "programming computers to optimize a performance criterion using example data or past experience" (Alpaydin 2014: 3). In short, ML involves developing machines that can learn from 'experience'. These systems are typically developed using several learning methods used to train algorithms to perform specific tasks through repeated exposure to relevant examples, including supervised learning, unsupervised learning, and reinforcement learning. Currently, supervised learning is the most common approach for developing ML systems (Esteva et al. 2019). It involves training an algorithm to accurately classify or predict outcomes using datasets containing 'labelled' images or examples. For example, suppose a developer wants to use supervised learning to develop an ML system that can reliably distinguish between images of motorbikes and images of stop signs. The developer would first prepare a training dataset by collecting a large number of example images that contain either motorbikes or stop signs. The developer would then ensure that each of these images are correctly labelled as containing either a motorbike or a stop sign. Finally, the developer would initiate the training process by feeding these training examples through the ML algorithm, whereby the algorithm would gradually improve its reliability in accurately classifying these images over time through trial and error.

Unsupervised learning and reinforcement learning are also being used to develop medical ML systems. Currently, however, they are substantially less popular than supervised learning approaches. In direct contrast to supervised learning, unsupervised learning involves training ML algorithms on *unlabelled* datasets to group data containing similar features or consolidate the number of features in a dataset. As Handelman and co-authors express:

> In unsupervised learning, the computer is provided with unclassified data records to recognize and determine whether any existing latent patterns are present, sometimes producing both answers and questions that may not have been conceived by the investigators (Handelman et al. 2018: 606).

For instance, unsupervised learning methods are used to develop ML systems that can detect similarities between groups of patients according to patterns in their disease characteristics, treatment responses, or disease progressions. Reinforcement learning, moreover, involves training an ML algorithm to maximise reward functions and "optimize sequences of decisions for long-term outcomes" (Gottesman et al. 2019: 16). In particular, reinforcement learning algorithms learn from data collected from their (real or simulated) environment and adjust their performance according to how well they achieve certain pre-defined objectives. For instance, reinforcement learning can be used to improve robotic-assisted surgery by learning from the movements of human surgeons (Esteva et al. 2019).

As noted previously, health data has never been more widely collected and readily available. Despite this, developers of medical ML systems face substantial challenges with respect to accessing and preparing data for training supervised learning systems. In particular, the heavy data labelling requirements of these systems pose a stubborn and persistent obstacle to the development of supervised learning systems. This is because their training datasets must often contain hundreds of thousands, if not millions, of accurately labelled training examples (Marcus 2018). However, manually labelling hundreds of thousands of training examples is time-consuming and expensive. While recruiting cheap labour through crowdsourcing platforms such as Mechanical Turk can ease the burden of data labelling, medical training examples must be labelled by persons with medical knowledge and domain expertise, which can greatly reduce the pool of qualified candidates.

As a result, developers of medical ML systems may turn increasingly toward unsupervised learning methods in the near to mid-future (see Marcus 2018). However, some experts argue advocate for the use of alternative learning methods that circumvent these obstacles without abandoning supervised learning entirely, such as transfer learning or self-supervised learning (Krishnan, Rajpurkar, and Topol 2022). For instance, transfer learning involves retraining existing ML systems to perform tasks in domains where large, high-quality datasets are unavailable (Pan and Yang 2009). In contrast to traditional supervised learning methods, in which developers create medical ML systems from scratch, transfer learning allows developers to leverage the prior learning of existing ML systems by fine-tuning the performance of these systems to adjacent tasks in new domains. For instance, transfer learning has been used to retrain an ML system designed to classify everyday objects into a system that detects breast cancer in histopathological images (Khan et al. 2019). Moreover, self-supervised learning incorporates both supervised and unsupervised learning methods into the development of medical ML systems. Specifically, unsupervised learning is used to automatically generate labels for training data which is then used to train a traditional supervised learning algorithm (Krishnan et al. 2022).

Despite the broad range of computational architectures used to develop medical ML systems (e.g. support vector machines, random forests, Bayesian models, etc.; see Handelman et al. 2018; Jiang et al. 2017), deep learning systems currently dominate the field. Deep learning refers to a subfield of ML that is concerned with developing "computational models that are composed of multiple processing layers [that can] learn representations of data with multiple levels of abstraction" (LeCun, Bengio, and Hinton 2015: 436). Deep learning systems are developed using a computational architecture known as deep neural networks. As Geoffrey Hinton, who is widely known as the 'godfather of deep learning', has expressed, deep neural networks are "inspired by the ability of brains to learn complicated patterns in data by changing the strengths of synaptic connections between neurons" (Hinton 2018: E1). In particular, deep neural networks consist of a series of layered and interconnected nodes, known as artificial neurons, that each hold statistical weightings, learns specific features from a dataset, and influence a system's final outputs. These artificial neurons are organised into at least three distinct layers – an input layer, a 'hidden' layer, and an output layer. What distinguishes 'deep' neural networks from their 'shallow' counterparts is that deep neural networks contain multiple hidden layers that improve their capacity to, "like the visual cortex, learn a hierarchy of progressively more complex feature detectors" (Hinton 2018: E1). Currently, deep neural networks are recognised as the dominant computational architecture for tasks involving image- and objection-detection (LeCun et al. 2015).

Deep learning systems generated enormous excitement amongst AI researchers following the 3rd Annual ImageNet Large Scale Visual Recognition Challenge (ILSVRC), in which participants compete to develop AI systems that can correctly label the highest number of images from a dataset of over 14 million examples. During this challenge, a deep learning system known as 'AlexNet' shattered previous records by over 10% "which blew every other competitor out of the water and shocked the computer-vision community" (Mitchell 2019: 129; see also Alom et al. 2018). While deep learning has proven to be a particularly popular method for developers of medical ML systems (Avati et al. 2018; Esteva et al. 2019, 2021; Miotto et al. 2017; Shickel et al. 2018), these systems are also 'opaque' insofar as clinicians cannot evaluate the systems' reasoning against their own knowledge and expertise. The popularity of deep learning in medicine presents major obstacles to understanding and communication between clinicians and patients, as I argue in chapter four.

## 2.  The current state of medical machine learning

A rapidly expanding arsenal of medical ML systems has recently emerged to assist clinicians in performing a wide range of clinical tasks. By the time this thesis is assessed, it is likely that the overview of clinical applications of medical ML systems provided in this section will be out of date. My aim in this section is to provide just a snapshot of the current state of ML systems in medicine. I discuss a range of popular clinical applications of ML systems that have been designed to assist in a variety of tasks including diagnosis, risk prediction, patient triage, and patient monitoring. I also discuss several non-clinical applications of ML systems in the domains of clinical research, medical education and training, and healthcare administration. While this overview of medical ML systems is not intended to be exhaustive, interested readers may wish to consult the regularly updated archive of cutting-edge advances in medical ML systems that is maintained by Emma Chen, Pranav Rajpurkar, and Eric Topol on their Doctor Penguin blog.[2] In addition, *The Medical Futurist* currently maintains an up-to-date categorised list of medical AI systems approved by the US FDA.[3]

"Currently," according to Kun Hsing Yu and co-authors, "automated medical-image diagnosis is arguably the most successful domain of medical AI applications" (Yu, Beam, and Kohane 2018: 722). An expansive range of diagnostic ML systems have been developed for use across the entire spectrum of medical specialisations. For instance, various diagnostic ML systems have been developed for use in ophthalmology to diagnose diabetic retinopathy from retinal fundus images (Beede et al. 2020; Gulshan et al. 2016; van der Heijden et al. 2018). Many diagnostic ML systems have also been developed for use in dermatology to diagnose keratosis, carcinomas, and melanomas from dermatoscopy images of skin lesions (Esteva et al. 2017;

---

[2] https://doctorpenguin.com/

[3] https://medicalfuturist.com/fda-approved-ai-based-algorithms/

Tschandl et al. 2020), and in oncology to diagnose malignant breast lesions from mammogram images (Bejnordi et al. 2017; Kooi et al. 2017; McKinney et al. 2020). Finally, diagnostic ML systems have been developed for use in radiology to diagnose pulmonary tuberculosis, pneumonia, COVID-19, and various other common lung diseases from chest x-ray images (Ozturk et al. 2020; Patel et al. 2019; Rajpurkar et al. 2017; Yu et al. 2018; Zech et al. 2018).

Indeed, diagnostic ML systems constitute the vast majority of medical ML systems that have already received regulatory approval by government agencies such as the US FDA (Lyell et al. 2021). For instance, IDx-DR by Digital Diagnostics, a DL system designed as a screening tool to detect more-than-mild diabetic retinopathy from diabetic patients' retinal fundus images, received FDA approval in 2018 (US Food and Drug Administration (2018a) and is credited as the first FDA-approved system "that provides a screening decision without the need for a clinician to also interpret the image or results" (US Food and Drug Administration 2018b: 1). FerriSmart Analysis System by Resonance Health Analysis Services, an ML system that automatically detects abnormalities in patients' liver iron concentration levels, also received FDA approval in 2018 (US Food and Drug Administration 2018c). More recently, PowerLook Density Assessment by iCAD, a DL system designed to measure the density of patients' breast tissue from mammography images, received FDA approval in 2021 (US Food and Drug Administration 2021).

Despite the popularity of diagnostic ML systems, a broad range of ML systems to assist clinicians in predictive tasks have also been developed for use in a variety of medical specialisations. For instance, various predictive ML systems have been developed for use in oncology, to predict short-term breast cancer risk from mammograph images and short-term risk of disease progression in existing breast cancer patients (Ferroni et al. 2019; Heidari et al. 2018). Predictive ML systems have also been developed for use in emergency medicine and intensive care units to predict patients' risk of short-term hospital readmission (Caruana et al. 2015; Jamei et al. 2017; Min, Yu, and Wang 2019) and short- to long-term mortality risk (Avati et al. 2018; Yue Gao et al. 2020; Thorsen-Meyer et al. 2020). In surgical operating rooms, ML systems have also been developed to predict the risk of intraoperative complications, postoperative complications, and 30-day mortality in patients undergoing major surgery (Bihorac et al. 2019; Lee et al. 2018; Lundberg et al. 2018).

ML systems have also been developed to assist clinicians and even automate tasks associated with triaging patients for urgent review. For instance, several ML systems have been designed for use in emergency medicine to assist in nurse triage by predicting patients' likelihood of hospital admission and risk of health complications (Goto et al. 2019; Hong, Haimovich, and Taylor 2018; Levin et al. 2018). Many triage ML systems designed for use in radiological settings have already received regulatory approval by the US FDA. For instance, Critical Care Suite by GE Medical Systems, a DL system designed to identify and prioritise patients with suspected indicators of pneumothorax, received FDA approval in 2019 (US Food and Drug Administration 2019b). CmTriage by CureMetrix, an ML system designed to automatically scan mammogram images and notify the clinician of cases with one or more suspicious findings, also received FDA approval in 2019 (US Food and Drug Administration 2019a). More recently, Briefcase by Aidoc, a suite of ML-based products designed to assist in prioritising time-sensitive cases for radiologists by identifying suspected instances of issues (e.g. acute cervical spine fracture, large vessel occlusion, intracranial haemorrhage, and pulmonary embolism), received FDA approval in 2022 (US Food and Drug Administration 2022).

There are also ML systems that automate patient monitoring tasks in both clinical and non-clinical contexts. For instance, a variety of ML systems have been developed to continuously monitor blood sugar levels and predict the onset of hypoglycemic events for diabetic patients as they go about their daily lives outside clinical settings (Porumb et al. 2020). Other ML systems have been created for use in aged care settings to detect when patients or residents have fallen (Hussain et al. 2019). Moreover, several ML systems have been designed for use in clinical settings to monitor patients' movements in order to promote early patient mobilisation (Haque, Milstein, and Fei-Fei 2020).

While FDA approvals of ML systems for continuous monitoring are relatively rare at present (Lyell et al. 2021), a variety of these systems have nevertheless received regulatory approval in the US. For instance, WAVE Clinical Platform by Excel Medical Electronics, an ML system designed to continuously monitor patient vital signs and predict the onset of heart attack or respiratory failure, received FDA approval in 2018 (US Food and Drug Administration 2018d). Biovitals Analytics Engine by Biofournis Singapore, an ML system designed to monitor the vital signs of heart failure patients and alert clinicians of changes from the patient's baseline measurements, also received FDA approval in 2019 (Reynolds 2019). Moreover, BodyGuardian Remote Monitoring System by Preventice, an ML system designed to enable remote monitoring and detection of atrial fibrillation in patients with cardiac arrythmias, received FDA approval in 2020 (US Food and Drug Administration 2020).

In passing, a variety of ambient sensing ML systems have also been designed to continuously monitor and evaluate the performance and behaviour of clinicians. For instance, ML systems have been developed for use in hospital settings to track and monitor clinicians' hand-washing practices (Haque et al. 2017) and to evaluate their competence in the performance of various clinical tasks (Dias, Gupta, and Yule 2018). While ML systems for patient monitoring typically require regulatory approval prior to implementation and use, ML systems designed to continuously monitor the behaviour and performance of clinicians do not face these regulatory hurdles because they do not directly influence or impact patients' treatment or care. These workplace monitoring technologies have important implications for clinician-patient relationships that I discuss in chapter six.

Finally, proofs-of-concept for ML systems designed to assist clinicians in ethical decision-making are increasingly being developed. For instance, Lukas Meier and co-authors (2022) argue that current ML systems have the technical capacity to provide ethical decision-making support for human clinicians, and suggest that these systems could be particularly useful in emergency settings where rapid decision-making in high-stakes scenarios occurs on a regular basis. Moreover, several writers have argued that ML systems could, and indeed, ought to be used as tools that assist clinicians and surrogate decision-makers in making medical decisions for non-autonomous patients (Biller-Andorno and Biller 2019; Lamanna and Byrne 2018; Rid and Wendler 2014). Camillo Lamanna and Lauren Byrne (2018), for instance, argue that ML systems ought to be used to predict the medical preferences of non-autonomous patients using data collected from their social media profiles.

Clinical applications of ML systems are numerous and varied. However, a broad range of ML systems have also been developed for use in a variety of non-clinical settings, including clinical research, medical education and training, and healthcare administration. In clinical research, for instance, ML systems have been developed to assist in various tasks associated with drug design, discovery, and development (Dara et al. 2022; Vamathevan et al. 2019). In particular, ML systems have been developed to predict the effects of cancer drugs based on analyses of patterns in gene expression, DNA methylation, gene copy number alterations, and somatic mutation data (Iorio et al. 2016). Moreover, ML systems such as AlphaFold by DeepMind have been developed to assist in the design of pharmaceuticals through predicting protein structure matter (Jumper et al. 2021).

Beyond drug development, ML systems are also being devised to assist in tasks associated with designing, recruiting for, and assessing the quality of clinical trials. For instance, ML systems that predict the risk of early trial termination using information about trials' study characteristics and eligibility criteria have been creased (Kavalci and Hartshorn 2023). ML systems have also been developed to identify potential candidates for randomised clinical trials by matching patient health data collected in EHR records to trial eligibility criteria (Zhang et al. 2020).

In medical education and training, ML systems have proven particularly popular in surgical settings. For instance, ML systems (in combination with virtual reality technologies) have been developed to evaluate trainee surgeons' performance in basic surgical tasks (Rogers et al. 2021). ML systems have also been created to measure the performance of surgeons in completing robotic surgical procedures (Hung, Chen, and Gill 2018), and to deliver real-time intraoperative feedback to surgeons (e.g. automatically counting the number of surgical objects used to ensure that no objects are left inside the patient; see Haque et al. 2020). Other ML systems have been designed to deliver personalised predictions of surgical learning curves for individual students and practitioners (Gao et al. 2020), and create 'virtual patients' on which trainee clinicians can practice 'soft skills' (e.g. empathetic communication with patients; see Isaza-Restrepo et al. 2018).

ML systems are also being applied to tasks associated with healthcare administration. For instance, ML systems have been developed to assist in the optimisation of patient appointment scheduling by identifying patients with the greatest risk of non-attendance and scheduling them into overbooked slots to minimise inefficiencies (Srinivas and Ravindran 2018). ML systems have also been created to assist in allocating healthcare resources in hospital environments (e.g. operating rooms) by predicting clinical case durations and patient cancellations, and to immediately identify the number of healthcare workers on a ward at any given time (Bellini et al. 2020; Haque, Milstein, and Fei-Fei 2020). Finally, ML systems are being designed to automatically transcribe patient consultations (van Buchem et al. 2021) and predict demand for healthcare resources (e.g. post-anaesthetic care resources, hospital beds, and ventilators) (Belciug and Gorunescu 2015; Fairley, Scheinker, and Brandeau 2019; Bednarski, Singh, and Jones 2021).

As discussed previously, non-clinical applications of ML systems face few, if any, regulatory barriers in comparison to their clinical counterparts. Consequently, these various non-clinical applications of ML systems are likely to receive faster and more widespread adoption than clinical applications.

The explosion of activity in the development of medical ML systems that I have described in this section has generated soaring expectations amongst patients, clinicians, and policy makers concerning the potential benefits of these systems. In the next section, I provide an overview of these anticipated benefits and the associated vision for the coming age of AI in medicine that currently prevails in the scientific literature on the topic.

## 3. The coming age of medical artificial intelligence

A variety of world-renowned experts in AI and medicine, including Eric Topol (2019), Geoffrey Hinton (2018), and Abraham Verghese (see Israni and Verghese 2019; Verghese, Shah, and Harrington 2018) currently anticipate that medical ML systems will generate a range of benefits for patients, clinicians, and health systems at large. According to Eric Topol, for instance, medical ML systems are already generating benefits "for clinicians, predominantly via rapid, accurate image interpretation; for health systems, by improving workflow and the potential for reducing medical errors; and for patients, by enabling them to process their own data to promote health" (Topol 2019: 44). According to the vision for the coming age of AI in medicine that currently prevails amongst experts in the field, the use of medical ML systems will improve patient health and safety, health equity, and efficiency and productivity in medicine.

Medical ML systems are anticipated to generate substantial improvements to patient health and well-being by reducing the current rate of iatrogenic error in medicine, recently estimated to be the third leading cause of death in the US (Makary and Daniel 2016). In particular, experts anticipate that clinicians could greatly reduce their risk of error by using medical ML systems to generate second opinions, allowing these systems to direct their attention to relevant features of a clinical case, or considering alternative clinical hypotheses identified by these systems (Esteva et al. 2019; Graber 2022; Liu et al. 2019; Miotto et al. 2017; Topol 2019; Yu et al. 2018). According to a recent scoping review, moreover, medical ML systems could improve patient health and safety by reducing healthcare-associated infections, adverse drug events, surgical complications, and incidences of venous thromboembolism and pressure ulcers (Bates et al. 2021). Alvin Rajkomar and co-authors (2019) also anticipate that medical ML systems will improve patient health and well-being by enabling clinicians to more quickly and accurately diagnose rare health conditions that are easily and often missed. Indeed, experts are especially optimistic about the potential impact of medical ML systems on patient health and safety given the impressive accuracy demonstrated by these systems in a variety of clinical tasks. The authors of a recent systematic review and meta-analysis of the performance of medical ML systems, for instance, "found deep learning algorithms to have equivalent sensitivity and specificity to health-care professionals" (Liu et al. 2019: e291).

Experts also anticipate that medical ML systems could improve patient health outcomes by facilitating substantial improvements in personalised medicine (Fröhlich et al. 2018;

Handelman et al. 2018; Meskó 2017; Sebastiani et al. 2022; Zhang et al. 2018). Personalised (or precision) medicine refers to a "field of health care that is informed by each person's unique clinical, genetic, genomic, and environmental information" (Chan and Ginsburg 2011: 217). Experts anticipate that medical ML systems will enable greater personalisation in risk prediction tasks and the recommendation of complex treatments and interventions due to their ability to detect subtle patterns in largescale, heterogeneous datasets (Khan et al. 2020; Ozer, Sarica, and Arga 2020; Price 2015; Weiss et al. 2012). Experts are particularly optimistic about the potential of medical ML systems to improve personalised medicine due to their capacity to tailor their outputs and recommendations to small clusters of patients with similar clinical, genetic, genomic, and environmental characteristics. Indeed, as Jack Wilkinson and co-authors observe, the "potential to revolutionise the individual tailoring of medical treatments is one of the most widely discussed and appealing promises of machine learning-powered precision medicine" (Wilkinson et al. 2020: 2).

In addition to improving patient health and safety, medical ML systems are also anticipated to improve health equity. In particular, Danton Char and co-authors (2018) suggest that medical ML systems could offset and compensate for known biases in human clinicians that contribute to current health disparities (Chapman, Kaatz, and Carnes 2013; Char et al. 2018; Hoffman et al. 2016; Salles et al. 2019). For instance, racial biases in the administration of pain medication often result in Black patients receiving comparatively less pain relief than white patients (see Hoffman et al. 2016). Char and co-authors suggest that medical ML systems could be designed to offset such biases in human clinicians, thereby reducing such disparities (see also Pierson et al. 2021).

Experts also anticipate that medical ML systems will improve health equity by augmenting the quality of medical care available to underserviced and resource-poor communities (Mehta, Katz, and Jha 2020; Scheetz et al. 2021; Wahl et al. 2018). According to Topol (2019a), for instance, this is because medical ML systems will improve the quality of treatment and care that clinicians can provide remotely, using telepresence technologies. Indeed, Topol even anticipates that medical ML systems could improve remote caregiving capabilities to the point that many hospitals could deliver care almost entirely remotely (Topol 2019; see also Wachter 2015; Allen 2017). Moreover, Jonathan Guo and Bin Li (2018) argue that medical ML systems will improve the quality of care available to underserviced communities by enabling non-specialist clinicians to perform specialist clinical services that are typically unavailable to underserviced populations (see also Susskind and Susskind 2015). Saurabh Jha and Eric Topol (2023) even suggest that medical ML systems may be implemented in lower and middle-income countries before they receive widespread adoption in high-income countries. "Perhaps one day," they claim, "US health care might implement AI for chest x-rays just to catch up with Africa" (Jha and Topol 2023: 1920).

Experts also anticipate that ML systems will generate substantial improvements in efficiency and productivity in medicine. In particular, medical ML systems are anticipated to improve time efficiency by allowing clinicians to speed up the performance of clinical tasks. For instance, according to Rajkomar and co-authors (2019), these systems could improve time efficiency by helping to "expose relevant information in a patient's chart for a clinician without multiple clicks" (Rajkomar et al. 2019: 1353). Moreover, experts anticipate that medical ML systems will improve time efficiency by allowing healthcare practitioners to speed up and even automate the performance of many administrative tasks. For instance, according to Rajkomar and co-authors:

> Data entry of forms and text fields can be improved with the use of machine-learning techniques such as predictive typing, voice dictation, and automatic summarization. Prior authorization could be replaced by models that automatically authorize payment based on information already recorded in the patient's chart (Rajkomar et al. 2019: 1353; see also Lenert, Lane, and Wehbe 2023).

In addition, medical ML systems are anticipated to improve cost efficiency in medicine. According to Topol (2019a), for instance, clinicians' use of medical ML systems could assist in optimising the use of scarce healthcare resources by reducing the high rate at which clinicians currently order unnecessary tests and scans. As Nittas and co-authors observe, experts in the scientific literature also anticipate that medical ML systems will "lower direct and indirect costs through time and diagnostic efficiency, automation, and enhanced workflows" (Nittas et al. 2023: 6).

According to this vision for the future of AI in medicine that currently prevails in the literature, medical ML systems will improve patient health and safety, health equity, and efficiency and productivity. As discussed above, however, Eric Topol argues that these are merely the "secondary gains" of the coming age of AI in medicine (Topol 2019a: 309). This is because, according to Topol, the coming age of AI is "our chance, perhaps the ultimate one, to bring back real medicine: Presence. Empathy. Trust. Caring. Being Human" (Topol 2019a: 309). In the remainder of this thesis, I argue that Topol's vision for the future of clinician-patient relationships is fundamentally misguided, and that medical ML systems are likely to negatively impact on these relationships in several ways. In particular, medical ML systems are likely to compromise the quality of trust, care, empathy, and understanding between clinicians and patients. To protect and preserve clinician-patient relationships in the coming age of AI in medicine, stakeholders must therefore resist being carried away by excitement over the anticipated benefits of these systems, and think substantially more about the costs they are likely to impose.

## 4. Chapter overview

The remainder of this thesis proceeds as follows. In chapter one, I argue that the prevailing vision for the age of AI in medicine, discussed in the previous section, exaggerates the likelihood that medical ML systems will deliver on their anticipated benefits, and discounts the risks generated by their use. I also suggest that many of the risks identified in this chapter are likely to have indirect negative effects on the quality of clinician-patient relationships, discussed at several points throughout the thesis.

In chapter two, I begin my analysis of the direct impact of medical ML systems on clinician-patient relationships. I argue in this chapter that the use of medical ML systems is likely to compromise the quality of trust between clinicians and patients. In particular, I suggest that this is due to the fact that medical ML systems are not the appropriate objects of trust, and because describing humans' relationships with medical ML systems using the language of trust is likely to interfere with the attribution of responsibility for patient harm resulting from the use of these systems. This chapter includes my sole-authored article, 'Limits of trust in medical AI', published in the *Journal of Medical Ethics*.

In chapter three, I address another way in which the use of medical ML systems is likely to impact negatively on the quality of clinician-patient relationships. In particular, I turn to discuss clinicians' communicative obligations with respect to medical ML systems in the context of clinician-patient interactions, and I consider whether clinicians are ethically obligated to disclose their use of medical ML systems to patients. I argue that clinicians *are* ethically obligated to disclose their use of medical ML systems for treatment recommendation to secure their patients' informed consent. I also suggest that clinicians are ethically obligated to disclose their use of medical ML systems, regardless of informed consent requirements, due to the risks these systems present to patient safety and patient privacy, and to enable patients to exercise their right to refuse diagnostics and treatment planning by these systems.

In chapter four, I turn to discuss how opacity in medical ML systems is likely to negatively impact the quality of clinician-patient relationships. In particular, I argue that the use of opaque medical ML systems is likely to compromise the quality of communication between clinicians and patients due to their impact on patient understanding and shared decision-making in medicine. I also argue that a blanket prioritisation of accurate but opaque medical ML systems over comparatively interpretable medical ML systems is unjustifiable, even where superior accuracy is demonstrated by the former. This chapter includes my article, 'The virtues of interpretable medical AI', co-authored with Robert Sparrow and Mark Howard and published in the *Cambridge Quarterly of Healthcare Ethics*.

In chapter five, I turn to discuss how medical ML systems that continue learning from new data even after being deployed in a clinical setting, otherwise known as 'adaptive' ML systems, are likely to impact negatively on the quality of clinician-patient relationships. In particular, I argue that the use of adaptive ML systems is likely to increase clinicians' hermeneutic and administrative labour, and expand existing risks to patient health and well-being that is likely to compromise patient trust. This chapter includes my article, 'Diachronic and synchronic variation in the performance of adaptive machine learning systems: the ethical challenges', co-authored with Robert Sparrow and published in the *Journal of the American Informatics Association.*

In chapter six, I turn to discuss the impact of medical ML systems on care and empathy in medicine, directly addressing Topol's arguments supporting the claim that medical ML systems will revolutionise clinician-patient relationships. *Contra* Topol, I argue that the use of medical ML systems is likely to compromise the quality of care and empathy in medicine by expanding the administrative responsibilities of human clinicians, promoting burnout and professional dissatisfaction amongst clinicians, and generating physical and psychological distance between clinicians and patients.

I conclude that, rather than facilitating a renaissance era in the clinician-patient relationships, it is currently more likely that medical ML systems will expand and intensify a host of obstacles and challenges that will further compromise the quality of these relationships. Consequently, I suggest that AI developers, healthcare organisations, and policy makers need to think more about the costs of medical ML systems on the relationship between clinicians and their patients to avoid being carried away by exaggerated assessments of their potential benefits, which could result in substantial harm.

## (1)   EXAGGERATED BENEFITS, DISCOUNTED RISKS

### 1.   Introduction

As discussed in the introduction to this thesis, experts anticipate that the coming age of AI in medicine will deliver substantial improvements to patient health and safety, health equity, and efficiency in medicine. For instance, experts anticipate that medical ML systems will reduce the current rate of medical error, increase access to high-quality resources and treatment amongst underserviced populations, and enable clinicians to perform clinical and administrative tasks more efficiently (Nittas et al. 2023; Rajkomar, Dean, and Kohane 2019 Rajpurkar et al. 2022; Topol 2019b). In this chapter, however, I suggest that there are reasons to doubt that this vision for the coming age of AI in medicine is likely to materialise.

While Topol (2019a) argues that the benefits of medical ML systems for patient health and safety, health equity, and efficiency are merely the "secondary gains" of the coming age of AI in medicine, they are nevertheless a critical factor in the current hype surrounding medical ML systems that this thesis aims to contest. Indeed, the risks that medical ML systems present (discussed later in this chapter) are likely to exacerbate the negative effects that these systems are likely to have on the quality of clinician-patient relationships. Before turning to address the impact of medical ML systems on clinician-patient relationships directly in the following chapters, therefore, I analyse this broader vision for the coming age of AI in medicine, according to which patients will be safer and healthier, healthcare more equitable, and the delivery of medical care more efficient. I argue that this vision for the coming age of AI in medicine exaggerates the likelihood that medical ML systems will deliver on these anticipated benefits. I also argue that this vision for the coming age of AI discounts several risks that medical ML systems present, which are likely to compromise the overall positive impact of these systems for patients.

The remainder of this chapter proceeds as follows. In section two, I suggest that the recent history of technological innovation in medicine (including previous iterations of medical AI systems themselves) inspire little confidence in the capacity for medical ML systems to deliver on their promise to revolutionise the practice of medicine and the delivery of healthcare services. In section three, I argue that medical ML systems themselves generate new or expanded risks to patient health and safety, health equity, and efficiency, i.e. the very benefits that experts anticipate medical ML systems will deliver. In section four, I argue that experts also discount the substantial and underappreciated threats that medical ML systems present to patient privacy, accountability for patient harm, and the integrity of decision-making that threaten to generate net harm to patients. Finally, in section five, I conclude that accurately assessing the risks and benefits of medical ML systems requires substantially greater engagement with the risks that these systems present.

## 2. Lessons from the recent history of technological innovation in medicine

Medical ML systems are far from the first technology to stoke hopes for a revolution in medicine and healthcare, nor will they be the last. However, the recent history of technological innovation inspires little confidence that medical ML systems will deliver on these promises due to a series of recently disappointed expectations.

Consider, for instance, that early advances in computer systems in the late 20th century inspired enormous enthusiasm about the benefits that these systems could deliver for patient health and safety, health equity, efficiency, and clinician-patient relationships. In a highly influential article published in 1970, William Schwartz claimed that:

> it seems probable that in the not too distant future the physician and the computer will engage in frequent dialogue, the computer continuously taking note of history, physical findings, laboratory data, and the like, alerting the physician to the most probably diagnoses and suggesting the appropriate, safest course of action. One may hope that the computer, well equipped to store large volumes of information and ingeniously programmed to assist in decision making, will help free the physician to concentrate on the tasks that are uniquely human such as the application of bedside skills, the management of the emotional aspects of disease, and the exercise of good judgement in the nonquantifiable areas of clinical care (Schwartz 1970: 1258).

Experts acknowledge that computer systems have since failed to deliver on these promises (Wears and Berg 2005; see also Topol 2019a: 16–17). However, medical ML systems now appear to have neatly replaced computers as 'the' technology that is pegged to achieve these objectives. Indeed, one could simply replace Schwartz's references to 'computer systems' with that of 'medical ML systems' to produce something that is eerily close to contemporary claims about the benefits that experts currently expect medical ML systems to deliver. Consider, for instance, this recent quote from Eric Topol:

> The promise of artificial intelligence in medicine is to provide composite, panoramic views of individuals' medical data; to improve decision making; to avoid errors such as misdiagnosis and unnecessary procedures; to help in the ordering and interpretation of appropriate tests; and to recommend treatment (Topol 2019a: 9).

Computer systems are not the only 'revolutionary' technology to disappoint expectations in medicine. Indeed, high hopes concerning previous iterations of medical AI systems themselves have also resulted in substantial disappointment in the recent past. For instance, expert systems – an early type of AI system developed by formalising the knowledge of human experts into a series of 'IF-THEN' rules – generated soaring expectations in medicine throughout the 1970s and '80s (Nilsson 2009). During this period, a wide range of expert systems were developed for use in medicine including INTERNIST-I and MYCIN, which were expert systems designed, respectively, to generate differential diagnoses and to determine the cause of severe infections and generate personalised treatment recommendations (Miller, Pople, and Myers 1982; Shortliffe and Buchanan 1975). Rather than causing a revolution in the practice of medicine, however, medical expert systems were either relegated to providing meagre decision-support assistance in drug prescribing tasks[1] or simply never implemented in clinical practice to begin with, due to a range of ethical concerns and practical obstacles (Nilsson 2009; Zhou and Sordo 2021). For instance, in her account of the history of medical expert systems, Heather Heathfield (1999) highlights that medical expert systems did not receive widespread uptake due to patient safety risks, difficulties establishing clinical efficacy, obstacles to administrative and workflow integration, maintenance costs, and excessive data entry demands generated by these systems (Heathfield 1999; Zhou and Sordo 2021). The performance of medical ML systems now exceeds that of medical expert systems in clinical tasks. However, each of the practical and institutional obstacles identified by Healthfield continue to pose substantial challenges for hospitals, AI developers, and clinicians with respect to medical ML systems (see Maddox, Rumsfeld, and Payne 2019; Morse, Bagley, and Shah 2020; Sandhu et al. 2020).

The cases of computer systems and medical expert systems each provide a cautionary tale for clinicians, hospitals, and AI developers with respect to the coming age of AI in medicine.

---

[1] Even in this narrow domain, recent assessments of the clinical impact of medical expert systems have been mixed at best (Black et al. 2011).

However, computer systems and expert systems are only two examples taken from an ever-expanding list of strongly hyped medical technologies that have failed to deliver on their promises. As Jianxiang He and co-authors observe, "gene therapy, genomic-driven personalized medicine, and EHRs are all technologies that were purported to deliver revolutionary improvements in the delivery of health-care, but thus far many have felt that their potential has exceeded their performance" (He et al. 2019: 33). As such, the recent history of technological innovation in medicine inspires little confidence that medical ML systems will deliver on their promise to revolutionise the practice of medicine and the delivery of healthcare services. With respect to these systems, it may be more prudent to take our cue from history rather than bet on a revolution (see Tabery 2023).

Returning to the present day, it is worth noting that there is currently little to no evidence that using medical ML systems generates improved patient health outcomes (Beam, Manrai, and Ghassemi 2020; He et al. 2019). Despite great excitement about the reported accuracy of these systems in clinical tasks, accuracy is simply not enough to improve patient health. As Jonathan Chen and Steven Asch observe, "even a perfectly calibrated prediction model may not translate into better clinical care. An accurate prediction of a patient outcome does not tell us what to do if we want to change that outcome" (Chen and Asch 2017: 2508). Despite this, most studies of medical ML systems have so far been limited to demonstrating proofs-of-concept rather than improved patient health. According to several recent systematic reviews, between 87%-97% of studies of medical ML systems have been conducted on retrospective datasets alone, which offer no concrete evidence that these systems improve patient health (Aggarwal et al. 2021; Ben-Israel et al. 2020; Liu et al. 2019; Nagendran et al. 2020; Wu et al. 2021). Few randomised controlled trials have also been conducted to date, and even amongst those that have been conducted, "most did not fully adhere to accepted reporting guidelines and had limited inclusion of participants from underrepresented minority groups" (Plana et al. 2022: 1). Consequently, as Nagendran and co-authors suggest:

> at present, many arguably exaggerated claims exist about equivalence with or superiority over clinicians, which presents a risk for patient safety and population health at the societal level, with AI algorithms applied in some cases to millions of patients. Overpromising language could mean that some studies might inadvertently mislead the media and the public, and potentially lead to the provision of inappropriate care that does not align with patients' best interests (Nagendran et al. 2020: 11; see also Wilkinson et al. 2020: 2).

One explanation for these exaggerated assessments of the accuracy of medical ML systems may be that recent assessments of the risks and benefits of medical ML systems place disproportionate emphasis on the anticipated benefits of medical ML systems over their risks. For instance, several recent studies of AI in the news media have found that journalists tend to detail the benefits of these technologies while engaging minimally, if at all, with their risks. Colin Garvey and Chandler Maskal (2020) found that news media coverage of AI strongly skews strongly towards positive over negative sentiment. Indeed, Ching-Hua Chuan and co-authors also found that even when journalists *do* discuss risks associated with the use of AI systems, they typically do "not discuss a particular ethical issue in-depth, but [raise] general questions about potential ethical concerns, such as privacy and misuse of AI in the title, introduction, or conclusion paragraph, without providing specific discussions" (Chuan, Tsai, and Cho 2019: 5).

Another explanation may be that individuals involved in assessing the risks of new technologies tend to treat the anticipated benefits of these technologies as concrete and assured, and their risks as merely speculative and hypothetical. As Sheila Jasanoff (2016) observes:

> the methods most commonly used to assess risk are not value-neutral but incorporate distinct orientations toward attainable and desirable human futures. One bias that risk assessment begins with is a tacit presumption in favor of change, that what is new should be embraced unless it entails insupportable harm as judged by the standards of today. Another is that good outcomes are knowable in advance, whereas harms are more speculative and hence can be discounted unless calculable and immediate (Jasanoff 2016: 35).

These tendencies to ought to be strongly resisted with respect to medical ML systems due to the threats these systems themselves present to the very benefits that they are anticipated to deliver, as I now discuss.

### 3. Threats to anticipated benefits

As discussed in the introduction, experts anticipate that medical ML systems will improve patient health and safety, health equity, efficiency, and most importantly, clinician-patient relationships. I delay further engagement with the impact of these systems on clinician-patient relationships to chapter two in order to evaluate the current hype surrounding their other anticipated benefits. In this section, I argue that experts exaggerate the likelihood that medical ML systems will deliver on these anticipated benefits since, paradoxically, medical ML systems themselves threaten to negatively impact patient health and safety, health equity, and efficiency in medicine.

Medical ML systems threaten patient health and safety due to a variety of technical weaknesses and limitations in the performance and reasoning capabilities of these systems. First, medical ML systems are insensitive to context in several ways that generate new and expanded risks of medical error and patient harm. For instance, as Robert Challen and co-authors observe, "ML systems can be poor at recognising a relevant change in context or data, and this results in the system confidently continuing to make erroneous predictions based on 'out-of-sample' inputs" (Challen et al. 2019: 232). This phenomenon is known as 'distributional shift', and it occurs due to changes in the statistical distribution of a system's target. For instance, distributional shift can occur due to changes in the patient demographic of a hospital over time or when a medical ML system is implemented in a new clinical environment. This is likely to result, and indeed has resulted (Lazer et al. 2014), in substantial deterioration in the performance of medical ML systems over time. Sub-optimal outcomes or patient harm may result where the outputs of such systems are acted on by clinicians.

Second, medical ML systems are also insensitive to the contextual risks of a clinical case and cannot adapt their reasoning accordingly. For instance, while a clinician may err on the side of caution when diagnosing particularly serious medical conditions (e.g. cancer) by diagnosing these conditions more liberally, a medical ML system's diagnostic threshold remains static and unchanging throughout its lifecycle, which may increase the current rate of false negatives in the diagnosis of high-stakes medical conditions. Medical ML systems also lack situational awareness of broader contextual details of a clinical case, which may lead them to generate misguided or erroneous outputs and recommendations. As Mark Sujan and co-authors express:

> An autonomous infusion pump needs to know if the patient receives other medications that might affect the patient's physiology and response. […] The saying 'if it's not documented, it didn't happen' applies here with critical consequence: if there are relevant activities going on that are not documented and communicated to the autonomous agent ([e.g.] infusion pump), then as far as the AI is concerned, these literally did not happen because the system has no way of knowing about it. The results could be catastrophic (Sujan et al. 2019: 4).

Medical ML systems are also solely reliant on correlation over causation, which can compromise the robustness of their overall performance and can lead them to generate dangerous outputs or recommendations. For example, a medical ML system designed to detect pneumonia from chest radiographs was found to have been generating outputs on the basis of the type of scanner being used rather than the content of the medical images themselves (Zech et al. 2018). Moreover, an ML system designed to predict the mortality risk of patients presenting to the emergency department with pneumonia was found to classify asthmatic patients as low risk, despite their objectively higher risk of mortality than other patient cohorts, since these patients are typically referred immediately to intensive care (Caruana et al. 2015). If implemented in practice, this latter system could have delayed urgently needed treatment for asthmatic patients, leading to substantial patient harm or even death.

Medical ML systems also threaten patient health and safety due to a range of risks in the training and development of these systems. For instance, data leakage and variable confounding often occur in the development of medical ML systems which can compromise the performance of these systems once implemented in a real-world setting. As Cynthia Rudin and Joanna Radin express, *data leakage* occurs when:

> information about the label *y* sneaks into the variables *x* in a way that you might not suspect by looking at the titles and descriptions of the variables: sometimes you think you are predicting something in the future but you are only detecting something that happened in the present. In predicting medical outcomes, the machine might pick up on information within doctors' notes that reveal the patients' outcome before it is officially recorded and hence erroneously claim these as successful predictions (Rudin and Radin 2019: 5).

In contrast, *variable confounding* occurs when dependent and independent variables in a causal relationship are both affected by a third variable that can artificially inflate the accuracy of a medical ML systems during the validation phase. For instance, the type of radiographical scanner used (e.g. portable or non-portable) is a confounding variable for medical ML systems designed to diagnose pneumonia from radiographical images. This is because medical ML systems can distinguish between radiographical images generated by each type of system, and pneumonia is more likely to be detected where portable scanners are used (Zech et al. 2018). Data leakage and variable confounding present serious risks to patient safety by allowing medical ML systems to appear accurate during the development and testing phases. However, their performance deteriorates once implemented in a real clinical setting, resulting in dangerous or misguided outputs.

Critics may object that clinicians will be able to overcome these technical weaknesses and limitations in medical ML systems by evaluating the system's reasoning against their own knowledge and expertise. As discussed in the introduction to this thesis, however, medical ML systems are often 'opaque' insofar as patients, clinicians, and even the designers of these systems themselves are unable to understand how a medical ML system arrives at its particular outputs. This characteristic of many ML systems impedes clinicians' ability to evaluate the reasoning of a medical ML systems against their own knowledge and expertise. These technical weaknesses and limitations in medical ML systems present acute threats to patient health and safety, since clinicians will often be unable to detect when errors occur as a result of them (He et al. 2019; Yoon, Torrance, and Schneiderman 2021). I discuss the further implications of this characteristic of medical ML systems in chapter four.

Medical ML systems also generate new and expanded risks to patient health and safety due to a range of common biases exhibited by human users of algorithmic and automated systems (see Kostick-Quenet and Gerke 2022; Sujan et al. 2019). Several of these biases may lead clinicians to over-rely on medical ML systems and their outputs. For instance, 'automation bias' refers to "omission and commission errors resulting from the use of automated cues as a heuristic replacement for vigilant information seeking and processing" (Mosier et al. 1998: 47; see also Lyell and Coiera 2017). Troublingly, several recent studies have found that clinicians of all levels of experience exhibit automation bias towards medical ML systems (Bond et al. 2018; Dratsch et al. 2023; Uyumazturk et al. 2019; Wang et al. 2023). In contrast, medical ML systems may also promote 'automation complacency' amongst clinicians, which occurs when an individual's performance deteriorates as their role shifts from performing a task themselves to supervising the task's automation (Bailey and Scerbo 2007). Conversely, other biases may lead clinicians to reject the outputs of medical ML systems over their own clinical judgements in instances where the system is correct. For instance, algorithmic aversion refers to "the reluctance of human decision makers to use superior but imperfect algorithms" (Burton, Stein, and Jensen 2020: 220). Concerningly, clinicians are particularly likely to exhibit algorithmic aversion. This is because clinicians, particularly experienced clinicians, have 'domain expertise' in the tasks medical ML systems are designed to perform and users with domain expertise have been found to exhibit algorithmic aversion more frequently than users without domain expertise. Psychologists, for instance, have recently been found to prioritise human-generated clinical support tools over AI-generated ones, even where the human-generated tools are incorrect (Maslej et al. 2023).

Medical ML systems could also trigger narrower cognitive biases in human clinicians that generate new or expanded risks to patient health and safety. For instance, medical ML systems are likely to expand the range of options that clinicians are prompted to consider in clinical decision-making scenarios. This is likely to trigger a cognitive bias in which clinicians make decisions that they would otherwise reject were they to consider fewer options. In one study, for instance, Donald Redelmeier and Eldar Shafir (1995) found that, when deciding between two treatment options ($x$ and $y$), the addition of a third option ($z$) would often lead participants to change their preference between the two previously available options (e.g. a participant who initially selected treatment $x$ is likely to select treatment $y$ when given the additional option of $z$). According to Redelmeier and Shafir:

> The increased tendency to select a previously available option when facing a greater number of competing alternatives appears to be a cognitive bias: preference between two options shifts due to the availability of a third option that increases the difficulty of making a choice but is itself not chosen (Redelmeier and Shafir 1995: 304).

Clinicians may therefore select treatment options that they would otherwise have rejected were they not prompted to consider more options by a medical ML system. Medical ML systems could also interact with various other narrow cognitive biases and heuristics that could generate further threats to patient health and safety (see Kahneman 2013). Moreover, these biases could have compounding negative effects on the quality of clinicians' judgements.

It is also worth noting that new medical technologies are, inevitably, implemented in an economic and institutional system characterised by resource scarcity and opportunity costs. The decision to devote time, financial, cognitive, and institutional resources to developing and implementing new medical technologies entails *not* devoting these resources to other projects, innovations, or improvements. This is a problem for medical ML systems because, as Adam Henschke (2015) has argued, the resource investment that new technologies typically require will often be unjustified when appraised against the opportunity costs associated with their implementation (I discuss the resource investments of medical ML systems in more detail shortly). The current hype surrounding medical ML systems may therefore draw attention and resources away from low-tech strategies that may have greater potential to, for instance, improve patient health and safety. "In an action-oriented society," as Emily Mumford and co-authors have expressed, "reports of the considerable effects of modest interventions may command less attention than reports of the modest effects of more flamboyant interventions" (Mumford, Schlesinger, and Glass 1982: 144). This is particularly true of new medical technologies due to the sense of wonder, fascination, and power that they tend to inspire in their users (see Cassell 1997; Leff and Finucane 2008).

Moreover, medical ML systems generate new and expanded threats to health equity. For instance, diversity in the AI and technology industries is notoriously poor. As Sarah Myers West and co-authors (2019) report, women comprise only 10-15% of AI research staff at Facebook and Google; black workers make up a mere 2.5-4% of staff at Google, Microsoft, and Facebook; and public data on gender minorities (e.g. transgender and non-binary people) is simply unavailable. Underrepresentation of marginalised communities in the AI and technology industries is likely to result in the interests of sociopolitically and socioeconomically powerful groups being prioritised in the design and development of medical ML systems (see Nature Machine Intelligence 2020). This is due to what D'Ignazio and Klein refer to as the 'privilege hazard', which refers to "the phenomenon that makes those who occupy the most privileged positions among us – those with good educations, respected credentials, and professional accolades – so poorly equipped to recognise instances of oppression in the world" (D'Ignazio and Klein 2020: 29). For instance, according to the World Health Organization (2022), medical ML systems threaten to exacerbate ageism in medicine through exclusionary design due to the tendency for ML systems to be designed "on behalf of older people instead of with older people" (World Health Organization 2022: 8).

Many developers of medical ML systems will also rely on funding bodies for medical research and innovation for financial backing and support. However, like AI and technology organisations, funding bodies for medical research and innovation tends to prioritise the interests of socioeconomically and sociopolitically advantaged patients. For instance, funding for research into diseases that predominantly impact Caucasian patients (e.g. cystic fibrosis) is disproportionately higher than that for diseases that predominantly impact black patients (e.g. sickle cell disease) in the US (Farooq and Strouse 2018). Medical conditions that predominantly affect women and people with uteruses (e.g. endometriosis) are also notoriously underfunded and under-researched (Chen et al. 2021). Indeed, global research and development expenditure for diseases that predominantly impact low- and middle-income countries (e.g. malaria, HIV, and tuberculosis) is five times smaller than their overall global disease burden (Von Philipsborn et al. 2015). Medical ML systems are likely to recapitulate, and perhaps even intensify, these trends.

The risks and benefits of medical ML systems are also likely to be distributed inequitably due to the 'digital divide' in medicine and healthcare. As Ranit Mishori and Brian Antono observe:

> The digital divide refers to difficulty by certain populations to use the Internet. It is related to the availability of and access to the hardware required to engage in activities like telehealth, such as computers, laptops, tablets, and smartphones. It is also related to the availability of and access to technology and infrastructure that enable cyber engagement: electricity, Wi-Fi, and high-speed Internet connections (Mishori and Antono 2020: 320).

Communities cannot benefit from medical ML systems if they lack the underlying technical infrastructure that is needed to implement or operate these systems. Nor can individuals benefit from medical ML systems that rely on smartphones, internet connections, or computers if they do not have access to these technologies. In lower and middle-income countries, for instance, women are 8% less likely to own a mobile phone than men, 20% less likely to own a smart phone in particular, and 20% less likely to use mobile internet (Rowntree and Shanahan 2020). These disparities also contribute to the problem of 'health data poverty', discussed further below, insofar as they reduce the amount of data available concerning certain minority populations (Malanga et al. 2018). The digital divide presents a significant obstacle to the goal of improving health equity through the development and implementation of new medical technologies, such as medical ML systems. Indeed, as Saeed and Masters observe, "because of these long-standing financial, social, and other socioeconomic disparities, the promise and potential that HIT [Health Information Technology] offers has not been materialized" (Saeed and Masters 2021: 2). Again, medical ML systems are likely to recapitulate, and perhaps even intensify, these trends.

As noted in the introduction to this thesis, some experts anticipate that medical ML systems will reduce health disparities by being designed to offset the biases of human clinicians that are often exhibited in clinical reasoning and decision-making (Char, Shah, and Magnus 2018). However, medical ML systems also threaten to reproduce and intensify existing human biases, and the health disparities to which these biases contribute, by virtue of their susceptibility to 'algorithmic bias'.

Algorithmic bias occurs when "the application of an algorithm compounds existing inequities in socioeconomic status, race, ethnic background, religion, gender, disability, or sexual orientation" (Panch, Mattie, and Atun 2019: 1). Over the past several years, notorious examples of algorithmic bias have occurred so frequently that it may not even be necessary to mention them here.[2] Briefly, nevertheless, racial biases have been detected in ML systems for recidivism prediction, search engine optimisation, child maltreatment prediction, and face and object recognition (Angwin et al. 2016; Barr 2015; Chouldechova et al. 2018; Hurley 2018; Noble 2019). Gender biases have also been detected in ML systems for targeted job advertising and facial recognition (Datta, Tschantz, and Datta 2015), and biases against homosexuality have been detected in ML systems for text classification and hate speech detection (Dixon et al. 2018), among various other examples.

---

[2] Many books have recently been published that document these many recent examples of algorithmic bias. See, for instance, Benjamin (2019), Eubanks (2018), Fry (2018), Noble (2018), O'Neill (2016) or Wachter-Boettcher (2018).

Algorithmic bias in medical ML systems is likely to recapitulate or increase current health inequities and disparities in several ways. In particular, biased medical ML systems can cause or promote disparities in the quality of clinical judgements and recommendations and the allocation of healthcare resources. For instance, medical ML systems for skin cancer diagnosis are typically trained on open access skin image datasets such as the International Skin Imaging Collaboration: Melanoma Project. However, patients with darker skin are notoriously underrepresented in these datasets (Wen et al. 2022), which has resulted in diagnostic ML systems that underperform on these patients (Adewole and Smith 2018). Moreover, medical ML systems could cause 'allocative harm' to patients, which refers to "the effects of AI systems that unfairly withhold services, resources, or opportunities for some" (AI Now Institute 2018: 25). For instance, medical ML systems have also been found to prioritise Caucasian patients over black patients in the allocation of healthcare resources and the prioritisation of medical appointments (Obermeyer et al. 2019; Samorani et al. 2021). Finally, biased medical ML systems may cause representational harm, which refers to "the harm caused by systems that reproduce and amplify harmful stereotypes, often doing so in ways that mirror assumptions used to justify discrimination and inequality" (AI Now Institute 2018: 25). Medical ML systems could cause representational harm by detecting benign medical anomalies, or 'pseudo-diseases', that result in new forms of discrimination. Historically, for instance, "Afro-Americans who are carriers of the sickle-cell trait have been discriminated against by life insurers, although their condition does not give rise to an increased risk of death" (Hansson 2009: 1279).

A variety of approaches for remedying algorithmic bias in medical ML systems have been advanced in the literature. For instance, experts argue that algorithmic biases can be reduced by ensuring that training data for these systems are adequately representative of all groups within a target population, by removing sensitive demographic characteristics (e.g. patients' race) from datasets used to train these systems, or by calibrating the performance of these systems using statistical fairness metrics (see Parikh, Teeple, and Navathe 2019; Rajkomar, Hardt, et al. 2018; Vokinger, Feuerriegel, and Kesselheim 2021). However, while these debiasing strategies can help to reduce algorithmic bias, they often struggle to eliminate it.

One reason for this is that algorithmic biases occur for a wide variety of reasons and can emerge at any point during a system's conception, development or implementation (Chen et al. 2021; Danks and London 2017; Suresh and Guttag 2021; Vokinger, Feuerriegel, and Kesselheim 2021). According to Harini Suresh and John Guttag (2021), for instance, there are at least seven distinct varieties of algorithmic bias that each become embedded in ML systems at various points in their development or implementation. For instance, 'historical bias', 'representation bias', and 'measurement bias' each occur during the data generation and collection phase of model development. During this phase, developers define a target population, gather data, and define features, labels, and measurement metrics. *Historical bias* occurs when the world as it is (or was) is biased in ways that are reflected in a dataset. For example, datasets used to train search engines often contain harmful stereotypes and representations that are reproduced in the resulting model, often despite how well this data is collected and prepared (Noble 2019). *Representation bias* occurs when training data is under-representative of certain population groups. For instance, training a machine learning algorithm on data collected from patients attending wealthy research institutions may not be generalisable to patients attending rural hospitals (Futoma et al. 2021). *Measurement bias* occurs when the features and labels being measured are proxies for the target construct. For instance, taking body mass index (BMI) as an accurate representation of true body fat percentage can be misleading without proper interpretation, since BMI is a proxy measure for body fat percentage. Measurement bias can be compounded if different measurement standards are applied to different cohorts, or if measurement accuracy differs between groups. For instance, underdiagnosis, overdiagnosis, and misdiagnoses occur with greater frequency amongst certain population groups.

Moreover, 'aggregation bias', 'learning bias', 'evaluation bias', and 'deployment bias' each occur during the model development and deployment phase. During this phase, developers train the learning algorithm, test the performance of the resulting model, and implement the system in clinical practice. *Aggregation bias* occurs when a model incorrectly assumes a consistent relationship between inputs and outputs across an entire population. For instance, some disease categories present differently, and occur with differing frequencies, between different population groups (e.g. men and women or white Americans and black Americans). As a result, applying a one-size-fits-all model of this sort is often unsuitable for certain patient cohorts. *Evaluation bias* occurs when the test and validation datasets are unrepresentative or under-representative of the target population, resulting in a failure to detect model underperformance for specific groups. Finally, *deployment bias* occurs "when there is a mismatch between the problem a model is intended to solve and the way in which it is actually used" (Suresh and Guttag 2021: 6). Each of these biases also interact with one another in complex ways that are difficult for developers to anticipate or fully understand.

De-biasing approaches also struggle to eliminate algorithmic biases due to a range of weaknesses and limitations. For instance, developers often struggle to obtain sufficiently representative training data due to 'health data poverty', which refers to "the inability for individuals, groups, or populations to benefit from a discovery or innovation due to a scarcity of data that are adequately representative" (Ibrahim et al. 2021: e260). High-quality data from socially disadvantaged groups is often difficult to acquire since, historically, data from these groups has been collected sporadically, if it has indeed been collected at all. For instance, socioeconomically disadvantaged patients typically attend teaching clinics with more fragmented data collection practices than wealthy research-centred hospitals (Gianfrancesco et al. 2018). Patient data can also only be collected if individuals seek out medical care. However, some minority populations are less likely to seek out medical treatment due to lower rates of health insurance coverage and distrust due to historical instances of abuse by the medical profession (e.g. the Havasupai diabetes project, the Tuskegee experiment, etc.) (Malanga et al. 2018). The practices used to collect data for socially disadvantaged groups can also themselves be biased. Even if socially disadvantaged groups are well-represented in a dataset, the data collected from these groups may exhibit certain patterns that could be used to discriminate against them, such as higher frequencies of low-resolution medical images. Access to patient data in general is also heavily restricted by privacy legislation such as HIPAA (Health Information Privacy and Accountability Act) 1996 in the US or the Patient Privacy Act 1988 in Australia. Consequently, developers hoping to obtain sufficiently representative datasets in medicine often face substantial obstacles.

Removing sensitive demographic characteristics (e.g. patients' race) from training datasets for ML systems is often also ineffective because medical ML systems can reliably predict a patient's race from medical images *even when the patient's race is withheld* (Banerjee et al. 2021; Gichoya et al. 2022). Moreover, withholding sensitive demographic characteristics can even *cause* algorithmic biases. For instance, withholding demographic characteristics may cause aggregation bias, discussed above, insofar as doing so may preclude medical ML systems from accounting for statistical differences in disease incidence, prognosis, biomarkers, treatment effectiveness and symptomatology between different sexes or ethnic groups (see Cirillo et al. 2020; Lee, Guo, and Nambudiri 2021; McCradden et al. 2020). As Anirban Basu observes, therefore, "failure to include race corrections will propagate systemic inequities and discrimination in any diagnostic model and specific prognostic models" (Basu 2023: 1). However, even 'race correction' measures of this sort are risky since, historically, they have tended to intensify health disparities rather than correct them (see Braun 2014). Race correction measures may also bolster discriminatory conceptions of the relationship between health factors and demographic characteristics. As Darshali Vyas and co-authors note, "when clinicians insert race into their tools, they risk interpreting racial disparities as immutable facts rather than as injustices that require intervention" (Vyas, Eisenstein, and Jones 2020: 880).

Optimising the performance of ML systems using statistical fairness metrics also struggles to eliminate algorithmic bias due to substantial disagreement about the meaning of 'fairness', and how it ought to be statistically evaluating and encoded in these systems. For instance, this can be seen from a recent debate between journalists at ProPublica and Northpointe, the developing organisation of a recidivism prediction algorithm known as COMPAS. In 2016, Julia Angwin and co-authors (2016) alleged that COMPAS exhibited a racial bias insofar is this system was more likely to incorrectly predict high-likelihoods of recidivism for black defendants in comparison to white defendants. In other words, COMPAS failed to achieve equal rates of false positives between black and white defendants. Northpointe, however, responded by arguing that the performance of COMPAS was not racially biased because the overall performance of the system was equalised across racial groups, otherwise known as 'predictive parity' (Courtland 2018). Notably, computer scientists have since found that it is mathematically impossible to satisfy equal false positive and predictive parity simultaneously (Kleinberg, Mullainathan, and Raghavan 2016). Moreover, as Andrew Selbst and co-authors highlight, it may also be "that *no* definition may be a valid way of describing fairness" since statistical definitions of fairness eliminate the "procedural, contextual and politically contestable" properties of the concept (Selbst et al. 2019: 62).

As noted in the introduction to this thesis, experts anticipate that medical ML systems will improve cost efficiency in medicine. Historically, however, the adoption and use of new medical technologies (and the increased use of existing technologies) has been the largest driver of inflated healthcare costs (Callahan 2009; Fuchs 2011; Gelijns and Rosenberg 1994). As Daniel Callahan has expressed, if "it is true that the road to hell is paved with good intentions, it is no less true that the road to higher long-term costs [in medicine] is paved with claims of the eventual savings to be achieved by the use of expensive technologies" (Callahan 2007: 146). For instance, calls for the adoption of EHRs were initially justified on the basis of cost-savings generated through improved productivity and efficiency. Indeed, in 2005, representatives of the RAND Corporation argued that the adoption of EHRs could generated cost-savings of between US$142–$371 billion in healthcare (Hillestad et al. 2005). In a follow-up article in 2013, however, the authors observed that:

> Although the use of health IT has increased, quality and efficiency of patient care are only marginally better. Research on the effectiveness of health IT has yielded mixed results. Worse yet, annual aggregate expenditures on health care in the United States have grown from approximately $2 trillion in 2005 to roughly $2.8 trillion today (Kellermann and Jones 2013: 63; see also Agha 2014).

Similar concerns are also raised about additional costs to patients and health systems resulting from the adoption and use other medical technologies, such as robotic surgical systems (Barbash and Glied 2010; Crew 2020; Lotan 2012).

The adoption and use of medical ML systems may continue these trends by contributing to increased cost inefficiencies in medicine. For instance, adopting medical ML systems into healthcare settings will require substantial financial investment into updating the underlying technological infrastructure of these organisations. This is because, as Tristan Panch and co-authors note:

> most healthcare organizations lack the data infrastructure required to collect the data needed to optimally train algorithms to (a) 'fit' the local population and/or the local practice patterns, a requirement prior to deployment that is rarely highlighted by current AI publications, and (b) interrogate them for bias to guarantee that the algorithms perform consistently across patient cohorts (Panch, Mattie, and Celi 2019: 1).

Moreover, as Keith Morse and co-authors (2020) highlight, implementing medical ML systems in clinical practice also carries a variety of hidden costs that are difficult for organisations to predict or estimate in advance. Indeed, a recent economic evaluation found that "even when AI can achieve better diagnostic capacities than the average physician, this may not directly translate to better or cheaper care" (Rossi et al. 2022: 1). Thomas Grote and Philipp Berens (2020, 2022) also note that medical ML systems threaten to increase time, resource, and cost inefficiencies in medicine by promoting the practice of defensive medicine, in which clinicians make medical decisions that protect themselves against the threat of litigation even where these decisions do not benefit patients.

Ultimately, therefore, experts exaggerate the likelihood that medical ML systems will deliver on their promises because these systems threaten the very objectives that they are anticipated to achieve. In particular, medical ML systems generate new and expanded threats to patient health and safety due to a variety of stubborn technical weaknesses in these systems, and common human biases exhibited by users of algorithmic systems. The use of medical ML systems also seems likely to recapitulate or even enhance current health disparities due to the susceptibility of these systems to algorithmic bias. In addition, incorporating medical ML systems into clinical practice may contribute substantially to inflated healthcare costs and cost inefficiencies in medicine. But these are not the only factors likely to compromise the overall positive impact that experts currently anticipate ML systems will have in medicine. In particular, experts also tend to discount several risks that these systems present to patients, clinicians, and health systems at large, as I now discuss.

## 4. Discounted risks

Medical ML systems generate new and expanded threats to patient privacy and confidentiality, accountability for patient harm, and the integrity of medical decision-making. For instance, ML systems enable government agencies and private companies to extract insights concerning individuals' health and well-being from publicly available (or otherwise easily accessible) data (e.g. social media data, purchasing history, census records, police records, location history, and so on; see Weber, Mandl, and Kohane 2014). The capacity to easily gain insight into individuals' health and well-being using ML systems threatens to harm patients in several ways.

In one famous case, Target identified that one of their teenage customers was pregnant by using an ML system to analyse their customer's online browsing patterns. Target then inadvertently revealed this information to the customer's family by mailing her advertisements for maternity items (Hill 2012). Health insurance companies also have strong financial incentives to use medical ML systems to deny coverage to individuals or increase their premiums on the basis of algorithmically generated risk scores (e.g. risk of developing certain diseases, risk of post-operative complications, etc.). As Cathy O'Neill (2016) observes, customers will often be unable to contextualise these risk scores since neither the customer nor the insurance organisation can assess the reasoning process through which these scores are generated.

ML systems may also enable predatory individuals or vigilante groups to extract health-related insights that could be weaponised against these individuals. For instance, ML systems may enable anti-abortion vigilante groups to identify individuals that may have received an abortion with the aim of having them convicted of a crime (Huq and Wexler 2022; Ohlheiser 2022). In some cases, moreover, ML systems will generate false or misleading outputs concerning individuals' health and well-being, who may be harmed by the inaccurate beliefs that result (e.g. inaccurate diagnoses of depression from social media data generated by online 'depression detectors'; see Islam et al. 2018).

Finally, individuals may be wronged by breaches of their sensitive information even if the breach does not cause them any direct harm (Price and Cohen 2019). For instance, suppose a medical practice were to share one's health information with an AI organisation for the purpose of developing a new medical ML system. The information shared with the AI organisation contains personal health information that one does not feel particularly sensitive or embarrassed about, and the organisation ultimately destroys the information once it has been used to train a supervised learning algorithm. A key function of privacy law and regulation is to enable individuals to exercise control over who has access to this data. According to Michael Froomkin, for instance, privacy itself ought to be understood as "the ability to control the acquisition or release of information about oneself" (Froomkin 2000: 1464). The patient in this situation may therefore be wronged simply because their degree of control over who has access to their personal information has been compromised.

Healthcare organisations already have a strong incentive to use patients' health data to improve hospital efficiency and cost-efficiency. This is problematic for patient privacy since, as Gina Neff (2013) observes, using patient health data in this way prioritises the interests of healthcare organisations over patients since there is typically no way for patients to opt-out. However, medical ML systems are likely to expand these existing financial and practical incentives for organisations to collect, store, analyse, and share patient health data across multiple institutions and organisations (He et al. 2019). This is because, in order to develop and maintain medical ML systems, healthcare organisations and AI developers will require the ongoing availability of largescale, high-quality datasets that largely consists of information relating to patients' bodies, medical histories, health, and well-being.

Moreover, while individuals' identifying characteristics are likely to be removed from these datasets that are shared between organisations, de-identification does not eliminate privacy risks. By combining de-identified datasets with other publicly available datasets, malicious actors can trace this data back to identifiable individuals (Narayanan and Shmatikov 2008). Moreover, even where de-identified data *does* preserve individuals' privacy, it may infringe 'group privacy rights' that "restrict the flow and acceptable uses of aggregated datasets and profiling" (Mittelstadt and Floridi 2016: 326). Ensuring the availability of these datasets may also increase the scope and intensity of current surveillance practices by governments and private organisations which, historically, have been biased against particular ethnic and soci-oeconomic groups (Browne 2015; Eubanks 2016).

Furthermore, medical ML systems threaten to reduce accountability in medicine since incor-porating these systems into clinical reasoning results in what Hannah Bleher and Matthias Braun refer to as 'diffused responsibility', i.e. "a phenomenon in which divergent attributions of responsibility to various different agents are possible, or in which attributions of responsi-bility are manifold, uncertain, or not consolidated in particular administrative, legal or social structures" (Bleher and Braun 2022: 748). For instance, suppose that a medical ML system generates a mistaken output due to a fault in the system. A clinician decides to accept this output and pursue a course of action that ultimately results in significant patient harm. Diver-gent attributions of responsibility are possible in this scenario due to the causal role that the clinician, the designers of the system, and the system itself played in generating this outcome. However, in many cases, no single individual will reasonably be able to shoulder the primary burden of responsibility. Indeed, automated medical ML system (e.g. remote monitoring sys-tems) will likely generate 'responsibility gaps' in medicine, which refer to instances in which "nobody has enough *control* over the machine's actions to be able to assume responsibility for them" (Matthias 2004: 177). The use of medical ML systems for automated patient mon-itoring may generate responsibility gaps insofar as healthcare practitioners, healthcare organ-isations, AI organisations, and so on, may be unable to achieve meaningful human oversight or control over the actions or outcomes of using these systems (see Hille, Hummel, and Braun 2023).

Medical ML systems could also interfere with cultures of accountability in medicine by ena-bling individuals to defer responsibility for patient harm resulting from the use of these sys-tems. Some clinicians, for instance, are likely to be tempted to use medical ML systems as 'moral buffers' to avoid being held accountable for their errors. A moral buffer refers to "an artifact or process, such as a computer interface or automated recommendations" that "adds an additional layer of ambiguity and possible diminishment of accountability and responsibil-ity" to an individual's actions (Cummings 2006: 26). Some clinicians are likely to use medical ML systems as moral buffers by deferring to the outputs of medical ML systems over their own clinical judgements to evade being held accountable for patient harm for which they are at least in part responsible. Indeed, according to Matthias Braun and co-authors (2020), trans-ferring decision-making authority to machines reduces the accountability of individual clini-cians. Data and scientists and AI developers, moreover, often refuse to see themselves as engaged in ethical or political action. In particular, data scientists often defer ethical or political responsibility for their work by arguing that engaging in ethics and politics is not the role or responsibility of data scientists, or that by attending to ethical and political risks of their technologies, they 'make the perfect the enemy of the good' (Green 2021; see also Nissenbaum 1994). This reluctance amongst data scientists and AI developers to accept the ethical and political consequences of their technologies further exacerbates the threats that medical ML systems present to responsibility and accountability in medicine.

Medical ML systems also generate new and expanded threats to the integrity of medical decision-making due to the increasing power and influence that medical ML systems grant to technology organisations in this domain. Recent advancements in AI have substantially contributed to the growing cultural authority and economic power of largescale technology corporations including Google, Facebook, Microsoft, Amazon, Apple, and IBM. Each of these organisations have already entered the market for AI in healthcare due to the enormous financial opportunities it offers (Meskó 2022). According to the McKinsey Institute, medical AI technologies are estimated to generate between US$200 - $300 billion of value in the sector (Chui et al. 2018). As James Bridle (2018) argues, however, new technologies such as medical ML systems are more likely to intensify existing inequalities and relations of power and domination rather than act as a force for egalitarian democracy. This is because laws and regulations typically lag behind the adoption of new technologies due to the difficulties associated with anticipated their challenges and impact. AI organisations themselves have also made many attempts to either delay or manipulate proposed laws and regulations in their favour (Benkler 2019; Foroohar 2019; Gibney 2016).

The tendency for law and regulation to lag behind technological innovation is concerning with respect to medical ML systems since the financial and political interests of private technology organisations are often misaligned with the interests of patients, clinicians, and health systems at large. As Shoshana Zuboff (2019) argues, private technology corporations have a strong economic interest in the prediction and modification of human behaviour, along with a powerful capacity to directly modify human behaviour through the design of their technologies. In particular, private technology organisations have a vested interest in expanding what Zuboff describes as their 'instrumentarian power', which refers to "the instrumentation and instrumentalization of behavior for the purposes of modification, prediction, monetization, and control" (Zuboff 2019: 352). For instance, data collected for the purpose of training medical ML systems, or through the use of medical ML systems, could be used by such organisations to predict and modify the behaviour of (groups of) patients and align their actions with the aims and objectives of the developing organisations. These organisations may have an economic incentive to influence patient behaviour to improve the predictive capacities of their systems and to ignore when their systems influence patient behaviour in ways that indirectly benefit the organisations.

Developers of medical ML systems may also be tempted to design these systems in ways that mislead or deceive users and purchasers or subvert regulatory standards due to political interests that are misaligned with the goals of public health and the values of patients. This is because these organisations have vested interests generate strong incentives to design their technologies in ways that intentionally mislead or deceive stakeholders. In 2015, for instance, the US Environmental Protection Agency found that Volkswagen had installed software into their diesel vehicles that severely underreported the nitrogen oxides emissions of these vehicles, thereby violating the Clean Air Act (Johnson and Verdicchio 2018). It would be naïve to think that similar attempts to subvert regulatory standards will not occur in the context of medicine. Current legal and regulatory gaps, for instance, have previously enabled private technology corporations to conceal their use of large datasets of patient health information to develop medical ML systems (see Barber and Molteni 2019; New Scientist 2016). Indeed, opacity in medical ML systems increases the likelihood of these systems being designed to promote economic and political incentives over patients' best interests insofar as it interferes with the capacity for regulators to assess the causal reasoning processes through which these systems generate their outputs.

Medical ML systems also risk increasing the incentives and capacity for healthcare administrators to compromise the integrity of medical decision-making. For instance, medical ML systems may be designed to prioritise the financial and institutional interests of healthcare organisations over the best interests of individual patients. As Danton Char and co-authors observe, for instance:

> Given the growing importance of quality indicators for public evaluations and determining reimbursement rates, there may be a temptation to teach machine-learning systems to guide users toward clinical actions that would improve quality metrics but not necessarily reflect better care (Char, Shah, and Magnus 2018: 982).

Jianxiang He and co-authors highlight the possibility that "clinical decision support systems could be programmed to increase profits for certain drugs, tests, or devices without clinical users being aware of this manipulation" (He et al. 2019: 33). Healthcare administrators and practice managers could also use medical ML systems as tools to standardise medical decision-making, thereby narrowing the discretion that clinicians can exercise in their clinical judgements and decisions and restricting their capacity to act in their patients' best interests.

According to the prevailing vision for the coming age of AI in medicine, these threats that medical ML systems present to patient privacy, accountability for patient harm, and the integrity of medical decision-making, do not overshadow the anticipated benefits of these systems for patient health and safety, health equity, and efficiency in medicine. As I argued in section three, however, there are reasons to doubt that medical ML systems will deliver on these promises, since medical ML systems themselves generate new and expanded threats to patient health and safety, health equity, and efficiency that are largely overlooked by advocates of these systems. To avoid overinvesting in medical ML systems on the basis of exaggerated assessments of their likely benefits, therefore, advocates of medical ML systems need to engage more deeply with the many risks presented by the use of these systems.

## 5. Conclusion

Recent developments in AI and ML have generated high hopes for the coming age of AI in medicine. In this chapter, however, I have argued that experts exaggerate the likelihood that medical ML systems will deliver on their anticipated benefits, and discount the risks that these systems present. I suggest that this is because advocates of medical ML seem to forget the recent history of failure and disappointment associated with previous attempts to revolutionise medicine through the implementation of new technologies, and discount the threats these systems themselves present to the very benefits they are anticipated to deliver. This problem is compounded by the tendency amongst advocates of medical ML systems to discount the new and expanded risks these systems present to patient privacy, accountability for patient harm, and the integrity of medical decision-making. Accurately assessing the risks and benefits of medical ML systems will thus require substantially greater engagement with the risks of these systems in order to ensure safe and effective deployment of ML in medicine.

As discussed in the introduction to this thesis, experts anticipate that medical ML systems are likely to have their most positive and substantial impact in the context of clinician-patient relationships. According to Eric Topol, for instance, the coming age of AI in medicine is "our chance, perhaps the ultimate one, to bring back real medicine: Presence. Empathy. Trust. Caring. Being Human" (Topol 2019: 309). It is necessary to investigate whether the tendency to exaggerate the benefits of medical ML systems and discount their risks, discussed in this chapter, has generated unrealistic expectations about the potential impact of medical ML systems on these relationships. I now return to this investigation in the following chapter, in which I focus on analysing the impact of medical ML systems on relations of trust between clinicians and patients.

## (2)   LIMITS OF TRUST IN MEDICAL MACHINE LEARNING

### A. INTRODUCTION

In the previous chapter, I argued that journalists and experts often discount the risks that medical ML systems present to patient privacy, accountability for patient harm, and the integrity of medical decision-making. I also argued that journalist and experts often exaggerate the likelihood that these systems will deliver on their anticipated benefits to patient health and safety, health equity, and efficiency in medicine. As I discussed in the introduction to this thesis, however, experts such as Eric Topol (2019a) argue that these anticipated benefits are merely the "secondary gains" of the coming age of AI in medicine due to the revolutionary impact that medical ML systems are likely to have on the quality of clinician-patient relationships. As Bertalan Meskó and co-authors express, medical ML systems are anticipated to "bring forward a renaissance era in the doctor-patient relationship" (Meskó, Hetényi, and Győrffy 2018: 3).

However, the tendency for experts and journalists to exaggerate the benefits and discount the risks of medical ML systems (discussed in the previous chapter) suggests that critical analysis of these high hopes for the future of clinician-patient relationships is urgently needed. In this chapter, therefore, I begin my analysis of the impact of medical ML systems on the quality of clinician-patient relationships, focusing on their impact on relations of trust between clinicians and patients. I argue that, rather than improving trust between clinicians and patients, the use of these ML system is likely to negatively impact the quality of these relations of trust.

The core argument of this chapter is given in part B, which consists of my article, 'Limits of trust in medical AI', published in the *Journal of Medical Ethics*. In this article, I argue that while medical ML systems can be relied upon, they cannot coherently be trusted.  This is because medical ML systems are unable to satisfy a number of basic preconditions for interpersonal trust under prevailing philosophical accounts. Consequently, I suggest that the more that clinicians rely on medical ML systems to inform their judgements and recommendations, the more their relations with patients will be that of mere reliance, rather than of trust.

Since the publication of 'Limits of trust in medical AI' in early 2020, several objections and alternative theories have been advanced in the literature that directly challenge the arguments presented in this article (see Ferrario, Loi, and Viganò 2021; Nickel 2022; Starke et al. 2021). In part C of this chapter, I respond to these objections and alternative theories. In particular, I argue that none of them succeed in defending a conceptually coherent account of trust in medical ML systems. I also expand my argument concerning the impact of medical ML systems on relations of trust between clinicians and patients. In particular, I argue that describing humans' relations with medical ML systems using the language of trust threatens to compromise *actual* relations of trust between clinicians and patients. By appealing to recent arguments from Mark Ryan (2020), I argue that using the language of trust to describe human relations with medical ML systems is likely to expand existing threats to accountability for patient harm in medicine, discussed in the previous chapter. Using the language of trust to describe human relations with medical ML systems risks compromising the quality of clinician-patient relationships since clinicians will struggle to develop trusting relationships with their patients if their patients perceive them to be unaccountable for harm that may result from their use of these systems. I conclude that preserving the quality of clinician-patient relationships in the coming age of AI in medicine requires stakeholders to resist the temptation to inappropriately anthropomorphise medical ML systems by describing these systems using familiar but misleading concepts.

## B. LIMITS OF TRUST IN MEDICAL AI

[PDF begins on next page].

## Original research

# Limits of trust in medical AI

Joshua James Hatherley

**Correspondence to**
Joshua James Hatherley, School of Historical, Philosophical, and International Studies, Monash University, Clayton, VIC 3194, Australia;
joshua.hatherley@monash.edu

## ABSTRACT
Artificial intelligence (AI) is expected to revolutionise the practice of medicine. Recent advancements in the field of deep learning have demonstrated success in variety of clinical tasks: detecting diabetic retinopathy from images, predicting hospital readmissions, aiding in the discovery of new drugs, etc. AI's progress in medicine, however, has led to concerns regarding the potential effects of this technology on relationships of trust in clinical practice. In this paper, I will argue that there is merit to these concerns, since AI systems can be relied on, and are capable of reliability, but cannot be trusted, and are not capable of trustworthiness. Insofar as patients are required to rely on AI systems for their medical decision-making, there is potential for this to produce a deficit of trust in relationships in clinical practice.

## INTRODUCTION
Artificial intelligence (AI) is expected to revolutionise the practice of medicine. Recent advancements in the field of deep learning have demonstrated success in variety of clinical tasks; for instance, detecting diabetic retinopathy from images,[1] predicting hospital readmissions[2] and aiding in the discovery of new drugs.[3] It has been suggested that AI will facilitate a variety of improvements in medical practice, ranging from economic savings to the improvement of empathetic communication between doctors and patients, from increased productivity to greater professional satisfaction and from improved health outcomes on an amplified rate of discovery in medical research.[4] AI's progress in medicine, however, has led to concerns regarding the potential effects of this technology on relationships of trust, particularly between doctors and patients.[5 6] In this paper, I will argue that there is merit to these concerns, since AI systems are not the appropriate objects of trust under any familiar philosophical accounts of trust. This is critical, since, as I will argue in section 3, AI systems are likely to displace the epistemic authority of human clinicians if they come to exceed them in performance. As such, I will argue that insofar as patients are required to rely on AI systems for their medical decision-making, AI threatens to produce a deficit in trusting clinical relationships between doctors and patients.

## TRUST IN MEDICINE
Trust has both intrinsic and instrumental significance in medicine.[i] Intrinsically, trust is what imbues the doctor-patient relationship with its uniqueness and importance. A patient comes to a physician in a state of sickness and vulnerability, and is thereby forced to place their trust in another person to treat them with competence and, ideally, empathy and care. This vulnerability of the patient is what imbues the relationship with inherent value, since 'trust is inseparable from *vulnerability*, in that there is no need for trust in the absence of vulnerability'.[7] The vulnerability of the patient, and the resulting power of the physician, imbue the physician with a fiduciary obligation to behave in a morally upright and appropriate manner, to use their authority in the service of the patient as opposed to themselves or some other end.

In contrast, trust also has instrumental value in medicine. First, because patients are more likely to accept and behave in accordance with their physician's judgement if they have a trusting relationship with them. They are more likely to demonstrate 'willingness to seek care, reveal sensitive information, submit to treatment, participate in research, adhere to treatment regimens, remain with a physician and recommend physicians to others'.[7] Second, it is speculated that trusting doctor-patient relationships have a number of therapeutically valuable effects on patients—improved patient outcomes and placebo effects, for example. Finally, a good physician is one that can demonstrate care for their patients, and patients are more likely to feel that they have been adequately cared for when they trust the person caring for them.

## AI IN MEDICINE
AI's effect on relations of trust between doctors and patients is bound up with the precise role that AI may come to occupy in medical practice and the epistemic authority that it comes to hold in clinical decision-making procedures. If AI systems are eventually adopted as merely another tool at the clinician's disposal—akin to a stethoscope, thermometer or blood pressure monitor—the effect of these systems on trust would likely be minimal. Patients, of course, would rely on the accuracy of these tools, but their trust would be staked in the judgement of the human physician who interprets their outputs and incorporates them into their own clinical judgements. However, recent developments in areas such as deep learning suggest that the epistemic authority of human clinicians in clinical decision-making will be challenged by the use of AI in medicine.

---

[i]Two kinds of trust are discussed in relation to medicine and clinical practice: interpersonal and social.[28] Interpersonal trust concerns trust between persons (eg, between doctors and patients), while social trust is more general and abstract, directed towards groups and institutions as opposed to individuals (eg, between a patient and a particular hospital or the medical institution more generally). In this paper, I leave the issue of social trust to one side in order to focus on medical AI and interpersonal trust. All references to trust will henceforth refer exclusively to interpersonal trust.

Researchers in AI are working busily to develop AI systems that can surpass the performance of human clinicians in diagnosis, prognosis and treatment selection[4]—three of the four fundamental tasks of the clinician, according to Eric Cassell.[8] [ii] Indeed, a recent systematic review and meta-analysis comparing the performance of deep learning AI systems to human clinicians found that deep learning AI systems already match the accuracy of human clinicians in the performance of certain diagnostic tasks.[9][iii] If AI succeeds in surpassing the performance of human clinicians in such principal medical tasks, how might this effect the epistemic authority of human clinicians in clinical practice?

The prospect gestures at an important problem currently faced in the sciences, which Paul Humphreys has called our 'anthropocentric predicament'. Humphreys argues that advanced technologies have produced a situation in which 'an exclusively anthropocentric epistemology is no longer appropriate because there now exist superior, non-human, epistemic authorities. So we are now faced with a problem, which we can call the *anthropocentric predicament*, of how we, as humans, can understand and evaluate computationally based scientific methods that transcend our own abilities'.[10]

There have been two principal kinds of response to medicine's anthropocentric predicament in the wake of medical AI, which I will refer to as substitutionism and extensionism. Substitutionists argue that advanced AI will eventually make doctors obsolete by surpassing them in the performance of key clinical tasks and roles.[11] Extensionists, in contrast, argue that AI will simply extend and improve on the capabilities and competencies of human clinicians without replacing them outright. In particular, this is because AI systems lack emotional intelligence and empathy, abilities that are essential in the delivery of healthcare, meaning that a human presence will still be essential.[12] Yet among both camps, the likely disruptive impact—what Liu and colleagues have labelled a 'seismic shift'[13]—that AI will have medicine widely undisputed. Although extensionists rally against the substitution of clinicians, the likelihood of their displacement in key clinical roles is often acknowledged. For instance, Eric Topol, a principal physician advocate for the use of AI in medicine, along with his colleague Saurabh Jha, claim that '(j)obs are not lost; rather, roles are redefined; humans are displaced to tasks needing a human element'.[14]

This displacement of the roles of human clinicians in the wake of advanced medical AI reflects a displacement of their epistemic authority. Indeed, if AI surpasses the performance of clinicians in key clinical tasks, doctors will have an epistemic obligation to defer to the judgements of the machine or align their judgements with the AI in their clinical decision-making.[15] As Bjerring and Busch have argued, 'if a practitioner knows of an epistemic source that is more knowledgeable, more accurate and more reliable in decision-making, she should treat it as an expert and align her verdicts with those of the source'.[16] This displacement of the epistemic authority of clinicians would be necessary to realise some of the goals of the introduction of AI in medicine. Aside from the possible reduction of burdensome administrative tasks and the improvement of cost-efficiency in medicine, a primary motivation for research into medical AI is the potential to reduce the alarming prevalence of wastefulness and human error in medical practice.[4] [17] In order to achieve this, it would be necessary in most instances for human clinicians to give more weight to the outputs of a supremely reliable AI system over their own clinical intuitions and judgements.

The displacement of clinicians from a position of epistemic authority in clinical decision-making has important implications for relations of trust between patients and doctors, since it implies a displacement of patient trust from human clinicians to AI systems. In the next section, I will argue that this displacement of trust from humans to machines could lead to shallow relations of trust in clinical practice that are lacking in important respects.

## TRUST IN AI

Trust has been a central topic of concern in the debate over AI and its many applications, with some private corporations and research organisations releasing guidelines for the development of trust and trustworthiness in AI.[18] [19] Concerns over the 'black box' nature of some AI systems—particular deep learning—along with the threat of algorithmic bias have pushed the issue of trust to the forefront of debate.[20] But what does it mean to say that one trusts an AI, or that an AI is trustworthy? A key response to this question has been to emphasise the centrality of reliability in trust. Alex John London claims that '(i)f the goal is to secure trust among stakeholders, then the accuracy of a system relative to viable alternatives must be a central concern'.[21] Similarly, Zachary Lipton claims that if trust is 'simply confidence that a model will perform well (… then) a sufficiently accurate model should be demonstrably trustworthy'.[22]

But is confidence in someone or something's accuracy or reliability sufficient for trust? According to many accounts of interpersonal trust have been proposed in the philosophical literature, the answer to this question is no. According to these accounts, trusting someone to do $x$ is more than merely relying on them to do $x$. Consider the following two scenarios:

1. Stan, a thief, is planning a burglary. He has observed a wealthy homeowner, Jane, leaving her home at 09:00 a.m. and returning at 19:00 p.m. every Monday for the past month. Stan is hoping to go through with his planned burglary next Monday, and is relying on Jane to continue her pattern in order for his burglary to be successful.

2. Brendan has a chronic illness that causes him significant pain and suffering. His illness is managed by his regular general practitioner, Dr Smith. Dr Smith has supported Brendan through his illness for 15 years. Brendan has recently been experiencing significantly more pain than usual, which is causing him extreme discomfort. He makes an appointment with Dr Smith, confident that she will be able to help him relieve this pain in some way.

In the scenario (1), although the thief relies on Jane to leave her house at 09:00 a.m., it seems inappropriate to say that the thief trusts Jane to do so in the same way that Brendan trusts Dr Smith to successfully treat his illness in scenario (2), despite the fact that Brendan also relies on Dr. Smith. How do we explain this intuition? What makes trusting someone more than merely relying on them?

[ii]The fourth is the identification of causes. Given that AI systems based on neural networks learn from correlations alone, their capacity to illuminate underlying causes of illness is limited.[29]

[iii]Importantly, the study identified a number of troubling methodological limitations in the broader literature comparing the performance of human clinicians to deep learning AI systems, so this finding ought to be taken with a grain of salt. Most alarmingly, of the 31 587 scholarly articles returned on a search for articles comparing the performance of deep learning systems and human clinicians, only 14 compared performance between the two groups on the same test data set.

## Original research

Russell Hardin argues that reliance is insufficient for trust because trusting someone also requires a belief that one's interests are encapsulated in the interests of the trusted person.[23] 'What matters', claims Hardin, '(…) is not merely my expectation that you will act in certain ways but also my belief that you have the relevant motivations to act in those ways, that you deliberately take my interests into account because they are mine'.[23] For Hardin, trust requires not only a predictive expectation on the part of the truster, but also a belief that one's interests are encapsulated in the interests of the trusted person and that the trusted person has the right motivations for action. Indeed, Hardin claims that 'I would not, in our usual sense, trust a fully programmed automaton, even if it were programmed to discover and attempt to serve my interests—although I might come to rely heavily on it'.[23]

Having the right kind of motivations for action is an important part of many other influential accounts of trust. Annette Baier, for instance, argues that reliance underdetermines trust because trust 'seems to be reliance on (the trusted person's) good will toward one, as distinct from their dependable habits, or only on their dependably exhibited fear, anger or other motives compatible with ill will toward one, or on motives not directed to one at all'.[24] This emphasis on the good will of the trusted person is also central to Karen Jones' account, wherein she claims that 'to trust someone is to have an attitude of optimism about her goodwill and to have the confident expectation that, when the need arises, the one trusted will be directly and favourably moved by the thought that you are counting on her'.[25] If the right kind of motivations are necessary for the kind of trust that we would usually recognise as interpersonal trust, then AI systems would not appear to be the appropriate objects of this kind of trust. Unlike a human clinician, AI systems have no goodwill towards us, nor any motivation to act in our interests. This may be at least part of the reason that some people may be uncomfortable with the idea of placing their trust in an AI for important medical decisions or tasks.

Additionally, other philosophical accounts of trust distinguish between trust and reliance on the basis of normative and descriptive expectations. I *rely* on you when I *predict* that you will behave in a certain way, though I *trust* you when I judge that you *ought* to behave in a certain way.[26] Trusting someone, that is, generates an obligation on behalf of the trusted person to (at least genuinely attempt to) do what one is trusting them to do. There are some important limitations to this claim, for example, in circumstances where the trust that one has in another is misguided or unwelcome. Suppose, for instance, that one were to place their trust in a friend who is a dermatologist to remove their wisdom teeth. Trusting the dermatologist for this procedure would appear quite mistaken, given that the dermatologist does not have the expertise or competency to perform this task. Nor, presumably, would the dermatologist welcome this trust in any way.

But outside of this and other somewhat fanciful scenarios, clinicians do in fact have an obligation to perform those tasks that have been entrusted to them, providing of course that this trust has been communicated to them. This is precisely the nature of fiduciary obligations in medicine. If this is true, another limitation of trusting AI would also be demonstrated, since AI systems are not the appropriate objects of moral responsibility. In order for an agent to be morally responsible for an action, they must be blameworthy when they fail to come through on that action. But if an AI system were to incorrectly diagnose a patient, leading to their avoidable death, it would appear misguided or inappropriate to blame the AI for its error. Rather, one would generally look to the designers, the supervising clinician, the hospital, etc, in order to apportion blame. Trusting a clinician generates a moral responsibility on behalf of the clinician, while trusting an AI system generates a moral responsibility on behalf of seemingly anyone but the AI system.

These considerations highlight two important deficits in relations between patients and medical AI systems that each stem from a lack of agency on the part of the AI. First, AI systems lack the right kind of motivation for trust—either in the form of encapsulated interest or a sense of good will—since they lack motivation entirely. Second, relations with AI systems cannot be said to be trusting relations, as one might have with a human clinician, since trust generates normative obligations that cannot be borne by an AI. To say that one can trust an AI is thus akin to saying that one can trust a naturally occurring phenomenon. Although I am supremely confident that tomorrow the sun will rise in the east and set in the west, there is not familiar sense in which I could reasonably said to *trust* the sun to do so. Trusting relations, in other words, are exclusive to beings with agency, meaning that the displacement of human clinicians from a position of epistemic authority and privilege in the clinical encounter threatens to lead to relations of trust that are shallow or deficient in important respects within medical practice.

## CONCLUSION

To say that one can trust an AI system, or that the AI is trustworthy, is merely to say that one can rely on the AI system, or that the system is reliable. Yet as we have seen, reliability is insufficient to generate a relation of trust under any of its familiar philosophical notions, which all require characteristics essential and exclusive to beings with a form of agency. What does this mean for the pursuit of 'Trustworthy AI' initiated by the European Union's High Level Expert Group on Artificial Intelligence (HLEG AI)?[18] Although valuable, the pursuit of trustworthy AI represents a notable conceptual misunderstanding, since AI systems are not the appropriate objects of trust or trustworthiness. Interestingly, this has also been suggested by a key member of the HLEG AI, Thomas Metzinger.[27] Rather than trustworthy AI, this pursuit may be better served by being reframed in terms of reliable AI, reserving the label of 'trust' for reciprocal relations between beings with agency.

In contrast to AI, therefore, human clinicians can offer their patients the kind of rich interpersonal trust that imbues the doctor-patient relationship with its uniqueness and significance. Insofar as patients come to rely on AI systems for important medical assessments and decisions as opposed to human clinicians, they may be sacrificing opportunities for trusting relationships in medicine. A more thoughtful engagement is needed with the potential effects of AI on medical practice to further understand the implications of this technology, so that it can be deployed is such a way as to reap its potential benefits while retaining those aspects of medicine—such as trust—that are particularly valuable for its functioning.

**Acknowledgements** Thanks are due to Rob Sparrow and an anonymous reviewer from the Journal of Medical Ethics for their helpful comments and insight, which assisted me greatly in the preparation of this paper.

**Contributors** JJH is the sole author.

**Funding** Research for this paper was funded through Australian Government Research Training Program Scholarship.

**Competing interests** None declared.

**Patient consent for publication** Not required.

**Provenance and peer review** Not commissioned; externally peer reviewed.

**ORCID iD**
Joshua James Hatherley http://orcid.org/0000-0002-8581-9669

## C. DISCUSSION AND RESPONSE TO OBJECTIONS

As noted in part A of this chapter, several objections and alternative theories have recently been advanced in the ethics literature that directly challenge my arguments in 'Limits of trust in medical AI'. In this section, I respond to these objections and alternative theories, and I argue that none of them succeed in defending a coherent account of trust in medical ML systems, for two main reasons. First, none of these objections or theories of trust in medical ML systems can maintain a meaningful distinction between trust and mere reliance. Second, none of these objections or theories of trust can establish that so-called trust in medical ML systems is anything more than trust in the developers of these systems. I also expand my argument for the claim that medical ML systems are likely to interfere with relations of trust between clinicians and patients because medical ML systems are not the appropriate objects of trust. In particular, drawing on arguments advanced by Mark Ryan (2020), I argue that continued use of the language of trust to describe users' relations with medical ML systems is likely to obfuscate the appropriate attribution of responsibility for harms that may result from the use of these systems.

The remainder of part C of this chapter proceeds as follows. In section one, I respond to Andrea Ferrario and co-authors' (2021) objection that, by taking interpersonal trust as the starting point for my argument against trust in medical ML systems, I beg the question against trust in these systems. I also respond to Ferrario and co-authors' objection that medical ML systems are the appropriate objects of what they refer to as 'simple trust'. In section two, I respond to Georg Starke and co-authors' (2021) objection that medical ML systems can coherently be trusted because the meanings of concepts are circumscribed by how they are used in everyday language, and because medical ML systems can satisfy some core preconditions for trust under prevailing philosophical accounts. In section three, I respond to Philip Nickel's (2022) objection that medical ML systems are the appropriate objects of what he refers to as 'discretionary trust'. In section four, I argue that using the language of trust to describe human relations with medical ML systems is likely to further compromise *actual* relations of trust between clinicians and patients by expanding existing threats to accountability for patient harm in medicine, discussed in the previous chapter. Finally, in section five, I offer some concluding remarks.

### 1. Simple trust

As noted above, several writers have contested my central thesis in this chapter by arguing that medical ML systems are in fact the appropriate objects of trust. The first set of objections are advanced by Andrea Ferrario and co-authors, who suggest that "the choice of applying human trust to describe human-AI interactions is not fully justified [because it] begs the question against AI" (Ferrario, Loi, and Viganò 2021: 437). This is because, according to Ferrario and co-authors, AI systems simply are not human beings, and it is therefore unreasonable to expect these systems to meet the conditions for trust between human beings. In other words, Ferrario and co-authors claim that, by setting up the problem of trust in medical ML systems as one of meeting the conditions for interpersonal trust between human beings, the possibility of trust in medical ML systems is unjustifiably precluded from the outset.

However, my decision to take interpersonal trust as a point of departure in my analysis of trust in medical ML systems is justified because it neatly aligns with the theory of classification known as prototype theory. Prototype theory "construes membership in a concept's extension as graded, determined by similarity to the concept's 'best' exemplar" (Osherson and Smith 1981: 35) and first found empirical support in Linda Coleman and Paul Kay (1981) influential study of the concept of 'lying'. In this study, Coleman and Kay found that participants classified scenarios as examples of lying according to their perceived 'closeness' to certain prototypical examples of lying. For instance, prototypical examples of lying include instances in which a person intentionally deceives their spouse about their whereabouts to pursue an affair, while non-prototypical examples include instances in which a person intentionally deceives their spouse about their whereabouts to organise their surprise birthday celebration. Coleman and Kay found that people tend to classify non-prototypical examples of lying, such as the latter, according to how much they overlap with paradigmatic exemplars, such as the former. As Mark Johnson has expressed, Coleman and Kay ultimately found that:

> [the concept of lying] is not defined by a set of fixed essential features but is rather a radially structured concept, with prototypical instances making up the center of the concept and nonprototypical instances radiating out at various (conceptual) distances from the central members (Johnson 1993: 92).

Prototype theory is relevant to the current debate over trust in medical ML systems because, like the concept of lying, the concept of trust has what Johnson refers to as an "internal prototype structure" (Johnson 1993: 189). This is because trust is a gradated notion with blurry conceptual boundaries, as demonstrated by the existence of persistent and ongoing debates about the possibility of trust between humans and animals, institutions, or objects to which the arguments of this chapter contribute. It is also because the concept of trust consists of prototypical examples that radiate outwardly toward non-prototypical examples. In particular, the radial core of the concept of trust consists of prototypical examples of inter-personal trust between human beings, e.g. trust between romantic partners, friends, col-leagues, and so on (Hardin 2002), while non-prototypical examples of trust – including previ-ously noted instances of trust between humans and animals, institutions, or objects – orbit around these prototypical examples at varying conceptual distances.

Prototype theory is relevant to *this* discussion in particular because it demonstrates the inac-curacy of Ferrario and co-authors' objection that "the choice of applying human trust to de-scribe human-AI interactions is not fully justified [because it] begs the question against AI" (Ferrario, Loi, and Viganò 2021: 437). This is for two reasons.

First, prototype theory demonstrates that the decision to analyse trust in medical ML systems by comparing and contrasting it with interpersonal relations of trust between human beings *is* justified. This is because instances of interpersonal trust between human beings represent *the* prototypical examples of trust; they form the radial core against which non-prototypical examples of trust must be compared in order to understand the boundaries of the concept. By denying that interpersonal trust is the prototypical example of trust against which non-prototypical examples must be compared, Ferrario and co-authors are simply no longer talk-ing about trust, but rather something else entirely, as I discuss further below.

Second, prototype theory demonstrates that taking interpersonal trust as a point of depar-ture does *not* beg the question against trust in medical ML systems. This is because, as Mark Johnson (1993) has argued, prototype theory allows for the creative extension of concepts based on imaginative reinterpretations of their prototypical exemplars. As Johnson has ex-pressed:

> A central part of our moral development will be the imaginative use of particular proto-types in constructing our lives. Each prototype has a definite structure, yet that structure must undergo gradual imaginative transformation as new situations arise. It thus has a dynamic character, which is what makes possible our moral development and growth (Johnson 1993: 192).

Taking interpersonal trust as a point of departure is therefore consistent with the fact that the conceptual parameters of trust are dynamic and open to reinterpretation in light of new exemplars. It does not beg the question against trust in medical ML systems.

While this argument allows for the possibility that the conceptual boundaries of trust ought to be imaginatively reinterpreted to allow for trust in medical ML systems, there are two reasons why this move ought to be rejected. First, as I have suggested in 'Limits of trust in medical AI', the notion of trust in medical ML systems overlaps only marginally with proto-typical examples of interpersonal trust, and therefore, is simply too far removed from these prototypical examples to warrant inclusion in the category of relationships that we currently describe using the language of trust. Second, what many currently describe as instances of trust between users and medical ML systems are more accurately and succinctly captured by the concept of 'mere' reliance, as I argue shortly.

In addition to their objection, Ferrario and co-authors (2021) advance an alternative theory of trust that purports to explain and account for relations of trust between human users and medical ML systems. In particular, Ferrario and co-authors argue that medical ML systems are the appropriate objects of 'simple trust', which they define as:

> a reliance property that describes the willingness of the physician to rely on the medical AI without intentionally generating and/or processing further information about the medical AI's capabilities to achieve the goal at hand (e.g. by monitoring the medical AI) (Ferrario et al. 2021: 437).

Under this account, a clinician trusts a medical ML system when they rely on its outputs without attempting to improve their understanding of the system by scrutinising its capabilities or operations. However, Ferrario and co-authors' (2021) account of simple trust in medical ML systems ought to be rejected since it cannot maintain a meaningful distinction between trust and mere reliance.

As noted in 'Limits of trust in medical AI', trust is distinct from mere reliance. This is because, as Annette Baier has argued, trust "can be betrayed, or at least let down, and not merely disappointed" (Baier 1986: 235). This point has been developed further by Richard Holton, who argues that "the difference between trust and reliance is that trust involves something like a participant stance towards the person you are trusting" (Holton 1994: 4). The participant stance refers to the general psychological standpoint that allows persons to exhibit what Peter Strawson (2008) famously refers to as 'reactive attitudes' toward other beings. Reactive attitudes refer to psychological responses – including betrayal, indignation, guilt, and resentment – that, as Bennet Helm has expressed, "are important to understanding not just our *holding* each other responsible but also our *being* responsible" (Helm 2014: 187). According to Holton, the capacity to trust another being depends on the capacity to hold the participant stance toward that being, and it is this stance that distinguishes instances of trust from those of mere reliance. In Holton's own words:

When you trust someone to do something, you rely on them to do it, and you regard that reliance in a certain way: you have a readiness to feel betrayal should it be disappointed, and gratitude should it be upheld. In short, you take a stance of trust towards the person on whom you rely. It is the stance that make the difference between reliance and trust. When the car breaks down we might be angry; but when a friend lets us down we feel betrayed (Holton 1994: 4).[1]

Ferrario and co-authors argue that simple trust maintains a meaningful distinction between trust and mere reliance by proposing an alternative approach to distinguishing between these two concepts. In particular, Ferrario and co-authors argue that trust is distinct from reliance by virtue of the fact that trust involves "rely[ing] on the medical AI without updating beliefs on its trustworthiness" (Ferrario, Loi, and Viganò 2021: 437). In contrast, under this account, clinicians merely rely on medical ML systems when they continue seeking out further information about these systems' performance and capabilities.

However, Ferrario and co-authors' (2021) claim that simple trust maintains a meaningful distinction between trust and mere reliance ought to be rejected, for two reasons.

First, Ferrario and co-authors (2021) provide no justification for revising the distinction between trust and reliance in terms of "economising on monitoring" (Ferrario, Loi, and Viganò 2021: 437). However, revising the distinction between trust and reliance in this manner is question-begging, since the only explicit justification for accepting this premise is that it supports Ferrario and co-authors' conclusion that medical ML systems can be trusted.

Second, distinguishing between trust and mere reliance on the basis of whether the truster continues to process information about the trusted's capacity to meet the truster's expectations is misaligned with prototypical exemplars of the two concepts.  For instance, if we return to scenario 1 discussed in 'Limits of trust in medical AI', suppose that Stan the thief stops seeking out further evidence that Jane's home will be unoccupied between 9am and 7pm on the Monday that he has decided to execute the burglary. Under Ferrario and co-authors' (2021) account, Stan no longer merely relies, but now *trusts* Jane to be out between these hours on this particular Monday. This is false, however, because Stan's decision to economise on monitoring Jane's behaviours is based on mere predictive expectations about how Jane is likely to behave. In other words, the normative and affective elements of trust that distinguish it from mere reliance are completely absent.

---

[1] A stronger version of this claim is also defended by Bennet Helm, who argues "that trust *is* a reactive attitude and, moreover, that trust doesn't simply presuppose the participant stance; it is an emotion without which the participant stance would be unintelligible" (Helm 2014: 187-188).

## 2. Intentional machines?

A second set of objections to my argument in 'Limits of trust in medical AI' have been advanced by Georg Starke and co-authors (2021).

Starke and co-authors' first object that:

> if we subscribe to a Wittgensteinian approach that (in most cases) the meaning of a word is its use in the language, trust factually describes a much broader phenomenon than mere interpersonal relationships. From trust in local governments to trust in healthcare systems, trust is commonly used to denote an attitude towards non-human or non-living entities, for instance towards bridges, cars, or institutions (Starke et al. 2021: 3).

Similar claims have also been advanced in less recent articles defending the notion of trust between humans and 'artificial agents' (AAs), and even between AAs themselves. For instance, Mariarosaria Taddeo has argued that:

> Trusting AAs to perform actions that are usually performed by human agents (HAs) is not science-fiction but a matter of daily experience. There are simple cases, such as that of refrigerators able to shop online autonomously for our food, and complex ones, such as that of Chicago's video surveillance network, one of the most advanced in the US. In the latter case, […] HAs – the entire Chicago police department – trust an artificial system to discern dangerous systems from non-dangerous ones (Taddeo 2010: 245).

However, Starke and co-authors' (2021) first objection ought to be rejected. This is because accepting the claim that the meanings of concepts are determined by how they are used in everyday language has the untenable implication that errors in the use of language simply cease to exist if they occurred frequently enough in everyday discourse.

If the meanings of concepts are determined by how they are used in everyday language, for instance, then 'jealousy' and 'envy' would therefore be synonymous due to the frequency with which these concepts are used interchangeably. That jealousy and envy are synonymous, however, is false, because the concepts of jealousy and envy have distinct meanings. In particular, one is jealous when they fear that someone will take something one has (e.g. a romantic partner), while one is envious when they desire something that someone else has (e.g. a new car) (Ben-Ze' 1990; Parrott and Smith 1993). Despite the frequency with which people use the concepts of jealousy and envy interchangeably by, for instance, claiming that they are jealous, rather than envious, of their friend's new car, the meaning of these concepts remains distinct. Rather than revising the meaning of jealousy and envy, therefore, speakers that use these concepts interchangeably simply conflate their meanings.

Ultimately, if the meanings of concepts are not entirely determined by how they are used in everyday language, then the fact that the concept of trust is often used in everyday language to describe humans' relationships with non-human beings, including AI systems, does not ipso facto demonstrate that these objects are the appropriate objects of trust. Indeed, just as a speaker that uses jealousy when they mean envy conflates two related concepts, a writer that describes users' relations with medical ML systems in terms of trust simply conflates trust with reliance, as – again – I argue further in section two below.

Starke and co-authors (2021) also raise a second objection to my argument in 'Limits of trust in medical AI'. In particular, Starke and co-authors argue that medical ML systems are the appropriate objects of trust because they do in fact exhibit a form of agency. Starke and co-authors appeal to actor-network theory and the arguments of Bruno Latour. Specifically, Starke and co-authors argue that medical ML systems are sociotechnical agents by appealing to Latour's example of a key design that was used in Berlin tenant houses during the first half of the 20th Century. Known as Berlin keys, these artifacts were:

> constructed in a way that […] compels [their] user[s] to re-lock the door of a building after entering: after unlocking a door, the key cannot be simply removed like a usual key but remains stuck in its position, unless it is pushed through the keyhole to the other side of the door. Only after locking the door from the other side can it be removed (Starke et al. 2021: 4).

Starke and co-authors echo Latour's claim that technical artifacts, such Berlin keys and medical ML systems, are agents in a sociotechnical system insofar as they play an active role in influencing the behaviour of their users toward bringing about certain ends or objectives. In short, medical ML systems are agents that can exhibit intentions insofar as they play this active, influencing role in a sociotechnical environment. "Under this premise," claim Starke and co-authors, "the concern that we *cannot* trust medical AI simply owing to its being non-human seems no longer convincing" (Starke et al. 2021: 4-5).

However, this second objection ought to also be rejected, for two reasons.

First, Starke and co-authors' (2021) objection oversimplifies my argument in 'Limits of trust in medical AI'. This is because my central claim is *not* that medical ML systems are not the appropriate objects of trust simply because they are non-human. Rather, my central claim is that medical ML systems are not the appropriate objects of trust because trust in medical ML systems fails to align with important features of interpersonal trust between human beings, i.e. being appropriately motivated by concern for the interests or well-being of another and being able to have obligations toward another.

Second, while Starke and co-authors (2021) suggest that medical ML systems do in fact overlap with these core features of interpersonal trust by virtue of their sociotechnical agency, this limited form of agency does little to increase the degree of overlap between interpersonal trust and so-called 'trust' in medical ML systems. This is because theories of sociotechnical agency are limited to demonstrating that medical ML systems generate affordances or embody intentions that can affect other agents (human and non-human) in the sociotechnical environment in which they are embedded. However, sociotechnical agency does not enable medical ML systems to form motivations toward other beings, nor does it enable these systems to hold normative obligations toward them.

Starke and co-authors (2021) anticipate and respond to the second of these objections by arguing that medical ML systems exhibit the technologically mediated intentions and motivations of their developers. In particular, Starke and co-authors claim that:

> In its direct, weaker sense, trust in AI does not require a fully independent agency of the program itself but rather ties trust to the intentions of its developers or those involved in its quality control […] For example, we may trust a system of medical AI because we trust the people who develop and regulate it (Starke et al. 2021: 4).

However, as I discuss further below, this argument only demonstrates that users of medical ML systems trust the human beings involved in the development and validation of medical ML systems, rather than the medical ML systems themselves. This is because intentions and motivations do not transfer from human beings to objects by virtue of the objects' mediating role in the achievement of certain ends. For instance, the fact that a carpenter is motivated to impress their family by building them a new dining table from scratch does not entail that the completed table has a mediated intention or motive to impress the carpenter's family. Rather, the motives and intentions remain those of the carpenter, and the carpenter alone. Similarly, the fact that AI developers develop medical ML systems with the motive and intention of improving patients' health and safety does not entail that medical ML systems themselves are motivated by concern for patient well-being. While users may trust the developers and validators of medical ML systems, they merely rely on medical ML systems.

## 3. Discretionary trust

A final challenge to my arguments in 'Limits of trust in medical AI' is advanced by Philip Nickel (2022). Nickel develops a theory of 'discretionary trust' that purports to justify and explain relations of trust between human users and medical ML systems. Nickel defines discretionary trust as occurring when "one entity is disposed to give a second entity *discretion* over some matter of value on the basis of normative and predictive expectations about that second entity" (Nickel 2022: 3). For instance, a clinician trusts a medical ML system in the discretionary sense when they give the system discretion over a designated clinical task due to their predictive expectation that the system *will* successfully perform this designated task, and their normative expectation that the system *ought* to perform this task successfully. Through this account of discretionary trust, Nickel aims to demonstrate that "there is a plausible notion of trust in medical AI that is grounded in reasonable, realistic attitudes of clinicians and explains the moral commitments of AI practitioners" (Nickel 2022: 3).

Nickel (2022) argues that this theory of discretionary trust is stronger than Ferrario and co-authors' account of simple trust for two reasons. First, while simple trust in medical ML systems is non-normative, discretionary trust contains a normative component. As previously noted, discretionary trust involves normative expectations on the part of the user that are directed toward the medical ML system. Second, unlike simple trust, discretionary trust "explains why inviting user trust entails moral commitments" (Nickel 2022: 2). In particular, Nickel argues that by generating predictive and normative expectations in users, the invitation for users to trust medical ML systems (in the discretionary sense) generates an ethical obligation for AI developers to ensure that these predictive and normative expectations are met.

However, Nickel's theory of discretionary trust fails to justify trust in medical ML systems, for two reasons.

First, the normative expectations that Nickel (2022) attributes to medical ML systems are substantially weaker than he suggests. This is because these normative expectations apply equally to various other objects that fall outside the reasonable scope of trust. For instance, like medical ML systems, automatic pencil sharpeners generate normative expectations in their users that these sharpeners ought to perform their designated function successfully.

However, it does not follow from this that one trusts an automatic pencil sharpener to sharpen their pencil.

Second, discretionary trust in medical ML systems is nothing more than trust in the developers of these systems. This is because, according to Nickel (2022), discretionary trust is a version of what he refers to as the 'reductive view' of trust in medical ML systems. According to the reductive view, "trust in AI is nothing more than trust in the designers, deployers, and overseers of the AI" (Nickel 2022: 5). Discretionary trust is a version of the reductive view because, under this account, "practitioners play an essential role in inviting and supporting trust in the technology, one layer removed from the experience of the user. They are the ultimate indirect object of user trust in the application" (Nickel 2022: 6). In other words, "clinicians trust the practitioners *through* the application" (Nickel 2022: 6). Nickel's account of trust is not an account of trust in medical ML *systems*, but merely an account of trust in medical ML *developers* and *regulators*. In short, while the clinician may rely on the medical ML system and its outputs, they do not trust the medical ML system at all.

Nickel appears to anticipate this second objection and attempts to avoid it by claiming that the reductive view is "not the view that trust in AI does not exist or is not explanatory, but rather the view that we can translate the moral content of statements about trust in AI into statements about human and institutional elements" (Nickel 2022: 5). However, this response fails to establish that discretionary trust provides a substantive account of trust in medical ML systems that is irreducible to trust in the developers of medical ML systems. This is because translating the moral content of statements about trust in AI into statements about trust AI developers is no different from accepting that so-called trust in AI simply *is* trust in AI developers.

## 4. Trust, responsibility, and clinician–patient relationships

Using of the language of trust to describe human relations with medical ML systems expands existing threats to accountability for harm in medicine. As I argued in the previous chapter, medical ML systems generate diffused responsibility, i.e. "a phenomenon in which divergent attributions of responsibility to various different agents are possible, or in which attributions of responsibility are manifold, uncertain, or not consolidated in particular administrative, legal or social structures" (Bleher and Braun 2022: 748). Medical ML systems for automated patient monitoring may also generate 'responsibility gaps', which occur when "nobody has enough *control* over the machine's actions to be able to assume responsibility for them" (Matthias 2004: 177). Clinicians are also likely to be tempted to use medical ML systems as 'moral buffers' which refer to technical artefacts that add "an additional layer of ambiguity and possible diminishment of accountability and responsibility" to individuals' actions (Cummings 2006: 26).

Using the language of trust to describe human relations with medical ML systems expands these existing threats to accountability in medicine by generating further complications and ambiguities for the attribution of responsibility for patient harm. As I argued in part B, trust requires the capacity to form motivations or hold normative obligations. Describing human relations with medical ML systems using the language of trust risks implicitly attributing these affective and normative characteristics to medical ML systems which they do not and cannot possibly possess. Insofar as these characteristics are implicitly attributed to medical ML systems, they risk obfuscating the responsibility of human clinicians and AI organisations for patient harm that results from their use. As Mark Ryan observes: "Referring to AI as trustworthy would inappropriately elevate AI, while disavowing the responsibility of those developing and implementing it" (Ryan 2020: 2763). Using the language of trust to describe human relations with medical ML systems increases the capacity for human clinicians, hospitals, or AI organisations to deny or distance themselves from moral or legal responsibility for patient harm that results from the use of these systems. It is perhaps no coincidence that AI organisations in particular have devoted special attention to the notion of 'trustworthy' AI systems.

Patients will be reluctant to place their trust in their clinicians if they perceive these clinicians (and also hospitals and ML developers) as unaccountable for harm that results from the use of medical ML systems. By expanding existing threats to accountability for patient harm that results from the use of medical ML systems, continued use of the language of trust to describe human relations with these systems is likely to further compromise relations of trust between clinicians and patients. Preserving the quality of clinician-patient relationships in the coming age of AI in medicine requires stakeholders to resist the temptation to inappropriately anthropomorphise medical ML systems by describing these systems using familiar but misleading concepts.

## 5. Conclusion

Medical ML systems cannot coherently be trusted, since they cannot satisfy several important preconditions for interpersonal trust under prevailing philosophical accounts. Several attempts have recently been made to defend coherent accounts of trust in medical ML systems. In this chapter, however, I have argued that none of these accounts succeed. This is either because they cannot maintain a meaningful distinction between trust and reliance on medical

ML systems, or because the account of trust they provide is reducible to trust in the *developers* of these systems. The more that clinicians rely on medical ML systems to inform their clinical judgements and recommendations, the more that their relations with patients will be that of mere reliance, rather than of trust. Despite this, human relations with medical ML systems continue to be described using the language of trust. For instance, trust and trustworthiness remain some of the most commonly discussed principles in the growing literature of AI ethics guidelines (Jobin, Ienca, and Vayena 2019), and the UK government has recently announced £54 million of research funding for projects concerning 'trustworthy' AI systems (UK Department of Science, Innovation, and Technology 2023). I have argued that this continued use of the language of trust presents further threats to *actual* relations of trust between clinicians and patients insofar as patients are unlikely to trust clinicians that they perceive as less-than-accountable for causing patient harm. This is because continued use of the language of trust to describe human relations with medical ML systems is likely to expand existing threats to accountability for patient harm that results from the use of medical ML systems, discussed in the previous chapter.

While Topol (2019a) anticipates that medical ML systems with substantially improve clinician-patient relationships by augmenting relations of trust, I have argued in this chapter that medical ML systems are more likely to compromise the quality of relations of trust between clinicians and patients than enrich them. However, medical ML systems are likely to not only impact negatively on the quality of clinician-patient relationships by virtue of their effects on relations of trust; they are also likely to impact negatively on the quality of these relationships by virtue of their implications for patient autonomy and the ethical ideal of shared decision-making in medicine. In the next chapter, I turn to analyse the effects of medical ML systems for these elements of the clinician-patient relationship, along with clinicians' ethical obligations with respect to communicating with patients about their use of these systems.

# (3)  MUST CLINICIANS DISCLOSE THEIR USE OF MEDICAL MACHINE LEARN-ING SYSTEMS?

## 1.  Introduction

Suppose you have been suffering from a debilitating chronic health condition for over a year. You have previously pursued multiple treatment strategies recommended by several physicians with little to no success. In some cases, these treatments have only enhanced your suffering. Now more desperate than ever before, you seek out a new specialist with a strong reputation as an expert in their field. After several consultations, your new specialist recommends a new treatment plan that is likely to be long, inconvenient, and painful. The outcome is uncertain, but your specialist is confident that your condition will improve as a result. You discuss the recommended treatment with the specialist in detail, and eventually give your consent to proceed. Soon before the treatment is scheduled to begin, however, you discover that your specialist has decided to pursue this treatment plan on the recommendation of a medical ML system. How would this make you feel? By failing to disclose this information to you, would your specialist have wronged you? Would knowing this information lead you to withdraw your consent?

In the previous chapter, I argued that the use of medical ML systems is likely to negatively impact clinician-patient relationships by compromising the quality of trust between clinicians and patients. In this chapter, I discuss another way in which the use of medical ML systems is likely to compromise the quality of these relationships. As noted in the introduction to this thesis, promoting patient autonomy and securing patients' informed consent become core duties of clinicians upon entering into these fiduciary relationships with their patients. In this chapter, I analyse the threats presented by medical ML systems to patient autonomy and informed consent, and evaluate clinicians' ethical obligations with respect to communicating with patients about their use of these systems. I argue that clinicians are ethically obligated to disclose their use of medical ML systems for treatment recommendation to secure their patients' informed consent due to the threats these systems present to patients' autonomous decisions. In addition, I argue that clinicians are ethically obligated to disclose their use of medical ML systems for several reasons other than protecting patients' autonomous decisions.

The remainder of this chapter proceeds as follows. In section two, I provide an account of the principle of respect for patient autonomy and the doctrine of informed consent in medicine. In section three, I critically analyse Glenn Cohen's (2020) recent argument against a legal obligation in US jurisdictions for clinicians to disclose their use of medical ML systems to patients, and I argue that the case for a legal obligation to disclose is stronger than Cohen suggests. In section four, I argue that clinicians are ethically obligated to disclose their use of medical ML systems to secure patients' informed consent because failure to disclose is likely to interfere with their patients' autonomous choices and the ethical ideal of shared decision-making in medicine. In section five, I argue that clinicians are ethically obligated to disclose their use of medical ML systems to patients for several reasons beyond the scope of the doctrine of informed consent. In particular, I argue that clinicians are ethically obligated to disclose their use of medical ML systems due to the risks these systems present to patient health and safety and patient privacy and confidentiality. I argue that clinicians are also ethically obligated to disclose their use of medical ML systems to enable patients to act on their right to refuse diagnostics and treatment planning by these systems, recently defended by Thomas Ploug and Søren Holm (2019). In section six, I offer some concluding remarks.

## 2. Informed consent and respect for patient autonomy

Recent discussions of whether clinicians are ethically or legally obligated to disclose their use of medical ML systems to patients are typically contextualised within the theoretical frameworks of informed consent and respect for autonomy. Before discussing the case against a legal obligation for clinicians to disclose their use of medical ML systems to patients, therefore, it is necessary to provide an overview of informed consent and the principle of respect for autonomy.

One of the most influential accounts of the doctrine of informed consent is advanced by Tom L. Beauchamp and James Childress, who define informed consent as "an individual's autonomous authorization of a medical intervention or of participation in research" (Beauchamp and Childress 2019: 78). According to Beauchamp and Childress (2019), seven conditions must be met for a clinician to secure informed consent from patients:

  i.   the patient must be competent to consent to treatment;
  ii.  the patient's consent must be voluntary, rather than (e.g.) coerced or manipulated;

iii.   the clinician must disclose all information that is material to the patient's decision;

iv.   the clinician must provide a recommended treatment plan;

v.   the patient must have a sufficient understanding of the clinician's recommendation and the material information disclosed to them;

vi.   the patient must make a clear decision; and

vii.   the patient must provide unambiguous authorisation to proceed with the treatment plan.

Informed consent is an ethical requirement, and legal requirement in most jurisdictions, in both clinical practice and clinical research settings. In clinical practice, clinicians are ethically, and usually legally, obligated to secure informed consent from patients to proceed with a recommended medical intervention. In clinical research, investigators are required to secure informed from human research subjects to participate in clinical studies. In this chapter, I am concerned with the impact of medical ML systems on informed consent to treatment in clinical practice.[1]

The duty for clinicians to secure informed consent from their patients flows directly from the principle of respect for autonomy. According to Beauchamp and Childress, patient autonomy refers to, "at minimum, self-rule that is free from controlling interference by others and from limitations, such as inadequate understanding, that prevent meaningful choice" (Beauchamp and Childress 2019: 58). Beauchamp and Childress suggest that autonomous decisions must satisfy three minimum conditions: the patient must decide intentionally, with adequate understanding, and without interference from controlling influences (e.g. coercion or manipulation) (Beauchamp and Childress 2019). These conditions for autonomous action are reflected in the conditions for informed consent, outlined above.

The principle of respecting patient autonomy imposes several duties on clinicians. In particular, Beauchamp and Childress suggest respecting patient autonomy requires that clinicians treat their patients as 'ends in themselves' rather than as means to an end. Beauchamp and Childress also suggest that respecting patient autonomy requires clinicians to uphold two basic obligations. First, a 'negative' obligation to avoid controlling, constraining, or otherwise interfering with the autonomous choices of their patients. Second, a 'positive' obligation to promote their patients' autonomous decision-making, by demonstrating proactive disclosure of information and respectful engagement with patients' values and preferences (Beauchamp and Childress 2019).

---

[1] For discussions of the impact of medical ML systems on informed consent to participate in clinical research, see Grote (2022) and McCradden and co-authors (2022).

Obligations for clinicians to disclose certain information to patients can be divided into reactive and proactive variants. Clinicians have a *reactive* obligation to disclose their use of medical ML systems to patients if this obligation only obtains under the condition that the patient asks about, for instance, the role of a medical ML system's outputs in the clinician's judgement or recommendation. In contrast, clinicians have a *proactive* obligation to disclose their use of medical ML systems to patients where this obligation obtains regardless of the information that patients do or do not request from their clinician. Before discussing the case against an obligation to disclose, therefore, it is necessary to specify the type of obligation that I aim to discuss.

That clinicians typically have a reactive ethical obligation to disclose their use of medical ML systems to patients is, I think, relatively uncontroversial. Indeed, although Glenn Cohen ultimately rejects a legal obligation for clinicians to disclose their use of medical ML systems (as I discuss further below), even he suggests that clinicians may breach a patients' informed consent if the patient "were to ask their physician 'is this what the AI/ML recommended?' or 'did you rely on an AI/ML?' and the physician were to mislead the patient by falsely denying that they did so" (Cohen 2020: 1442). A reactive obligation to disclose is also significantly narrower and less consequential than a proactive obligation to disclose, as it applies only in circumstances where patients explicitly request information about their clinicians' use of a medical ML system. Given these limiting factors, my aim in this chapter is to argue that clinicians have a *proactive* ethical obligation to disclose their use of medical ML systems to patients. All future references to an obligation to disclose will refer to this proactive obligation to disclose.

Obligations for clinicians to disclose certain information to patients can also be divided into broad and narrow variants. Clinicians have a *broad* obligation to disclose their use of medical ML systems to patients where this obligation pertains under all conditions, irrespective of, for example, the type of medical ML system being used, the context in which it is used, or the way it is used. In contrast, clinicians have a *narrow* obligation to disclose where this obligation only applies under certain conditions, e.g. where a particular type of medical ML system is used, or where a medical ML system is being used in a particular way. Throughout this chapter, I discuss both broad and narrow obligations to disclose and specify when I refer to each. In section four, I argue that clinicians have a *narrow* ethical obligation to disclose their use of medical ML systems *for treatment recommendation* to secure their patients' informed consent. As noted above, however, clinicians may be ethically obligated to disclose their use of medical ML systems even if failing to do so does not violate patients' informed consent. In section five, I argue that clinicians have a *broad* ethical obligation to disclose their use of medical ML systems for several reasons other than protecting their patients' autonomous decisions and securing their informed consent.

## 3.  Medical machine learning and US informed consent law

Currently, there are few direct or detailed discussions of whether clinicians are ethically obligated to disclose their use of medical ML systems to patients to secure informed consent in the ethics literature.[2] Recently, however, Glenn Cohen (2020) has provided an extensive discussion of whether clinicians are *legally* obligated, under current US case law, to disclose their use of medical ML systems to secure their patients' informed consent. While my primary aim in this chapter is to defend an *ethical*, rather than legal, obligation that clinicians disclose their use of medical ML systems to patients, Cohen's discussion offers a useful starting point for my analysis as many of the arguments he advances are relevant to both the legal and ethical obligations to disclose.

In this section, I critically analyse Cohen's arguments against a legal obligation that clinicians disclose their use of medical ML systems to secure their patients' informed consent. In particular, I analyse Cohen's discussion of four sets of arguments supporting both broad and narrow legal obligations to disclose. These four sets of arguments are built on the legal standards of material information, empirical assessments of material information, common law reasoning, and normative reasoning, respectively. I argue that the case for several narrow legal obligations to disclose is stronger than Cohen suggests.

### a.  Standards of material information

Whether a clinician is obligated to disclose a certain piece of information is often thought to depend on whether that information is considered 'material' to a patient's medical decision (Faden and Beauchamp 1986; Sawicki 2016). Roughly, information is material if it meets a certain minimum standard of relevance with respect to a patient's decision, formally known as minimum standards of 'materiality'. In many jurisdictions, clinicians will have a legal obligation to disclose their use of medical ML systems if it can be established that their use of these systems is material to their patients' medical decisions.

---

[2] Exceptions to this include Astromskė, Peičius, and Astromskis (2019), Kiener (2020), and Schiff and Borenstein (2019). Compared to these articles, however, Cohen's discussion is significantly more detailed and comprehensive.

Cohen identifies two standards of materiality that currently prevail in US courtrooms: the physician-based standard and the patient-based standard. According to the physician-based standard, material information is defined as "what a reasonable physician would customarily disclose – or in the words of the 1960 Kansas Supreme Court decision in *Natanson v. Kline* only 'those disclosures which a reasonable medical practitioner would make under the same or similar circumstances'" (Cohen 2020: 1443). Material information under this standard is thus dictated by the behaviour of a hypothetically reasonable clinician, along with the dominant set of norms and customs within the professional community of doctors (see Beauchamp and Childress 2019). By contrast, the patient-based standard defines material information as "information which the physician knows or should know would be regarded as significant by a reasonable person in the patient's position when deciding to accept or reject a recommended medical procedure" (*Wheeldon v. Madison* 1985, cited in Cohen 2020: 1433-1434). Under this standard, the preferences and interests of a hypothetically reasonable patient take precedence over the behaviour of a hypothetically reasonable clinician or the prevailing norms and customs within the professional community of doctors.

According to Cohen, the physician-based standard cannot justify a broad legal obligation in US jurisdictions for clinicians to disclose their use of medical ML systems to patients. This is because norms and customs surrounding the use of medical ML systems have not yet been established in the professional community of doctors. As a result, Cohen suggests that the physician-based standard faces a bootstrapping problem with respect to these systems. As Cohen himself expresses, "given a brand new technology, and what is more a brand new way of using a technology in medical practice, how can we say what reasonable medical practitioners in fact do?" (Cohen 2020: 1425).

Cohen also argues that a broad legal obligation to disclose cannot be justified using the patient-based standard. This is because, according to Cohen, the outputs of medical ML systems are analogous to various other informational sources that the patient-based standard does not legally obligate clinicians to disclose, including "vague memories from a medical school lecture, what the other doctors during residency did in such cases, the latest research in leading medical journals, the experience with and outcomes of the last 30 patients the physician saw, etc." (Cohen 2020: 1442). Unless courts come to accept that clinicians are legally obligated to disclose each and every one of these inputs that influenced a clinician's reasoning, which is highly unlikely, Cohen suggests that they will not (and indeed ought not) accept that clinicians have a broad legal obligation to disclose their use of medical ML systems to secure their patients' informed consent.

I agree that neither standard supports a broad legal obligation that clinicians disclose their use of medical ML systems to patients. As Nadia Sawicki observes, US courts typically restrict the scope of the patient-based standard of materiality to information "about the patient's diagnosis and proposed treatment; the treatment's risks and benefits; alternative procedures and their risks and benefits; and the risks and benefits of taking no action" (Sawicki 2016: 831). As Cohen notes, US courts understand information "about the patient's diagnosis and proposed treatment" (Sawicki 2016: 831) as information about the diagnosis and treatment themselves, rather than *how* a clinician arrived at a particular diagnosis or treatment recommendation. This limitation excludes information about medical ML systems for diagnosis and treatment recommendation from its scope. However, some medical ML systems designed to treat patients, or assist clinicians in treating them (e.g. ML-assisted surgical systems), will generate risks to patients that clinicians will need to disclose under this standard of materiality. Thus, while a broad legal obligation to disclose cannot be justified under either standard of materiality, the patient-based standard nevertheless supports a narrow legal obligation that clinicians disclose their use of medical ML systems designed to treat (or assist in treating) patients, and the risks associated with using these systems.

### b. Empirically assessing materiality

In practice, the ability to resolve disputes about materiality is sometimes limited by the fact that the behaviour of hypothetically 'reasonable' clinicians or patients in any given scenario is perennially open to debate and difficult to establish conclusively. However, we can gain some insight into the behaviour of hypothetically reasonable persons by investigating how real people actually behave. For instance, it seems likely that a hypothetically reasonable person in scenario *S* would perform action *A* if it were demonstrated that *actual* people in *S* tend to *A*. Clinicians' use of medical ML systems may therefore be material if it were demonstrated that *real* clinicians and patients judge this information to be material to patients' decisions.

Unfortunately, using empirical evidence to gain insight into the behaviour of hypothetically reasonable clinicians and patients toward medical ML systems faces certain practical and methodological challenges. For instance, as Cohen observes, empirical data concerning patients' and clinicians' expectations and behaviours toward medical ML systems is currently unavailable. Consequently, the search for empirical evidence in support of physician-based standard of materiality appears to face an identical bootstrapping problem to that discussed in the previous subsection. According to Cohen, however, this obstacle may be overcome by considering the preferences and behaviours of clinicians and patients toward technologies that are similar or adjacent to medical ML systems. For instance, with respect to the physician-based standard, Cohen claims that "the use of 'dumb' (relatively speaking) computer decision aids by physicians is probably the closest to how AI/ML is likely to be used in the foreseeable future" (Cohen 2020: 1450). However, Cohen observes that "few physicians explicitly disclose that they have used a computer decision aid 'in the background' in deciding on a course of treatment" (Cohen 2020: 1450). Currently, therefore, a broad legal obligation that clinicians disclose their use of medical ML systems under the physician-based standard is unsupported by empirical evidence.

However, Cohen (2020) neglects recent empirical evidence that offers some support for a broad legal obligation to disclose the use of medical ML systems under the patient-based standard of materiality. In particular, Jessica Findley and co-authors (2020) have recently argued that it is likely that disclosing the use of a medical ML systems will lead a substantial number of patients to withdraw their consent to a recommended treatment. To support this claim, Findley and co-authors appeal to recent evidence of the phenomenon of algorithmic aversion, wherein people tend to avoid relying on algorithmic systems over their own judgements, or the judgements of other human beings, even when these systems perform better than human decision-makers (see Burton, Stein, and Jensen 2020; Dietvorst, Simmons, and Massey 2015). Findley and co-authors suggest, following Roy Spece and co-authors (2014), that a piece of information (e.g. the use of a medical ML system) is likely to be material (under the patient-based standard) if disclosing it to actual patients causes a substantial proportion of them to change their consent decisions. Given this, Findley and co-authors conclude that evidence for algorithmic aversion suggests that clinicians' use of medical ML systems is material to a patient's decision, and thus, that clinicians have a broad legal (and ethical) obligation to disclose their use of these systems to secure their patients' informed consent.

It is possible that actual patients (and clinicians) may consider the use of medical ML systems material to their decisions. However, it is not evident that patients will exhibit algorithmic aversion toward clinicians whose judgements and recommendations are influenced by medical ML systems. This is because, while algorithmic aversion has been observed toward the outputs of algorithmic systems, it has not yet been observed toward the judgements of human beings that have been informed by the outputs of an algorithmic system. It is not clear, moreover, that a *reasonable* patient would withdraw their consent due to algorithmic aversion, since algorithmic aversion exhibits an irrational bias against algorithmic systems that generates worse outcomes overall. While empirical evidence may emerge that provides empirical support for the materiality of clinicians' use of medical ML systems, such empirical evidence is not yet available.

This concludes my discussion of arguments for a legal obligation to disclose the use of medical ML systems that rely solely on the notion of materiality and material information. In the next two sections, I turn to discuss several arguments for a legal obligation to disclose that involve drawing analogies to prior legal precedents (section 3c), and engaging with the underlying normative justifications for informed consent requirements (section 3d).

### c. Common law reasoning

Cohen (2020) identifies four rulings that have been made in US court rooms that one could use to support several narrow legal obligations to disclose based on what Cohen refers to as 'common law reasoning'.

First, Cohen (2020) suggests that one might argue for a legal obligation to disclose on the basis of precedents set in *Moore v. Regents, University of California*. In this case, the California Supreme Court ruled that:

> a physician who is seeking a patient's consent for a medical procedure must, in order to satisfy his fiduciary duty and to obtain a patient's informed consent, disclose personal interests unrelated to the patient's health, whether research or economic, that may affect his medical judgement (*Moore v. Regents, University of California* 1990, cited in Sawicki 2016: 842).

Cohen suggests that one might use this precedent to argue that clinicians are legally obligated to disclose their use of a medical ML system where the system has been designed or implemented to promote the purchasing organisation's financial interests (e.g. hospital, insurance agency, managed care organisation) over those of the patient. For instance, he suggests that this obligation may apply where:

> a hospital system adopts medical AI/ML to reduce costs after a study that shows it does not affect patient care one way or the other or improves some patient care and worsens other patient care or (most cynically) leads to a small diminution in the quality of patient care that is cost-justified (Cohen 2020: 1445-1446).

Second, Cohen (2020) suggests that one might argue for two narrow legal obligations to disclose on the basis of precedents set in *Johnson v. Kokemoor*. In this case, a patient sued their surgeon after being rendered quadriplegic, alleging that the surgeon violated the patient's informed consent by failing to disclose that they were relatively inexperienced in performing the procedure. The Wisconsin Supreme Court ruled that "a reasonable person in the plaintiff's position would have considered such information material in making an intelligent and informed decision about the surgery" (*DeGennaro v. Tandon* 2005, cited in Cohen 2020: 1447). Cohen suggests that one might use this precedent to argue that clinicians have narrow legal obligations to disclose their use of medical ML systems in at least two circumstances.

Cohen suggests that clinicians may have a narrow legal obligation to disclose the 'qualifications' of a medical ML system where the system itself is understood as another member of the patient's care team. Second, Cohen suggests that non-specialist clinicians may have a narrow legal obligation to disclose their use of medical ML systems that enable them to perform clinical tasks that are typically reserved for specialists. For instance, non-ophthalmologists may be legally obligated to disclose their use of IDx-DR by Digital Diagnostics to patients since the diagnosis of diabetic retinopathy is typically restricted to ophthalmologists. Indeed, Digital Diagnostics themselves advise clinicians to inform patients "that their images are analyzed to determine whether further examination is needed by an eye care provider" (US Food and Drug Administration 2018a: 2).

Cohen (2020) also suggests that one might argue for a narrow legal obligation to disclose on the basis of precedents set in *Hurley v. Kirk* and in *Perna v. Pirozzi*. In each of these cases, the courts rules that information about *who* performs a clinical intervention is material to a patient's decision. For instance, in *Hurley v. Kirk*, the Supreme Court of Oklahoma ruled that

> the doctrine of informed consent requires a physician to obtain the patient's consent before using a non-doctor to perform significant portions of a surgery for which the physician was engaged to perform thereby subjecting the patient to a heightened risk of injury (*Hurley v. Kirk* 2017, cited in Cohen 2020: 1436).

Moreover, in *Perna v. Pirozzi*, the New Jersey Supreme Court ruled that a "patient has a right to choose the surgeon who will operate on him and refuse to accept a substitute" (*Perna v. Pirozzi* 1983, cited in Cohen 2020: 1437).[3] Cohen suggests that one might use this precedent to argue that, where patients requests a particular clinician to perform their diagnostics or treatment planning, clinicians have a narrow legal obligation to disclose their use of medical ML systems.

---

[3] This case was not tried as a breach of informed consent. However, Cohen (2020) argues that informed consent law provides a superior explanation for the breach.

However, I suggest that several further narrow legal obligations to disclose may also be supported by each of these precedents. For instance, the precedent set in *Moore v. Regents, University of California* provides strong support for a legal obligation to disclose under circumstances in which a clinician has financial stakes in the organisation responsible for the development of medical ML systems that they use to inform their clinical judgements or recommendations. This financial conflict of interest could interfere with the clinician's judgement when interpreting the outputs of these systems, or when deciding if a medical ML system ought to be used in a patient's care. For instance, vested financial interests in the developing organisation may cause a clinician to be more susceptible to automation bias, or prompt a clinician to use a medical ML system inappropriately (e.g. in ways that are misaligned with the patient's best interest).

The precedent set in *Johnson v. Kokemoor* also provides strong support for a legal obligation to disclose under circumstances in which the clinicians is inexperienced in using medical ML systems. This is because such clinicians are at heightened risk of misinterpreting the outputs of these systems in ways that may cause patient harm, e.g. giving them too little or too much weight in their clinical judgements. In some cases, inappropriate treatments may be recommended by clinicians whose judgements are significantly influenced by the outputs of medical ML systems with which they have little experience. This risk is also heightened by the use of *adaptive* ML systems that change over time and between clinical sites, as I discuss further in chapter five. A reasonable person in such patients' positions may find this information material when deciding whether they will act on or accept a clinician's judgements or recommendations.

In some circumstances at least, the common law reasoning recounted by Cohen can be used to argue for a legal obligation that clinicians disclose their use of medical ML systems to patients. In particular, one could argue that clinicians have a narrow legal obligation to disclose their use of medical ML systems where these systems are optimised to prioritise financial austerity over patients' best interests, where these systems enable generalist clinicians to perform specialist tasks, or where specific clinicians are requested to perform a particular clinical task. While Cohen ultimately expresses doubt about these arguments being accepted in US courts, I have argued that common law reasoning may nevertheless support a narrow legal obligation for clinicians to disclose their use of medical ML systems where the clinician has a financial stake in the developing organisation of the system, or where a clinician is relatively inexperienced in using a medical ML system to inform their clinical judgements. Further support for a broad or narrow legal argument to disclose could be generated by appealing to the underlying normative justifications for informed consent requirements, as I now discuss.

### d. Normative reasoning

Philosophers defend the need for informed consent requirements in medicine on several distinct bases. Nir Eyal (2019), for instance, distinguishes between seven primary justifications of informed consent requirements that have been advanced in the philosophical literature. According to these justifications, informed consent requirements are needed to:

i.    protect the medical and non-medical interests of patients;

ii.   protect and promote patient autonomy in medical decision-making;

iii.  prevent patients against being subject to abusive conduct by clinicians (e.g. deceit or exploitation);

iv.   protect and restore relations of trust between patients and clinicians;

v.    protect patients against being subject to the arbitrary control of another's will.

vi.   protect the self-ownership rights of patients;

vii.  preserve the personal integrity of patients;

Broad or narrow legal (and ethical)[4] obligations that clinicians disclose their use of medical ML systems could be defended if disclosing this information were to align with one or more of these justifications for informed consent.

For instance, if disclosing the use of medical ML systems were to protect the medical and non-medical interests of patients, then clinicians may have a broad legal obligation to disclose this information. However, Cohen argues that medical ML systems do not threaten the medical interests of patients due to the capacity of medical ML systems to tailor their outputs to each individual patient. As Cohen himself expresses, a medical ML system "is *more* particularized in its chance of success for this patient than either the patient or the physician might choose unaided" (Cohen 2020: 1445). Cohen also argues that medical ML systems do not threaten patients' non-medical interests so long as "the physician's recommendation [is] discussed with the patient and a decision [is] made in a shared decision-making framework that is also sensitive to non-medical interests" (Cohen 2020: 1445). In the following sections, however, I contest these claims by arguing that medical ML systems do in fact threaten to compromise the medical and non-medical interests of patients.

---

[4] Cohen's primary purpose throughout this argument remains that of assessing the case for a legal obligation that clinicians disclose their use of medical ML systems to patients. At this point in his argument, however, by turning to the underlying ethical and normative justifications for informed consent requirements, Cohen also implicitly evaluates several arguments for an *ethical* obligation to disclose.

Clinicians may also have a broad legal obligation to disclose their use of medical ML systems if doing so is found to protect patients against interference with their autonomous decisions. However, Cohen argues that medical ML systems do not threaten patients' autonomous decision since, as discussed above, these systems provide only provide inputs for a clinician to consider in their reasoning process. Cohen thus suggests that failure to disclose the use of a medical ML system presents no greater threat to patients' autonomous decisions than failing to disclose the influence of various other inputs in a clinician's reasoning process (e.g. memories from medical school lectures). I have already highlighted some problems with this analogy above. Again, however, I also contest this claim of Cohen's in the following subsection by arguing that medical ML systems present greater threats to patient autonomy than he suggests.

By disclosing their use of medical ML systems, clinicians may also be protecting their patients against threats of abuse. According to Cohen, however, "it is hard to see why informed consent in the AI/ML case creates a *protection against abuse*, in the sense of preventing such 'offenses as assault, deceit, coercion, and exploitation" (Cohen 2020: 1447). He does not elaborate further, except to note the possible exception of "cases where AI/ML is implemented not in the patient's interest but in tension with that interest – for example, an AI/ML that is designed to recommend cheaper but less efficacious treatments than the physician otherwise would recommend" (Cohen 2020: 1447).

Disclosing their use of medical ML systems may also allow clinicians to protect and restore relations of trust with their patients. Cohen argues that patients are unlikely to fear medical ML systems enough to pose a serious threat to trust in their clinicians. However, if my arguments in chapter two are correct, the use of medical ML systems poses a serious threat to the quality of relations of trust in clinician-patient relationships. As Eyal (2019) observes, however, informed consent requirements typically cannot be justified on the basis of trust alone. While my arguments in chapter two provide some support for a broad legal obligation to disclose, therefore, they are unlikely to be sufficient.

Finally, by disclosing their use of medical ML systems, clinicians may protect patients' personal integrity or self-ownership rights, or protect patients against being subject to the arbitrary control of another's will. However, Cohen (2020) largely passes over these three justifications for informed consent on the grounds that they have little to say about, and no straightforward application to, the use of medical ML systems.

Ultimately, therefore, Cohen concludes that none of these seven justifications can support a broad legal (or, indeed, ethical) obligation to disclose. As signalled above, however, I contest some of Cohen's objections to these normative arguments in the following sections.

This concludes my critical analysis of Cohen's discussion of broad and narrow legal obligations that clinicians disclose their use of medical ML systems to patients to secure their informed consent in the US. While Cohen considers a broad range of arguments that could be advanced to support both broad and narrow legal obligations to disclose, he suggests that none of them succeed under US informed consent law, or are likely to succeed in US courts. I have argued, however, that Cohen's analysis overlooks several justifications that make the case for a legal obligation that clinicians disclose their use of medical ML systems in US jurisdictions stronger than Cohen suggests. In particular, clinicians may be legally obligated to disclose when they have financial stakes in the developing organisations of a medical ML system that they use. Clinicians that are inexperienced in using medical ML systems may also be legally obligated to disclose this information to patients.

However, law and ethics are far from coextensive. Even if failing to disclose their use of medical ML systems does not violate a clinician's legal obligation, at least in the US, it thus may nevertheless violate their ethical duties to patients. In the next section, therefore, I turn to argue directly for an ethical obligation that clinicians disclose their use of medical ML systems for treatment recommendation to secure patients' informed consent. In section five, moreover, I argue that clinicians have a broad ethical obligation to disclose for several reasons beyond protecting their patients' autonomy and securing their informed consent.

## 4. The case for an ethical obligation to disclose

As discussed in the introduction to this thesis, experts anticipate that medical ML systems for treatment recommendation will support the aims of personalised medicine by recommending treatments that are individually tailored to each patients' clinical, genetic, genomic, and environmental characteristics (see Khan et al. 2020; Sebastiani et al. 2022). In this section, however, I argue that using these systems in clinical practice will require that clinicians communicate with their patients about their use of these systems. This is because medical ML systems for treatment recommendation will often contain embedded ethical values that threaten to compromise patients' autonomous decisions and the ethical ideal of shared decision-making. Protecting patients against such interferences will require clinicians to disclose their use of these systems, and the embedded values contain within them, to secure their patients' informed consent.

Patients have an interest in ensuring their medical decisions and decision-making processes are guided by, and aligned with, their own personal systems of values. Protecting this interest is a core justification for current professional standards mandating 'shared decision-making' in medicine (Elwyn et al. 2012). In particular, shared decision-making requires that clinicians treat patients as equal partners in the clinical decision-making process, and clinicians must ensure that medical decision-making is guided by the patients' own values (Brock 1991). Clinicians may thus be ethically obligated to disclose their use of medical ML systems if doing so protects patients against decisions or decision-making processes that are misaligned with the patient's own system of values, or guided by someone (or something) else's system of values entirely.

Medical ML systems for treatment recommendation often contain embedded ethical values. For instance, as Rosalind McDougall (2019) has argued, IBM's Watson for Oncology implicitly prioritises certain ethical values over others by its very design. Watson for Oncology is an ML system designed to generate ranked lists of personalised treatment recommendations for cancer patients. Watson for Oncology ranks treatments according to how likely they are to maximise the duration of a patient's life. However, this value system is misaligned with the preferences of patients who value the quality of their life over its duration. Consequently, using this system risks sidelining the values of patients in clinical decision-making and compromising shared decision-making practices in medicine (McDougall 2019).

DreaMed Advisor Pro is another medical ML system for treatment recommendation that is likely to contain embedded ethical values. This system is designed to recommend insulin doses for diabetic patients using data collected from glucose monitors and insulin pumps (US Food and Drug Administration 2018e). The system recommends dosages that aim to achieve glycemic control in diabetic patients, which refers to the optimal serum glucose concentration to prevent or delay microvascular complications including retinopathy, nephropathy, and neuropathy (Cryer, Davis, and Shamoon 2003). However, pursuing glycemic control can cause recurrent morbidity and potential mortality in diabetic patients as a result of iatrogenic hypoglycemia (Cryer 2014), which occurs when diabetic patients' blood glucose levels drop below a healthy threshold as a result of receiving insulin therapy. As a result, the embedded values of DreaMed Advisor Pro may be misaligned with those of diabetic patients who value reducing the risk of iatrogenic hypoglycemia over achieving strong glycemic control. Failing to disclose the use of medical ML systems for treatment recommendation may conceal the influence of embedded ethical values on patients' medical decisions and decision-making processes. Clinicians, therefore, are ethically obligated to disclose their use of medical ML systems for treatment recommendation to ensure that the decision-making process is guided by the patients' own values from start to finish.

However, disclosing that a clinician has used a medical ML system containing embedded ethical values to generate patients' treatment recommendations may not be sufficient to ensure that patients' own values continue to drive the decision-making process. This is because, by initiating these discussions after having received the recommendations of the system, or in response to them, clinicians may continue to sideline the values of patients. As McDougall has expressed:

> The patient's own values should be overtly shaping treatment decision making as a primary parameter, not a secondary consideration. Patient values should not be discussed as a reaction to an already ranked list. Such an approach diminishes the patient's role and represents a backward step in respecting patient autonomy (McDougall 2019: 158).

Responding to this concern, McDougall (2019) suggests that medical ML systems must be 'value-flexible' to avoid embedded values that are misaligned with the patient's own. She defines value-flexible medical ML systems as those that "allow for diversity among the values of individual users and can incorporate different values into decision making based on the specific user" (McDougall 2019: 158). Value-flexibility could be achieved in medical ML systems by translating information about patients' values into input data for medical ML systems to use in their analyses and outputs. Currently, data concerning patients' values are not currently used as input data for medical ML systems. But some argue that translating information about patients' values into input data for medical ML systems is both technically feasible and ethically desirable (Meier et al. 2022; Di Nucci 2019). For instance, incorporating patients' values into medical ML systems could enable these systems to generate outputs that are personalised to the values of each individual patient. Moreover, ML systems may more reliably predict the preferences of non-autonomous patients than human decision-makers (Lamanna and Byrne 2018).

However, using patient values as input data for medical ML systems also generates risks. For instance, while some argue that medical ML systems could predict patient values and preferences from social media data (Lamanna and Byrne 2018), patient values and preferences are often subject to change over time. Data collected from patients' past may not reliably predict their values and preferences in the present (Arnold 2021). As noted earlier, the outputs of medical ML systems are often seen as more objective and impartial than human judgements. Patients may mistakenly believe that medical ML systems know more about their own values and preferences than they do themselves. Clinicians may also be inclined to accept the medical ML systems assessment of the patients values over the patient's own account, or unquestioningly assume that the recommendations of a medical ML system automatically align with their patients' values. Clinicians may also rely on medical ML systems to understand patients' values and preferences rather than do the cognitive and emotional work involved in conversing with patients, eliciting their values and preferences, and coming to a shared understanding of treatment objectives. Promoting value-flexibility by attempting to predict or use information about patients' values and preferences risks being a solution worse that the problem it is trying to solve.

A more appropriate solution to the threats medical ML systems present to patient autonomy may be for clinicians to explicitly disclose the values that have been embedded in the system and their influence over the system's outputs. This would require both AI developers and clinicians to engage directly with the ethical values embedded in the technologies they create or use. In particular, developers and clinicians would need to explicitly identify these embedded values and understand their impact on their systems' recommendations to communicate this impact to patients. Clinicians may also protect patients' autonomous choices by explicitly discussing the recommendations of medical ML systems, as distinct from their own recommendation, at specific stages of the decision-making process. For instance, Glyn Elwyn and co-authors (2012) develop a model of shared decision-making in medicine that consists of three distinct phases: choice talk, option talk, and decision talk. *Choice talk* is a planning stage in which clinicians discuss available treatment options with their patients, emphasising the uncertainty of medicine and medical knowledge and deferring closure of the decision until later in the process. During *option talk*, clinicians discuss the available treatment options with patients in greater detail, outlining their anticipated benefits and harms. During *decision talk*, clinicians offer support to patients when deciding which treatment option is most suitable for them, eliciting the patients' values and preferences and reviewing the decision and decision-making process once a decision is made. To protect the integrity of patients' autonomous decisions, clinicians may delay explicit discussion of the recommendations of a medical ML system until the decision talk phase of the shared decision-making progress.

Ultimately, clinicians have a narrow ethical obligation to disclose their use of medical ML systems for treatment recommendation, and the embedded values contained within these systems, to secure their patients' informed consent. Failure to do so risks allowing these embedded values to drive the decision-making progress rather than ensuring the patient's own values remain at the centre of this process. While McDougall (2019) advocates for value-flexibility in medical ML systems to keep the patient's own values at the centre of medical decision-making, this approach risks generating costs to the quality of clinician-patient communication and clinician-patient relationships. Rather than changing the technology themselves, therefore, it is more suitable that clinicians adapt to this technology for the benefit of their patients. However, clinicians may also have a broad ethical obligation to disclose their use of medical ML systems to patients for reasons other than protecting their autonomy and securing their informed consent, as I now argue.

## 5. Beyond patient autonomy

Clinicians are also ethically obligated to disclose their use of medical ML systems to patients for reasons other than protecting patients' autonomous decisions. In particular, I argue in this section that clinicians have a broad ethical obligation to disclose their use of medical ML systems due to the notable risks these systems present to patients' health and safety, privacy, and confidentiality. I also argue that clinicians have a broad ethical obligation to disclose their use of medical ML systems to enable patients to exercise their right to reject diagnostics and treatment planning by medical ML systems, recently advanced by Thomas Ploug and Søren Holm (2019).

Maximilian Kiener (2020) has recently argued that clinicians have a broad ethical obligation to disclose their use of medical ML systems to patients in order to protect patients' medical and non-medical interests. In particular, Kiener suggests this is because clinicians are ethically obligated to disclose three specific risks associated with their use of medical ML systems to patients: the risks of cyberattack, algorithmic bias, and 'mismatch'. In this subsection, I argue that, while the risks associated with algorithmic bias and mismatch ought to be disclosed to patients, the risks associated with cyberattacks are too negligible to warrant disclosure. I also build upon Kiener's arguments for the disclosure of algorithmic bias.

First, Kiener suggests that clinicians have a broad ethical obligation to disclose their use of medical ML systems because these systems are especially vulnerable to 'adversarial attacks'. Adversarial attacks refer to a particular type of cyberattack that can be performed without infiltrating or interfering with the inner workings of an ML system. For instance, during an adversarial attack, so-called 'adversarial examples' – i.e. normal looking inputs with human imperceptible distortions – are inputted into an ML system that manipulate it into generating erroneous classifications or predictions. Medical adversarial examples are particularly easy to generate and appear to threaten patient harm by virtue of their ability to manipulate the outputs of medical ML systems, thereby interfering with the quality and safety of clinicians' judgements and recommendations.

However, the principal risk of adversarial attacks is that they will be used to perpetrate health insurance fraud rather than interfere with patient diagnostics or treatment planning (Finlayson et al. 2019). Moreover, while it is true that medical adversarial examples are particularly easy to generate, they are also particularly easy to detect using simple feature-based detectors (Ma et al. 2021). Currently, therefore, adversarial attacks do not appear to pose a serious threat to patient safety. An obligation for clinicians to disclose the threat of adversarial attacks may therefore be setting the threshold of risk disclosure too low. For if clinicians are ethically obligated to disclose the risk of adversarial attacks, they may also be obligated to disclose innumerable other marginal risks that may simply confuse and disorient patients rather than protect their interests.

Second, Kiener suggests that clinicians have a broad ethical obligation to disclose their use of medical ML systems due to the risk of 'mismatch', which Kiener describes as follows: "Since these AI systems are still insufficiently sensitive to causation as opposed to correlation, they may sometimes recommend courses of action that *do not match* the background situation of the individual patient, potential leading to great harm" (Kiener 2020: 710). I have previously discussed the risk of mismatch in chapter one. For instance, recall the medical ML system designed to predict mortality risk in patients presenting to an emergency department with pneumonia, developed by Rich Caruana and co-authors (2015). Despite asthmatic patients' high risk of mortality, this system began classifying these patients as low risk as a result of misinterpreting a correlative relationship as a causal one. The risk of mismatch presents significant risks to patient safety that clinicians ought to therefore disclose to patients.

Finally, as Kiener argues, clinicians have a broad ethical obligation to disclose their use of medical ML systems due to the risk of algorithmic bias. As discussed in chapter one, medical ML systems often adopt the biases of their designers or the societies in which they are developed, leading to substandard care, safety risks, and the inequitable distribution of healthcare resources. Kiener highlights that one may object that human clinicians are also biased (see Hoffman et al. 2016; Salles et al. 2019), yet they are not ethically obligated to disclose these risks to patients. For instance, as John Banja has expressed, "[e]xcessive attention paid to AI's errors and their implications for fairness ignores the ubiquity of human error, thus holding AI technologies to an unfairly high bar" (Banja 2019: 34). However, as Kiener notes, failure to disclose the risks of algorithmic bias in medical ML systems is likely to promote false beliefs in patients due to automation bias, in which individuals "display an unjustified reliance on machines over human decisions and are likely to think that AI is free from the frailties of human choice" (Kiener 2020: 709).

Further support for an ethical obligation to disclose the risks associated with algorithmic bias is provided by considering the practice of off-label drug prescribing, which refers to the practice of prescribing a drug for purposes or patient groups that lie beyond its approved scope. Off-label prescribing is both commonplace and unavoidable since, historically, drug trials have been conducted using ancestrally homogenous cohorts of research subjects. Certain patient groups are also systematically excluded from drug trials (e.g. pregnant people, palliative care patients, and children) that has further restricted diversity in participant cohorts. While clinicians are not legally obligated in the US and other jurisdictions to disclose instances of off-label drug prescribing to patients at present (Meadows and Hollowell 2008; Mithani 2012), Margaret Johns (2006) argues that this ought to change due to, among other things, their risks to patient health and safety (see also Wilkes and Johns 2008).

As Nicholson Price (2019) observes, using medical ML systems to treat patients from historically marginalised demographics is analogous to off-label drug prescribing because, just as the efficacy and side effects of certain drugs are often worse amongst patients outside the ancestrally homogenous cohort of research subjects on which they were tested, medical ML systems may perform worse on patients from socially disadvantaged communities. If clinicians are legally or ethically obligated to disclose off-label drug prescribing to patients, therefore, then they are also legally or ethically obligated to disclose their use of medical ML systems to patients from socially disadvantaged groups. Indeed, the need for labelling templates and standards for medical ML systems supporting such an obligation have recently been advanced in the literature (Gerke 2023; Mitchell et al. 2019).

Clinicians also have a broad ethical obligation to disclose their use of medical ML systems due to the risks these systems present to patient privacy and confidentiality. For instance, patient privacy may be infringed if data inputted into a medical ML system is shared with the system's developers to update the performance of the system or develop new systems from scratch. Indeed, maintaining continuous learning in adaptive ML systems may require that patient data is shared on an ongoing basis with the developers of these systems. Even where patient data is anonymised, the potential for data breaches suggests that patients have a right to opt-out of having their health information used as input data for medical ML systems.

Finally, clinicians are (broadly) ethically obligated to disclose their use of medical ML systems to allow patients to exercise their right to refuse diagnostics and treatment planning by medical ML systems. This right has recently been defended by Thomas Ploug and Søren Holm (2019), who advance three distinct arguments to support it.

First, Ploug and Holm observe that the outputs of medical ML systems are currently insensitive to patients' own values and preferences. Consequently, Ploug and Holm suggest that so long as medical ML systems cannot engage in meaningful, open-ended dialogue with patients about these values and preferences, then patients are entitled to refuse diagnostics and treatment planning by these systems to protect themselves against interventions that are misaligned with their values.

While it is possible that AI systems could extract insights about a patient's values from social media data and data from electronic health records, Ploug and Holm argue that data pertaining to patients' medical and non-medical values are limited. Electronic health records seldom contain information about patients' values. Social media data is often subject to privacy restrictions that interfere with their availability to incorporate into the AI system. Moreover, this data may be unreliable because of 'impression management' on online platforms. Moreover, preferences and values are often unstable and subject to change. Even when information about a patient's values can be acquired, this data may be out of date.

Second, AI systems are vulnerable to 'algorithmic bias'. Moreover, the opacity of machine learning AI systems makes these biases difficult for human clinicians to detect. Ploug and Holm suggest that patient's ought to have the option to opt-in to the risk of being affected by algorithmic bias rather than have these risks forced upon them. As a result, they suggest that algorithmic bias in medical AI constitutes a second reason in favour of a right to refuse. Like Kiener, Ploug and Holm consider the objection that biases are not unique to algorithmic systems in that human clinicians are prone to exhibit biases as well. However, they suggest that biases in human clinicians are regulated by a variety of formal and informal social mechanisms including team-based decision-making, informed consent requirements, education and training in medical ethics and law, and the potential for reputational harm. They conclude that so long as patients have good reason to believe that the regulatory mechanisms for avoiding and detecting biases in medical AI systems is weaker in comparison to that of human clinicians, they are justified and entitled to exercise a right to refuse.

Third, patients may have rational concerns about the undesirable societal implications of AI. Many patients express rational concerns about the risks and potential negative implications of medical ML systems (Richardson et al. 2021). Ploug and Holm define rational concerns as beliefs about the future state of the world that are minimally plausible and are reasonable according to the standards of public reason. They argue that rational concerns about the undesirable societal implications of medical AI include reductions in human care and concern in medicine, deskilling of clinicians, monopolisation of diagnostics and treatment planning by AI systems, and visions of AI-induced dystopias. They identify five reasons that a right to refuse based on rational concern ought to be granted. First, because democratic societies ought to be sensitive to the rational concerns of groups of citizens, and ought to grant citizens the capacity to express and act on their rational concerns. Second, liberal societies promote self-determination and the exercise of personal autonomy, and the capacity to act on one's rational concerns promotes the exercise of personal autonomy. Third, the healthcare system often accommodates niche concerns for the sake of solidarity (e.g. Jehovah's Witnesses). Rational concerns ought to equally be subject to accommodation. Fourth, a right to refuse could have the consequentialist benefit of reducing or avoiding the anticipated societal harms caused by AI in the future. Finally, because rational concerns are sensitive to evidence, there is an ethical and epistemological imperative to allow patients to act upon their rational concerns.

These concerns can relate to the impact of using medical ML systems upon one's own medical treatment and care. However, patients also have a right to act on rational concerns about the broader impact of medical technologies. These include concerns about their negative impact on the medical profession, on public health, or on society in general. As discussed in chapter one, incorporating medical ML systems generates risks in medicine. Some patients may hold rational concerns about the impact of ML systems in medicine. Clinicians may therefore be ethically obligated to disclose their use of medical ML systems to enable patients to exercise this right.

Notably, Ploug and Holm (2019) distinguish between weak and strong variants of the right to refuse. The *weak variant* asserts that patients have a right to refuse diagnostics and treatment planning performed entirely by an AI system. The *strong variant* asserts that patients have a right to refuse any and all involvement of AI systems in diagnostics and treatment planning.

According to Ploug and Holm, the justificatory power of the arguments from both autonomy and discrimination are limited to the weak variant of the right to refuse, i.e. "a right to demand that physicians are actively engaging with patients about their preferences, and that any output from an AI system is scrutinised by physicians prior to implementation" (Ploug and Holm 2019: 112). However, this is consistent with the objectives of legally marketed ML systems, which are typically designed to assist human clinicians in the performance of clinical tasks, largely designed to provide clinical decision support. The prospect of ML systems performing diagnostics and/or treatment recommendations from start to finish is unfeasible at present and for the near to mid-future. Hence, the arguments from autonomy and anti-discrimination cannot justify a broad ethical obligation for clinicians to disclose their use of medical ML systems to their patients so long as clinicians actively engage with their patients' preferences and scrutinise the outputs of these systems prior to acting on them. In contrast, however, the argument from rational concern supports the strong variant of the right to refuse, as it "is not primarily based on a worry about AI involvement in 'my' particular treatment, but a concern about the systemic effects of AI introduction and use in the health care system" (Ploug and Holm 2019: 112). While the autonomy and anti-discrimination justifications does not justify a broad ethical obligation to disclose, therefore, the right to act on rational concerns does.

Clinicians that communicate proactively with patients about their use of medical ML systems, the risks that these systems present, and their role in clinicians' judgements and recommendations, may enable patients to maintain a sense of control and empowerment during the transition to ML-enabled medicine. As Ploug and Holm (2020) have argued elsewhere, patients must be able to effectively contest the use of medical ML systems in practice, including the use of their personal health information as input data for medical ML systems, the potential biases that may be embedded in these systems, their performance characteristics, and the division of labour between medical ML systems and human clinicians. Ensuring that patients can effectively contest these various factors associated with clinicians' use of medical ML systems will require that clinicians communicate sensitively and discerningly to patients about their use of these systems and their various risks.

Ultimately, clinicians have a broad ethical obligation to disclose their use of medical ML systems to patients due to the risks these systems present to patient health and safety and the threat of algorithmic bias. Clinicians also have a broad ethical obligation to disclose their use of medical ML systems to enable patients to exercise their right to refuse diagnostics and treatment planning by medical ML systems.

## 6. Conclusion

In this chapter, I have argued that clinicians are ethically obligated to disclose their use of medical ML systems for treatment recommendation to patients to secure their informed consent. This is because medical ML systems contain embedded ethical values that threaten to interfere with patients' autonomous decision-making and compromise their informed consent. However, clinicians are also ethically obligated to disclose their use of medical ML systems to patients for reasons other than protecting patients' autonomous decisions and securing their informed consent. This is because medical ML systems present a variety of safety risks that warrant disclosure to patients. Using patient data as input data for medical ML systems also risks sharing patient data with the developers of these systems or making them vulnerable to data breaches which warrant disclosure to patients. Patients also have a right to act on rational concerns about the current and future impact of new medical technologies. Clinicians are ethically obligated to disclose their use of medical ML systems to patients to enable them to exercise this right to act on rational concerns about the current or future impact of medical ML systems.

Over the past two chapters, I have been concerned with analysing the impact of medical ML systems in general on clinician-patient relationships. I have argued that medical ML systems are likely to compromise the quality of relations of trust between clinicians and patients, and interfere with patients' autonomous decisions, thereby generating new communicative obligations for clinicians. Over the next two chapters, however, I turn to analyse the distinctive impact of two specific types of medical ML systems on clinician-patient relationships. In particular, I analyse the impact of 'opaque' medical ML systems (chapter four) and 'adaptive' medical ML systems (chapter five).

## (4)  THE DARK SIDE OF ACCURACY

### A.  OPACITY IN MEDICAL MACHINE LEARNING

#### 1.  Introduction

Over the past two chapters, I have argued that medical ML systems are likely to negatively impact clinician-patient relationships by compromising the quality of trust and patient autonomy in medicine. Over the next two chapters, I turn to evaluate the impact that two specific types of medical ML systems – opaque and adaptive systems – are likely to have on the quality of clinician-patient relationships.

As discussed in the introduction to this thesis, medical ML systems are increasingly being developed using deep learning and deep neural networks. However, these approaches to developing medical ML systems are notoriously opaque insofar as clinicians, patients, and even the designers of these systems themselves cannot understand how they reach their outputs. Several other computational architectures that are commonly used to develop medical ML systems also exhibit opacity (as cognitive mismatch; see part A, section two), including support vector machines, non-linear models, and tree ensembles (see Guidotti et al. 2018). In this chapter, I argue that using these types of opaque ML systems in medicine is likely to negatively impact trust in clinician-patient relationships by interfering with clinicians' capacity to communicate appropriately with patients about their ML-informed judgements and recommendations. I also argue that accepting these costs to clinician-patient relationships cannot be justified by appealing to the ostensibly superior accuracy of these systems.

This chapter is divided into three parts. In the remainder of part A, I argue that using opaque ML systems in medicine is likely to compromise the quality of communication and understanding between clinicians and patients. This is because opacity in medical ML systems undermines clinicians' capacity to test the outputs of these systems against their own knowledge and experience. Insofar as clinicians cannot test the outputs of medical ML systems in this manner, they will often be unable to answer basic questions about the reasons supporting patients' diagnoses or recommended treatments.

part B of this chapter consists of my article, 'The virtues of interpretable medical artificial intelligence', co-authored with Robert Sparrow and Mark Howard and published in the *Cambridge Quarterly of Healthcare Ethics*. Recently, several writers have argued that opaque ML systems ought to be used in medicine because these systems demonstrate superior accuracy and reliability (Durán and Jongsma 2021; Topol 2019; Wang, Kaushal, and Khullar 2020). As Alex John London has expressed:

> Any preference for less accurate models – whether computational systems or human decision-makers – carries risks to patient health and welfare. Without concrete assurance that these risks are offset by the expectation of additional benefits to patients, a blanker preference for simpler models is simply a lethal prejudice (London 2019: 18).

In this article, however, we argue that prioritising the use of opaque ML systems in medicine exhibits a "lethal prejudice" of its own. This is because the impact of using these systems on patient health and safety depends not only on the accuracy and reliability of these systems, but also on how users respond to the outputs of these systems. We highlight several underappreciated threats that the use of opaque medical ML systems present to patient health and welfare. We also note that the superior accuracy of opaque ML systems over interpretable systems appears to be exaggerated. We conclude that the use of less accurate but interpretable medical ML systems may in some cases generate better patient health outcomes more accurate but opaque systems. Finally, in part C of this chapter, I conclude that the anticipated benefits associated with using highly accurate, but opaque, ML systems is unable to justify the costs these systems are likely to impose on the quality of clinician-patient relationships.

The remainder of part A of this chapter proceeds as follows. In section two, I distinguish between three varieties of opacity in medical ML systems, and I highlight what Jenna Burrell (2016) refers to as 'opacity as cognitive mismatch' as my primary concern in this chapter. In section three, I argue that using opaque medical ML systems is likely to interfere with clinicians' ability to test the outputs of these systems against their own knowledge and experience, and reduce patients' understanding of the reasons behind their clinicians' judgements and recommendations.

## 2. What is opacity?

Before analysing the impact that opaque medical ML systems are likely to have on clinician-patient relationships, it is important to specify the particular type of opacity under consideration in this chapter. This is because opacity in ML systems comes in a variety of different forms. For instance, as Jenna Burrell (2016) observes, ML systems exhibit at least three distinct types of opacity: 'opacity as intentional concealment', 'opacity as technical illiteracy', and 'opacity as cognitive mismatch'.

*Opacity as intentional concealment* occurs when developers of an ML system restrict the availability of some (or all) of the system's internal components to those inside an organisation (e.g. the training data, training algorithm, test data, loss function, etc.). Typically, such information is withheld to retain a competitive advantage over rival organisations. Occasionally, however, it can be used to conceal unethical or illegal organisational practices (see Pasquale 2015). In contrast, *opacity as technical illiteracy* occurs when a user lacks the requisite technical knowledge to understand or interpret a model's inner workings or operations. Just as a motorist without a detailed knowledge of automotive engineering may inspect the engine of their vehicle and have no understanding of how it works, a user without a background in computer programming may try to look 'under the hood' of an ML system and be similarly bewildered and perplexed. Finally, *opacity as cognitive mismatch* occurs when the causal reasoning process of an ML system cannot be explained in terms that fall within the parameters of human understanding. More specifically, Burrell claims that opacity as cognitive mismatch occurs because of a "mismatch between mathematical optimization in high-dimensionality [that is] characteristic of machine learning and the demands of human-scale reasoning and styles of semantic interpretation" (Burrell 2016: 2).

Each of these types of opacity present distinctive challenges and concerns in their own right. In this chapter, however, my aim is to analyse the effects of using medical ML systems that exhibit *opacity as cognitive mismatch* on the quality of clinician-patient relationships. This is because, as I noted above, opacity as cognitive mismatch is a distinctive characteristic of several powerful computational architectures that are frequently used to develop medical ML systems (e.g. deep neural network, tree ensembles, support vector machines, and non-linear models; see Guidotti et al. 2018). All further references to opacity ought therefore to be understood to refer specifically to opacity as cognitive mismatch.

## 3. Costs to clinician–patient relationships

In this section, I argue that the use of opaque ML systems is likely to negatively impact patient understanding and communication between clinicians and patients in medicine. The main reason for this is that opacity in medical ML systems generates substantial obstacles for clinicians with respect to testing the outputs of these systems against their own knowledge and expertise. In particular, opacity in medical ML systems precludes clinicians from evaluating the reasoning process these systems take to arrive at each of their outputs. Consequently, as Nicholson Price observes:

> Once [a physician] has decided to use a particular black-box algorithm – itself a complex choice – he or she cannot understand and thus verify the algorithm's recommendation against his or her body of substantive expertise; the physician can only accept what the algorithm recommends or not (Price 2018: 300-301).

The inability to scrutinise the outputs of these systems generates substantial obstacles for communication between clinicians and patients. This is because clinicians simply cannot communicate what they do not understand. As Jens Christian Bjerring and Jacob Busch note:

> since black-box AI systems do not reveal to practitioners how or why they reach the recommendations that they do, then neither can practitioners who rely on these black-box systems in decision-making […] explain to patients how and why they give the recommendations that they do (Bjerring and Busch 2021: 361).

Where clinicians rely on the outputs of these systems to inform their judgements and recommendations, they will often also be unable to answer basic questions that their patient may have about their health and medical treatment (e.g. Why is treatment $x$ more likely than treatment $y$ to improve my condition? Why am $I$ particularly at risk of post-operative complications from this surgical procedure?). Clinicians that rely on the outputs of medical ML systems will often be unable to meaningfully or reliably answer these questions since they cannot understand how these systems reached their outputs.

These obstacles to communication and understanding generate further threats to relations of trust between clinicians and patients, adding to existing threats discussed in chapter two. In particular, the inability to answer basic questions about their patients' health and medical treatment will likely lead some patients to doubt the authority of their clinician, perceive their clinician as less-than-competent, or withdraw their reliance on them. This is not only because such patients may be unsettled by their clinicians' inability to answer such questions, but also because patients already tend to perceive clinicians who use decision-support systems as less professional, less thorough and systematic, and less competent than clinicians who do not use clinical decision-support tools (Arkes, Shaffer, and Medow 2007; Shaffer et al. 2013). Where clinicians are unable to answer basic questions and communicate with their patients about the underlying rationale for their judgements and decisions, this tendency to derogate clinicians that use these systems is likely to be exacerbated, further compromising relations of trust in medicine.

Opacity in medical ML systems also provides further support for a broad ethical obligation that clinicians disclose their use of these systems to patients. This is because, as I argued in the previous chapter, clinicians are ethically obligated to disclose certain risks that these systems present to patient health and well-being. However, opacity in medical ML systems substantially increases the severity of many of these risks. In particular, the inability of developers and regulators to evaluate the reasoning of these systems presents major obstacles to detecting technical weaknesses and errors in their development and operation, such as those discussed in chapter one (e.g. distributional shift, variable confounding, data leakage, etc.). For instance, as discussed in chapter one, Rich Caruana and co-authors developed an ML system to predict mortality risk for hospital in-patients presenting with pneumonia. Concerningly, this system learned to classify asthmatic patients as low-risk due to a statistical misinterpretation. When deciding between proceeding with an interpretable, rule-based system or an opaque artificial neural network, Caruana and co-authors decided "to not use the neural nets not because the asthma problem could not be solved, but because the lack of intelligibility made it difficult to know what *other* problems might also need fixing" (Caruana et al. 2015: 1722). Clinicians, regulators, and developers will also struggle to detect and remove algorithmic biases from medical ML systems whose inner workings cannot be scrutinised or understood (see Cabitza, Rasoini, and Gensini 2017; He et al. 2019; Terrasse, Gorin, and Sisti 2019; Yoon, Torrance, and Scheinerman 2021; Watson et al. 2019).

To protect and preserve the quality of clinician-patient relationships in the coming age of AI in medicine, it seems that we ought to prioritise the use of medical ML systems that are 'interpretable' to clinicians and patients. Recently, however, critics such as London have argued that prioritising interpretability in medical ML systems entails compromising on the accuracy of these systems due to what is known as the 'accuracy-interpretability' trade-off in ML, discussed further in part B (see Burrell 2016; Defense Advance Research Projects Agency 2016; Mori and Uchihira 2019), as I now discuss.

## B. THE VIRTUES OF INTERPRETABLE MEDICAL ARTIFICIAL INTELLIGENCE

[PDF begins on next page].

# The Virtues of Interpretable Medical Artificial Intelligence

Joshua Hatherley* , Robert Sparrow and Mark Howard

School of Philosophical, Historical, and International Studies, Monash University, Clayton, Victoria 3168, Australia
*Corresponding author. Email: joshua.hatherley@monash.edu

**Abstract**

Artificial intelligence (AI) systems have demonstrated impressive performance across a variety of clinical tasks. However, notoriously, sometimes these systems are "black boxes." The initial response in the literature was a demand for "explainable AI." However, recently, several authors have suggested that making AI more explainable or "interpretable" is likely to be at the cost of the accuracy of these systems and that prioritizing interpretability in medical AI may constitute a "lethal prejudice." In this article, we defend the value of interpretability in the context of the use of AI in medicine. Clinicians may prefer interpretable systems over more accurate black boxes, which in turn is sufficient to give designers of AI reason to prefer more interpretable systems in order to ensure that AI is adopted and its benefits realized. Moreover, clinicians may be justified in this preference. Achieving the downstream benefits from AI is critically dependent on how the outputs of these systems are interpreted by physicians and patients. A preference for the use of highly accurate black box AI systems, over less accurate but more interpretable systems, may itself constitute a form of lethal prejudice that may diminish the benefits of AI to—and perhaps even harm—patients.

## Introduction

Deep-learning artificial intelligence (AI) systems have demonstrated impressive performance across a variety of clinical tasks, including diagnosis, risk prediction, triage, mortality prediction, and treatment planning.[1,2] A problem, however, is that the inner workings of these systems have often proven thoroughly resistant to understanding, explanation, or justification, not only to end-users (e.g., doctors, clinicians, and nurses) but also to the designers of these systems themselves. Such AI systems are commonly described as "opaque," "inscrutable," or "black boxes." The initial response to this problem in the literature was a demand for "explainable AI." However, recently, several authors have suggested that making AI more explainable or "interpretable" is likely to be achieved at the cost of the accuracy of these systems and that a preference for explainable systems over more accurate AI is ethically indefensible in the context of medicine.[3,4]

In this article, we defend the value of interpretability in the context of the use of AI in medicine. We point out that clinicians may prefer interpretable systems over more accurate black boxes, which in turn is sufficient to give designers of AI reason to prefer more interpretable systems in order to ensure that AI is adopted and its benefits realized. Moreover, clinicians may themselves be justified in this preference. Medical AI should be analyzed as a sociotechnical system, the performance of which is as much a function of how people respond to AI as it is of the outputs of the AI. Securing the downstream therapeutic benefits from diagnostic and prognostic systems is critically dependent on how the outputs of these systems are interpreted by physicians and received by patients. Prioritizing accuracy over interpretability overlooks the various human factors that could interfere with downstream benefits to patients. We argue that, in some cases, a less accurate but more interpretable AI may have better effects 2    Joshua Hatherley et al.

on patient health outcomes than a "black box" model with superior accuracy, and suggest that a preference for the use of a highly accurate black box AI systems, over less accurate but more interpretable systems, may itself constitute a form of lethal prejudice that may diminish the benefits of AI to patients— and perhaps even harm them.

### The Black Box Problem in Medical AI

Recent advances in artificial AI and machine learning (ML) have significant potential to improve the current practice of medicine through, for instance, enhancing physician judgment, reducing medical error, improving the accessibility of medical care, and improving patient health outcomes.[5,6,7] Many advanced ML AI systems have demonstrated impressive performance in a wide variety of clinical tasks, including diagnosis,[8] risk prediction,[9] mortality prediction,[10] and treatment planning.[11] In emergency medicine, medical AI systems are being investigated to assist in the performance of diagnostic tasks, outcome prediction, and clinical monitoring.[12] A problem, however, is that ML algorithms are notoriously opaque, "in the sense that if one is a recipient of the output of the algorithm […], rarely does one have any concrete sense of how or why a particular classification has been arrived at from inputs."[13] This can occur for a number of reasons, including a lack of relevant technical knowledge on the part of the user, corporate or government concealment of key elements of an AI system, or at the deepest level, a cognitive mismatch between the demands of human reasoning and the technical approaches to mathematical optimization in high-dimensionality that are characteristic of ML.[14] Joseph Wadden suggests that the black box problem "occurs whenever one tries to explain why an AI decision-maker has arrived at its decision are not currently understandable to the patient or those involved in the patient's care because the system itself is not understandable to either of these agents."[15]

A variety of related concerns have been raised over the prospect of black box clinical decision support systems being operationalized in clinical medicine. Some authors worry that human physicians may act on the outputs of black box medical AI without a clear understanding of the reasons behind them,[16] or that opacity may conceal erroneous inferences or algorithmic biases that could jeopardize patient health and safety.[17,18,19] Others are concerned that opacity could interfere with the allocation of moral responsibility or legal liability in the instance that patient harm results from accepting and acting upon the outputs of a black box medical AI system,[20,21,22] or that the use of black-box medical AI systems may undermine the accountability that healthcare practitioners accept for AI-related medical error.[23] Still, others are concerned that black box medical AI systems cannot, will not, and perhaps ought not to be trusted by doctors or patients.[24,25,26] These concerns are especially acute in the context of emergency medicine, where decisions need to be made quickly and coordinated across teams of multispecialist practitioners.

Responding to these concerns, some authors have argued that medical AI systems will need to be "interpretable," "explainable," or "transparent" in order to be responsibly utilized in safety-critical medical settings and overcome these various challenges.[27,28,29]

### The Case for Accuracy

Recently, however, some authors have argued that opacity in medical AI is not nearly as problematic as critics have suggested, and that the prioritization of interpretable over black box medical AI systems may have several ethically unacceptable implications. Critics have advanced two distinct arguments against the prioritization of interpretability in medical AI.

First, some authors have highlighted parallels between the opacity of ML models and the opacity of a variety of commonplace medical interventions that are readily accepted by both doctors and patients. As Eric Topol has noted, "[w]e already accept black boxes in medicine. For example, electroconvulsive therapy is highly effective for severe depression, but we have no idea how it works. Likewise, there are many drugs that seem to work even though no one can explain how."[30] The drugs that Topol is referring to here include aspirin, which, as Alex John London notes, "modern clinicians prescribed […] as an analgesic for nearly a century without understanding the mechanism through which it works," along with lithium, which "has been used as a mood stabilizer for half a century, yet why it works remains uncertain."[31] Other authors have also highlighted acetaminophen and penicillin, which "were in widespread use for decades before their mechanism of action was understood," along with selective serotonin reuptake inhibitors, whose underlying causal mechanism is still unclear.[32] Still, others have highlighted that the opacity of black-box AI systems is largely identical to the opacity of other human minds, and in some respects even one's own mind. For instance, John Zerilli and coauthors observe that "human agents are […] frequently *mistaken* about their real (internal) motivations and processing logic, a fact that is often obscured by the ability of human decision-makers to invent post hoc rationalizations."[33] According to some, these similarities imply that clinicians ought not to be anymore concerned about opacity in AI than they are about the opacity of their colleagues' recommendations, or indeed the opacity of their own internal reasoning processes.[34]

This first argument is a powerful line of criticism of accounts that hold that we should entirely abjure the use of opaque AI systems. However, it leaves open the possibility that, as we shall argue below, interpretable systems have distinct advantages that justify our preferring them.

The second argument assumes—as does much of the AI and ML literature—that there is an inherent trade-off between accuracy and interpretability (or explainability) in AI systems. In their 2016 announcement of the "Explainable AI (XAI)" project, for instance, the United States Defence Advanced Research Projects Agency claims that "[t]here is an inherent tension between ML performance (predictive accuracy) and explainability; often the highest performing methods (e.g., deep learning) are the least explainable, and the most explainable (e.g., decision trees) are less accurate."[35] Indeed, attempts to enhance our understanding of AI systems through the pursuit of intrinsic or ex ante interpretability (for instance, by restricting the size of the model, implementing "interpretability constraints," or using simpler, rule-based classifiers over more complex deep neural networks) are often observed to result in compromises to the accuracy of a model.[36,37,38] In particular, the development of a high-performing AI system entails an unavoidable degree of complexity that often interferes with how intuitive and understandable the operations of these systems are in practice.[39]

Consequently, some authors suggest that prioritizing interpretability over accuracy in medical AI has the ethically troubling consequence of compromising the accuracy of these systems, and subsequently, the downstream benefits of these systems for patient health outcomes.[40,41] Alex London has suggested that "[a]ny preference for less accurate models—whether computational systems or human decision-makers—carries risks to patient health and welfare. Without concrete assurance that these risks are offset by the expectation of additional benefits to patients, a blanket preference for simpler models is simply a lethal prejudice."[42] According to London, when we are patients, it is more important to us that something works than that our physician knows precisely how or why it works.[43] Indeed, this claim appears to have been corroborated by a recent citizen jury study, which found that participants were less likely to value interpretability over accuracy in healthcare settings compared to non-healthcare settings.[44] London thus concludes that the trade-off between accuracy and interpretability in medical AI ought therefore to be resolved in favor of accuracy.

### The Limits of Post Hoc Explanation

One popular response to these concerns is to hope that improvements in post hoc explanation methods could enhance the interpretability of medical AI systems without compromising their accuracy.[45] Rather than pursuing ex ante or intrinsic interpretability, post hoc explanation methods attempt to extract explanations of various sorts from black-box medical AI systems on the basis of their previous decision records.[46,47,48] In many cases, this can be achieved without altering the original, black-box model, either by affixing a secondary explanator to the original model or by replicating its statistical function and overall performance through interpretable methods.[49]

The range of post hoc explanation methods is expansive, and it is beyond the scope of this article to review them all here. However, some key examples of post hoc explanation methods include sensitivity 4    Joshua Hatherley et al.

analysis, prototype selection, and saliency masks.[50] *Sensitivity analysis* involves "evaluating the uncertainty in the outcome of a black box with respect to different sources of uncertainty in its inputs."[51] For instance, a model may return an output with a confidence interval of 0.3, indicating that it has produced this output with low confidence, with the aim of reducing the strength of a user's credence. *Prototype selection* involves returning, in conjunction with the output, an example case that is as similar as possible to the case that has been entered into the system, with the aim of illuminating some of the criteria according to which output was generated. For instance, suppose a medical AI system, such as IDx-DR,[52] was to diagnose a patient with diabetic retinopathy from an image of their retina. A prototype selection explanator might produce, in conjunction with the model's classification, a second example image that is most similar to the original case, in an attempt to illustrate important elements in determining its output. Lastly, *saliency masks* highlight certain words, phrases, or areas of the image that were most influential in determining a particular output.

Post hoc explanation methods *have* demonstrated some potential to minimize some of the concerns of the critics of opacity discussed in section "The Black Box Problem in Medical AI," while also side-stepping the objections of critics of interpretability discussed in section "The Case for Accuracy." However, post hoc explanation methods also suffer from a number of significant limitations, which preclude them from entirely resolving this debate.

First, the addition of post hoc explanation methods to "black box" ML systems adds another layer of uncertainty to the evaluation of their outputs and inner workings. Post hoc explanations can only offer an approximation of the computations of a black-box model, meaning that it may be unclear how the explanator works, how faithful it is to the model, and why its outputs or explanations ought to be accepted.[53,54,55]

Second, and relatedly, such explanations often only succeed in extracting information that is highly incomplete.[56] For example, consider an explanator that highlights the features of a computed breast tomography scan that were most influential in classifying the patient as high-risk. Even if the features highlighted were intuitively relevant, this "explanation" offers a physician little reason to accept the model's output, particularly if the physician disagrees with it.

Third, the aims of post hoc explanation methods are often under-specified, particularly once the problem of agent-relativity in explanations is considered. Explanations often need to be tailored to a particular audience in order to be of any use. As Carl Zednik has expressed, "although the opacity of ML-programmed computing systems is traditionally said to give rise to the Black Box Problem, it may in fact be more appropriate to speak of many Black Box Problems—one for every stakeholder."[57] An explanation that assumes a background in computer science, for instance, may be useful for the manufacturers and auditors of medical AI systems, but is likely to deliver next to no insight for a medical professional that lacks this technical background. Conversely, a simple explanation tailored to patients, who typically lack both medical and computer science backgrounds, is likely to provide little utility to a medical practitioner. Some post hoc explanations may prove largely redundant or useless, while others may influence the decisions of end-users in ways that could reduce the clinical utility of these systems.

Finally, the focus on explanation has led to the neglect of *justification* in explainable medical AI.[58,59] Explanations are descriptive, in that they give an account of why a reasoner arrived at a particular judgment, but justifications give a normative account of why that judgment is a *good* judgment. There is a significant overlap between explanations and justifications, but they are far from identical. Yet within the explainability literature in AI, explanations and justifications are rarely distinguished, and when they are, it is the former that is prioritized over the latter.[60]

Consequently, despite high hopes that explainability could overcome the challenges of opacity and the accuracy-interpretability trade-off in medical AI, post hoc explanation methods are not currently capable of meeting this challenge.

## Three Problems with the Prioritization of Accuracy over Interpretability

In this section, we highlight three problems underlying the case for accuracy, concerning (1) the clinical objectives of medical AI systems and the need for accuracy maximization; (2) the gap between technical accuracy in medical AI systems and their downstream effects upon patient health outcomes; and (3) the reality of the accuracy-interpretability trade-off. Both together and separately, these problems suggest that interpretability is more valuable than critics appreciate.

First, the accuracy of a medical AI system is not always the principal concern of human medical practitioners and may, in some cases, be secondary to the clinician's own ability to understand and interpret the outputs of the system, along with certain elements of the system's functioning, or even the system as a whole. Indeed, the priorities of clinicians are largely dependent upon their conception of the particular aims of any given medical AI system. In a recent qualitative study, for instance, Carrie Cai and coauthors found that the importance of accuracy to medical practitioners varies according to the practitioners' own conception of a medical AI system's clinical objectives.[61] "To some participants, the AI's objective was to be as accurate as possible, independent of its end-user. […] To others, however, the AI's role was to merely draw their attention to suspicious regions, given that the pathologist will be the one to make sense of those regions anyway: "It just gives you a big picture of this is the area it thinks is suspicious. You can just look at it and it doesn't have to be very accurate".[62] In these latter cases, understanding a model's reasons or justifications for drawing the clinician's attention to a particular area of an image may rank higher on the clinicians' list of priorities than the overall accuracy of the system, in order that they may reliably determine why a model has drawn the clinician's attention to a particular treatment option, piece of information, or area of a clinical image. This is not to deny the importance of accuracy in, say, a diagnostic AI system for the improvement of patient health outcomes, but rather to suggest that, in some cases, and for some users, the accuracy of an AI system may not be as critical as London and other critics have supposed, and may rank lower on the clinician's list of priorities than the interpretability of the system. Depending upon the specific performance disparities between black box and interpretable AI systems, there may be cases where clinicians prefer less accurate systems that they can understand over black-box systems with superior accuracy. If users prefer interpretable models over "black box" systems, then the potential downstream benefits of "black box" AI systems for patients could be undermined in practice if, for instance, clinicians reject them or avoid using them. Implementing "black box" systems over interpretable systems without respect for the preferences of the users of these systems may result in suboptimal outcomes that could have otherwise been avoided through the use of less accurate but more interpretable AI systems. Even if clinicians' preference for interpretability *is* a prejudice, if it is sufficiently widespread and influential, it may be sufficient to justify designers of AI to prioritize interpretability in order to increase the likelihood that AI systems will be adopted and their benefits realized.

Second, *contra* London, clinicians may themselves be justified in this preference. The case for accuracy appears to erroneously assume a necessary causal link between technical accuracy and improved downstream patient health outcomes. While diagnostic and predictive accuracies are certainly important for the improvement of patient health outcomes, they are far from sufficient. Medical AI systems need to be understood as intervening in care contexts that consist of an existing network of sociotechnical relations, rather than as mere technical "additions" to existing clinical decision-making procedures.[63,64,65] How these AI systems will become embedded into these contexts, and alter existing relations between actors, is crucially important to the extent to which they will produce downstream health benefits. As Sara Gerke and coauthors argue, the performance of medical ML systems will be influenced by a variety of broader human factors beyond the system itself, including the way that clinicians respond to the outputs of the systems, "the reimbursement decisions of insurers, the effects of court decisions on liability, any behavioral biases in the process, data quality of any third-party providers, any (possibly proprietary) ML algorithms developed by third parties, and many others."[66] Thus, as Thomas Grote observes in his recent discussion of clinical equipoise in randomized clinical trials of diagnostic medical AI systems, "even if the AI system were outperforming clinical experts in terms of diagnostic accuracy during the validation phase, its clinical benefit would still remain genuinely uncertain. *The main reason is that we cannot causally infer from an increase in diagnostic accuracy to an improvement of patient outcome*" (emphasis added).[67] There is a gap, in other words, between the accuracy of medical AI systems and their effectiveness in clinical practice, insofar as improvements in the accuracy of a technical system do not automatically translate into improvements in downstream health outcomes. Indeed, this observation is borne out by the current lack of evidence of downstream patient benefits generated from even the most technically accurate of medical AI systems. Superior accuracy is, in short, insufficient to demonstrate superior outcomes.

One reason for this gap comes from the fact that human users do not respond to the outputs of algorithmic systems in the same way that we respond to our own judgments and intuitions, nor even to the recommendations of other human beings. Indeed, according to one recent systematic review in human–computer interaction studies, "the inability to effectively combine human and nonhuman (i.e., algorithmic, statistical, and machine) decision making remains one of the most prominent and perplexing hurdles for the behavioral decision making community."[68] Prevalent human biases affect the interpretation of algorithmic recommendations, classifications, and predictions. As Sara Gerke and coauthors observe, "[h]uman judgement […] introduces well-known biases into an AI environment, including, for example, inability to reason with probabilities provided by AI systems, over extrapolation from small samples, identification of false patterns from noise, and undue risk aversion."[69] In safety-critical settings and high-stakes decision-making contexts such as medicine, these sorts of biases could pose significant risks to patient health and well-being.

Moreover, some of these biases are more likely to occur in cases where the medical AI system is opaque, rather than interpretable. Algorithmic aversion, for instance, is a phenomenon in which the users of an algorithmic system consistently reject the outputs of an algorithmic system, even when the user has observed the system performance to a high standard consistently over time, and when following the recommendations of the system would produce better outcomes overall.[70,71] Algorithmic aversion is most commonly observed in cases where the users of the system have expertise in the domain for which the system is designed (e.g., dermatologists in the diagnosis of malignant skin lesions);[72] in cases where the user has seen the system make (even minor) mistakes;[73] but most importantly for our purposes, *in cases where the algorithmic system is perceived to be opaque by its user.*[74] "Thus," claim Michael Yeomans and coauthors, "it is not enough for algorithms to be more accurate, they also need to be understood."[75]

Finally, in passing, it is worth noting that some authorities have begun to contest the reality of the accuracy-interpretability trade-off in AI and ML. In particular, Cynthia Rudin has recently argued that the accuracy-interpretability trade-off is a myth, and that simpler, more interpretable classifiers can perform to the same general standard as deep neural networks after preprocessing, particularly in cases where data are structured and contain naturally meaningful features, as is common in medicine.[76,77] Indeed, Rudin argues that interpretable AI can, in some cases, demonstrate *higher* accuracy than comparatively black box AI systems. "Generally, in the practice of data science," she claims, "the small difference in performance between ML algorithms can be overwhelmed by the ability to interpret results and process the data better at the next iteration. In those cases, the accuracy/interpretability trade-off is reversed—more interpretability leads to better overall accuracy, not worse."[78] In a later article coauthored with Joanna Radin,[79] Rudin highlights a number of studies that corroborate the comparable performance of interpretable and black box AI systems across a variety of safety-critical domains, including healthcare.[80,81,82] Rudin and Radin also observe that even in computer vision and image-recognition tasks, in which deep neural networks are generally considered the state of the art, a number of studies have succeeded in implementing interpretability constraints to deep learning models without significant compromises in accuracy.[83,84,85] Rudin concludes that the uncritical acceptance of the accuracy-interpretability trade-off in AI often leads researchers to forego any attempt to investigate or develop interpretable models, or even develop the skills required to develop these models in the first place.[86] She suggests that black box AI systems ought not to be used in high-stakes decision-making contexts or safety-critical domains unless it is demonstrated that no interpretable model can reach the same level of accuracy. "It is possible," claim Rudin and Radin, "that an interpretable model can always be constructed—we just have not been trying. Perhaps if we did, we would never use black boxes for these high-stakes decisions at all."[87] While, as we have argued here, it will, at least in some circumstances, be defensible to prioritize interpretability at the cost of accuracy, if Rudin is correct, the price of the pursuit of interpretability may not be as high as critics—and our argument to this point—have presumed.

Superior accuracy is, therefore, not enough to justify the use of black-box medical AI systems over less accurate but more interpretable systems in clinical medicine. In many cases, it will be genuinely uncertain *a priori* whether a more accurate black-box medical AI system will deliver greater downstream benefits to patient health and well-being compared to a less accurate but more interpretable AI system. Indeed, under some conditions, less accurate but more interpretable medical AI systems may produce better downstream patient health outcomes than more accurate but nevertheless opaque systems.

## Conclusion

The prioritization of accuracy over interpretability in medical AI, therefore, carries its own lethal prejudices. While proponents of accuracy over interpretability in medical AI are correct to emphasize that the use of less accurate models carries risks to patient health and welfare, their arguments overlook the comparable risks that the use of more accurate but less interpretable models could present for patient health and well-being. This is not to suggest that the use of opaque ML AI systems in clinical medicine is unacceptable or ought to be rejected. We agree with proponents of accuracy that "black box" AI systems could deliver substantial benefits to medicine, and that the risks may eventually be reduced enough to justify their use. However, a blanket prioritization of accuracy in the technical trade-off between accuracy and interpretability itself looks to be unjustified. Opacity in medical AI systems may constitute a significant obstacle to the achievement of improved downstream patient health outcomes, despite how technically accurate these systems may be. More attention needs to be directed toward how medical AI systems will become embedded in the sociotechnical decision-making contexts for which they are being designed. The downstream effects of medical AI systems on patient health outcomes will be mediated by the decisions and behavior of human clinicians, who will need to interpret the outputs of these systems and incorporate them into their own clinical decision-making procedures. The case for prioritizing accuracy over interpretability pays insufficient attention to the reality of this situation insofar as it overlooks the negative effects that opacity could have upon the hermeneutic task of physicians in interpreting and acting upon the outputs of black box medical AI systems and subsequent downstream patient health outcomes.

**Acknowledgments.** Earlier versions of this article were presented to audiences at Macquarie University's philosophy seminar series, the University of Wollongong's seminar series hosted by the Australian Centre for Health Engagement, Evidence, and Values (ACHEEV), the University of Sydney's Conversation series hosted by Sydney Health Ethics, and the Ethics in AI Research Roundtable sponsored by Facebook Research, and the Centre for Civil Society and Governance of the University of Hong Kong. The authors would like to thank the audiences of these seminars for comments and discussion that improved the article.

**Funding Statement.** This research was supported by an unrestricted gift from Facebook Research via the Ethics in AI in the Asia-Pacific Grant Program. The work was conducted without any oversight from Facebook. The views expressed herein are those of the authors and are not necessarily those of Facebook or Facebook Research.

**Conflicts of Interest.** R.S. is an Associate Investigator in the ARC Centre of Excellence for Automated Decision-Making and Society (CE200100005) and contributed to this article in that role. J.H. was supported by an Australian Government Research Training Program scholarship.

## Notes

8    Joshua Hatherley et al.

## C. CONCLUSION

Clinicians' use of opaque medical ML systems is likely to negatively impact clinician-patient relationships because the use of opaque medical ML systems interferes with clinicians' capacity to test the outputs of these systems against their own knowledge and experience. Moreover, opaque medical ML systems are likely to negatively impact clinician-patient relationships because they are likely to compromise patients' capacity to understand the reasons behind clinicians' judgements and recommendations. While some argue that opaque medical ML systems ought nevertheless to be used in medicine on the basis that prioritising interpretable over opaque medical ML systems exhibits a "lethal prejudice," these critics overlook the substantial risks that clinicians' use of opaque medical ML systems present to patient health and safety. For instance, interfering with clinicians' capacity to evaluate the outputs of ML systems against their own knowledge and expertise is likely to compromise the quality of clinicians' ML-informed judgements and recommendations. In addition, clinicians are more likely to exhibit algorithmic aversion towards opaque medical ML systems compared to interpretable systems. Superior accuracy in medical ML systems, therefore, is insufficient to justify the costs these systems will have for the quality of clinician-patient relationships.

In this chapter, I have addressed the impact that the use of opaque medical ML systems is likely to have on clinician-patient relationships. In the next chapter, I turn to analyse the impact that another specific type of ML system is likely to have on clinician-patient relationships. In particular, I argue that the use of medical ML systems that continue learning from new data even after being deployed in a clinical setting, otherwise known as *adaptive* ML systems, are likely to negatively impact clinician-patient relationships by increasing clinicians' hermeneutic and administrative labour, and expanding existing risks to patient health and well-being that are likely to compromise patients' trust in their clinicians.

# (5)  THE COSTS OF CONTINUAL LEARNING

## A. INTRODUCTION

In the previous chapter, I argued that clinicians' use of opaque ML systems threatens to compromise the quality of communication and understanding between clinicians and patients. In this chapter, however, I turn to evaluate the impact of clinicians' use of 'adaptive' ML systems – i.e. systems that continue learning from new data even after being deployed in a clinical setting – on clinician-patient relationships. I argue that clinicians' use of adaptive ML systems is likely to negatively impact clinician patient relationships by expanding existing risks to patients' understanding (as I discussed in the previous chapter) and clinicians' administrative workloads (as I discuss further in chapter six). I also argue that adaptive ML systems generate distinctive risks of discrimination and inequity in medicine that may further compromise patient's trust in their clinicians.

Critical analysis of the ethical implications of clinicians' use of adaptive ML systems in medicine is urgently needed, for two reasons. First, while adaptive ML systems are not currently eligible for approval by regulatory agencies such as the US FDA, recent developments suggest that these systems are likely to become eligible in the near future. In particular, the US FDA has recently released multiple reports outlining a proposed framework for the regulation of adaptive ML systems in medicine. Second, writers in the ethics literature have focused almost exclusively on challenges associated with clinicians' use of 'locked' ML systems in medicine – i.e. systems that *do not* continue learning from new data after being implemented in a clinical setting – with few exceptions (see Babic et al. 2019; Gerke et al. 2020; Minssen et al. 2020; Smith and Severn 2022). Even in these exceptional circumstances, however, writers typically focus on regulatory, rather than ethical, challenges presented by adaptive ML systems.

My aim in this chapter is to remedy this gap in the literature by identifying some of the distinctive ethical risks presented by clinicians' use of adaptive ML systems in medicine. In particular, I argue that the use of adaptive ML systems generates new risks to patient health and safety, new hermeneutic challenges for clinicians trying to interpret and evaluate the outputs of medical ML systems, expanded risks of increased administrative labour for clinicians, new threats to patients' informed consent, and new threats of discrimination and inequity in medicine. I argue that these various risks are likely to compromise the quality of clinician-patient relationships by reducing patients' understanding, clinicians' capacity to care for and empathise with their patients, and patients' perceptions of the skill and competency of their clinicians.

The remainder of this chapter proceeds as follows. The main argument of this chapter is presented in part B, which consists of my article, 'Diachronic and synchronic variation in the performance of adaptive machine learning systems: the ethical challenges', co-authored with Robert Sparrow and published in the *Journal of the American Medical Informatics Association*. In this article, we argue that the performance of adaptive ML systems exhibits two types of variation, evolution over time ('diachronic evolution') and variation between clinical sites ('synchronic variation'). We argue that diachronic evolution and synchronic variation in adaptive ML systems generate several risks to clinicians and patients that have thus far been overlooked in the literature. In part C of this chapter, I discuss the implications of the challenges identified in part B for clinician-patient relationships, and identify future directions for further research into the ethical implications of MAMLS.

B. DIACHRONIC AND SYNCHRONIC VARIATION IN THE PERFORMANCE OF ADAPTIVE MACHINE LEARNING SYSTEMS IN MEDICINE: THE ETHICAL CHALLENGES

[PDF begins on next page].

## Review

# Diachronic and synchronic variation in the performance of adaptive machine learning systems: the ethical challenges

Joshua Hatherley ![ORCID]* and Robert Sparrow

Philosophy Department, School of Philosophical, Historical and International Studies, Monash University, Clayton, Victoria 3800, Australia

*Corresponding Author: Joshua Hatherley, MBioethics, Philosophy Department, School of Philosophical, Historical and International Studies, Monash University, Level 6, 20 Chancellor's Walk (Menzies Building), Wellington Road, Clayton, VIC 3800, Australia; joshua.hatherley@monash.edu

### ABSTRACT

**Objectives:** Machine learning (ML) has the potential to facilitate "continual learning" in medicine, in which an ML system continues to evolve in response to exposure to new data over time, even after being deployed in a clinical setting. In this article, we provide a tutorial on the range of ethical issues raised by the use of such "adaptive" ML systems in medicine that have, thus far, been neglected in the literature.
**Target audience:** The target audiences for this tutorial are the developers of ML AI systems, healthcare regulators, the broader medical informatics community, and practicing clinicians.
**Scope:** Discussions of adaptive ML systems to date have overlooked the distinction between 2 sorts of variance that such systems may exhibit—diachronic evolution (change over time) and synchronic variation (difference between cotemporaneous instantiations of the algorithm at different sites)—and underestimated the significance of the latter. We highlight the challenges that diachronic evolution and synchronic variation present for the quality of patient care, informed consent, and equity, and discuss the complex ethical trade-offs involved in the design of such systems.

**Key words:** artificial intelligence, bioethics, update problem, medicine, federated learning

## INTRODUCTION

Machine learning (ML) has the potential to facilitate "continual learning" in medicine, in which an ML system continues to adapt and evolve in response to exposure to new data over time, even after being deployed in a clinical setting. Leveraging this "adaptive" potential of medical ML could generate significant benefits for patient health and well-being. Recent engagements with the ethical issues generated by the use of adaptive ML systems in medicine have typically been limited to discussions of "the update problem": how should systems that continue to change and evolve postregulatory approval be regulated? In this article, we draw attention to an important set of ethical issues raised by the use of adaptive ML systems in medicine that have, thus far, been neglected and are highly deserving of further attention.

Discussions of adaptive ML systems to date have overlooked the distinction between 2 sorts of variance that such systems may exhibit—diachronic evolution (change over time) and synchronic variation (difference between cotemporaneous instantiations of the algorithmic system at different sites)—and underestimated the significance of the latter. Both diachronic evolution and synchronic variation will complicate the hermeneutic task of clinicians in interpreting the outputs of AI systems, and will therefore pose significant challenges to the process of securing informed consent to treatment. Equity issues may occur where synchronic variation is permitted, as the quality of care may vary significantly across patients or between hospitals. However, the decision as to whether to allow or eliminate synchronic variation involves complex trade-offs between accuracy and generalizability, as well as a number of other values, including justice and nonmaleficence. In some contexts, preventing synchronic variation from emerging may only be possible at the expense of the wellbeing, and the quality of care available to, particular patients or classes of patients. Designers and regulators of adaptive ML systems will need to confront these issues if the potential benefits of adaptive ML in medical care are to be realized.

## ADAPTIVE MACHINE LEARNING IN MEDICINE

ML is a form of AI that involves "programming computers to optimize a performance criterion using example data or past experience".[1] The application of ML in medicine could significantly improve the delivery of medical care, and expand the availability of medical knowledge and expertise, among other benefits.[2–5] ML systems can be either "locked" or "adaptive". *Locked* ML systems have parameters fixed prior to clinical deployment, and do not continue to learn from new data over time. While, to date, regulatory approvals of medical AI systems have been limited to locked systems the U.S. Food and Drug Administration (FDA) is considering regulatory approval for *adaptive* ML systems, which evolve as they are exposed to new data ("continuous learning"), even after the system has been deployed in a clinical setting.[6,7] We will refer to these sorts of ML devices as (medical) adaptive machine learning system(s) (MAMLS).

The use of MAMLS could have a number of benefits for patients. In some applications, MAMLS can continuously "tune" their algorithms to individual patients' physiology, along with any changes that occur in a patient's physiology over their lifetime, thereby contributing to the realization of "personalised medicine". The use of ML to deliver personalized medicine is already being explored via the combination of ML with a variety of other new and emerging technologies.[8] For example, ML-enabled wearables and implantables have been developed to enable personalized identification of ventricular arrythmias and hypoglycemic events for diabetic patients, and also to predict the onset of seizures in patients with drug-resistant epilepsy.[9–12]

Additionally, MAMLS could be trained on data collected from particular cohorts of patients to tune their performance to the features of the cohorts of each particular clinical site or institution.[13] For example, MAMLS could be used to predict risk of hospital readmission for outpatients, or to identify patients at a high-risk of heart attack within particular communities.[14] Some researchers are already seeking to enable such site-specific training of medical ML systems by making the source codes of their algorithms freely available online.[15]

## THE UPDATE PROBLEM

While there has been some engagement with the ethical issues raised by MAMLS, it has mostly been confined to discussions of "the update problem". Existing regulatory approaches in healthcare and medicine were designed to address products that do not evolve over time, such as pharmaceuticals. Consequently, the capacity for ongoing evolution in MAMLS presents a serious challenge for regulators. As Babic and coauthors have written: "After evaluating a medical AI/ML technology and deeming it safe and effective, should the regulator limit its authorization to market only the version of the algorithm that was submitted, or permit marketing of an algorithm that can learn and adapt to new conditions?".[16] If they approve MAMLS, regulators may be exposing patients to risks that have developed in the system postdeployment. However, restricting regulatory approvals to locked systems places a strong limit on the potential benefits that ML could generate for patient health outcomes.

In their recent *Proposed Regulatory Framework for Modifications to ML-Based Software as a Medical Device (SaMD)*[6] and subsequent *Artificial Intelligence/Machine Learning (AI/ML)-Based Software as a Medical Device Action Plan,*[7] the US FDA has attempted to address the update problem. A key feature of the FDA's preferred approach is the requirement for manufacturers of MAMLS to provide algorithmic change protocols (ACPs) as part of their applications for premarket approval. ACPs are supposed to outline how a MAMLS will change over time and what the limits of these changes will be. They will also require manufacturers to state how they will mitigate any risks that these changes will present. The FDA suggests that this approach could allow MAMLS to be approved and deployed in clinical settings without a need for ongoing regulatory review.

A number of serious criticisms have been raised against the FDA's proposed framework.[16,17] For instance, the proposed gives little indication as to how the performance of MAMLS will be monitored in practice, even suggesting that manufacturers could monitor these systems themselves. We are sympathetic to many of the concerns that have been expressed in the literature. However, we believe that the current focus on regulatory challenges that MAMLS present has led researchers to overlook the broader set of ethical issues that the use of these systems in medicine will present.

## TWO TYPES OF VARIATION: DIACHRONIC AND SYNCHRONIC

The literature on the ethics of MAMLS is cognizant that these systems will evolve over time—a phenomenon that we shall call *diachronic evolution*. As MAMLS continue learning from new data, their parametric weightings will change from update to update. They will respond differently to identical input data at different times. Their accuracy and performance will evolve over time, for better or worse. They may even adopt different classes of algorithmic bias as they continue to learn and evolve.

However, it is less often recognized that variation will emerge between copies of a MAMLS that have been implemented across different sites.

*Synchronic variation* refers to the differences that will emerge between copies of a MAMLS implemented at different sites or in different patients. MAMLS will be deployed across diverse clinical settings with different data collection policies, organizational procedures, user behaviors, data infrastructures, and patient demographics, each of which will affect the datasets upon which these systems learn. Even small variations in the datasets on which an algorithm learns can have significant effects on what it learns. If each copy of a MAMLS learns from data collected from the site at which it has been deployed, either exclusively, or even just to fine tune its parameters after initial learning from a training dataset, then these differences between site-specific datasets mean that copies of a MAMLS deployed at different sites (or devices implanted in different individuals) are likely to diverge over time. Eventually, identical data entered into different copies of a MAMLS will likely cause these systems to generate different outputs.

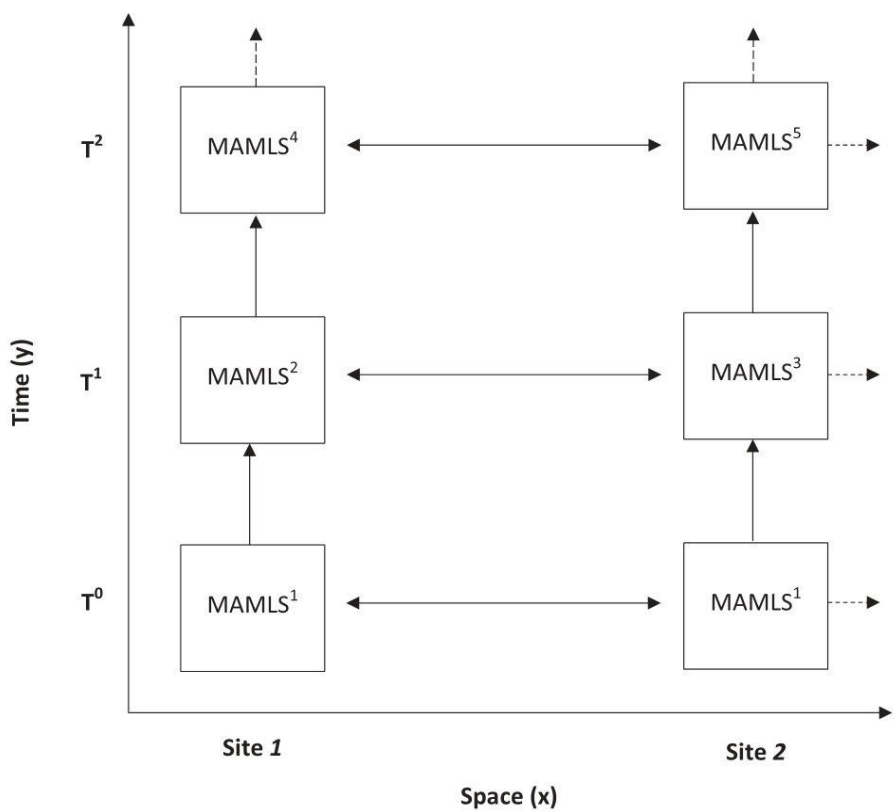

**Figure 1.** Schematic representation of diachronic evolution (y-axis) and synchronic variation (x-axis) in a MAMLS deployed across 2 sites.

Figure 1 illustrates the relation between these 2 types of variation.

Diachronic evolution alone would seem to require that regulators monitor the evolution of each MAMLS over time and conduct postmarket surveillance of the product's performance, errors, and instances of patient harm as it evolves. The possibility of synchronic variation suggests that, in addition regulators should monitor each *copy* of a MAMLS, due to the gradual divergence that may occur *between* each copy of the product over time. The additional administrative burden that this could entail would be expensive and time-consuming for both manufacturers, purchasers, and regulatory agencies. Moreover, the more instantiations there are, the more likely it is that one will go catastrophically wrong and undercut support for all of them. These factors could increase costs for manufacturers and chill the incentive to develop these systems in the first place.

## FEDERATED LEARNING TO PREVENT SYNCHRONIC VARIATION

In some cases, however, manufacturers may have the option of eliminating synchronic variation entirely via the adoption of "federated learning" (FL), which "involves training statistical models over remote devices or siloed data centers, such as mobile phones or hospitals, while keeping data localized".[18] To date, interest in FL in healthcare has mostly been driven by their potential to maintain privacy.[4,19] However, if FL can be used to train MAMLS, they could allow each iteration of a MAMLS to learn on the same pool of data, which would preclude the emergence of synchronic variation.

Eliminating synchronic variation would have a number of benefits for stakeholders. For instance, it would reduce the burden of reg-ulators and manufacturers by eliminating divergence and variation between copies of a MAML and thus the risk that different copies might require distinct regulatory evaluation and approval. In some cases, it might improve the generalizability of MAMLS by enabling these systems to learn from larger, more heterogenous datasets collected across multiple clinical sites. As we will argue below, it could also eliminate the potential for inequities in standards of care to emerge across clinical sites.

However, the decision to allow or eliminate synchronic variation carries practical trade-offs (along with some ethical trade-offs: see "Costs to particular cohorts" in the following section). Implementing FL in MAMLS may require updates to existing digital infrastructure that may be prohibitively expensive for many clinics and hospitals. Federated learning, for instance, "require[s] investment in on-premise computing infrastructure or private-cloud service provision and adherence to standardised and synoptic data formats so that ML models can be trained and evaluated seamlessly".[20] Moreover, if data collected at one hospital cannot easily be transferred to new sites without first being processed, this processing may itself introduce a form of synchronic variation by virtue of adding different extra layers of code to the AI system at different sites. Where implanted medical devices include MAMLS, FL will only be possible if these devices can transmit the result of training on local data back to other instantiations of the learning algorithm *and* can update the local algorithm in the light of the results of the training of other versions there-of, which increases the risk to patient privacy and iatrogenic harm, including as a result of hacking.

In an important set of cases, then, users will face a choice between either allowing synchronic variation to occur, or not using MAMLS at all.

## ETHICAL CONSIDERATIONS

The deployment and use of MAMLS generates a number of ethical concerns relating to the quality of patient care and doctor–patient relations, informed consent to treatment, threats to health equity and problems of obsolescence, and harms to particular cohorts of patients.

### Impact on quality of care

The use of MAMLS presents a number of risks to the quality of patient care.

Left to continue learning postdeployment MAMLS may adopt erroneous and potentially dangerous associations from new input data that could jeopardize patient health. In one well-known case, a mortality-prediction ML system learned to classify asthmatic patients presenting to the emergency department with pneumonia as "low risk" due to a true but misleading correlation in the training data.[21] If operationalized, the system would have presented critical risks to patient safety. With MAMLS there is a risk that such errors will emerge postdeployment as a result the process of continual learning. Moreover, MAMLS are susceptible to the phenomenon of "catastrophic forgetting", in which a MAMLS overwrites what it has previously learned during the process of learning from new data, leading to sudden poor performance that could significantly jeopardize the quality of physician judgments and the health and safety of patients.[22] Finally, MAMLS are susceptible to hacking and adversarial attacks, including by "data poisoning". In locked ML systems, adversarial attacks can only affect individual outputs.[23] In MAMLS, however, adversarial attacks could interfere with the performance of the system (or systems) in all future uses. These possibilities highlight the urgency of the "update problem".

### Challenges to clinical interpretation

Furthermore, achieving downstream benefits from the use of ML in medicine is critically dependent upon clinicians' ability to understand, interpret, and act on the outputs of these systems.[24] For instance, clinicians must decide how much epistemic weight they ought to give the outputs of algorithms in their clinical decision-making. Placing too much or too little weight on the outputs of an algorithmic system can result in patient harm, even death. An example of clinicians placing too little epistemic weight in the output(s) of an algorithmic system is "alert fatigue", which refers to "declining clinician responsiveness to a particular type of alert as the clinician is repeatedly exposed to that alert over a period of time, gradually becoming 'fatigued' or desensitized to it".[25] Alert fatigue can lead clinicians to ignore important alerts, potentially resulting in patient harm or death (for a particularly egregious instance of patient harm caused by alert fatigue, see reference [26]). An example of patient harm caused by clinicians placing too much epistemic weight in the outputs of an algorithmic system is "automation bias", which "refers to errors resulting from the use of automated cues as a heuristic replacement for vigilant information seeking and processing".[27] The presence of diachronic evolution and synchronic variation in MAMLS will pose a significant challenge to clinicians being able to reliably interpret and act upon the outputs of these systems. If every time clinicians encounter a MAMLS it is a subtly (or occasionally not so subtly) different system—different both to previous iterations, and between patients and across clinical sites—it may be exceedingly difficult for them to be confident how it is functioning and how much they should trust it. These challenges are further complicated by the opacity of ML systems, which make it

difficult to understand how or why a system works or has produced a certain output.[28]

Admittedly, that the performance of MAMLS will change over time, and will differ between sites and/or patients, does not in-and-of-itself distinguish them radically from other systems with which clinicians must engage in the course of their professional practice. Clinicians who work across different institutions often have to take account of differences in the way things are done, or particular devices are set up, in each institution. The fact that diagnostic tools and treatments are evolving all the time is, after all, why continuing medical education is so important. However, the key virtue of MAMLS is their ability to continuously improve at a faster rate than existing diagnostic tools without human intervention or oversight. The speed with which MAMLS evolve may outpace clinician's abilities to adapt to these changes.

### Impact on doctor–patient relations

Where clinicians make use of MAMLS for the purpose of clinical decision-support, both diachronic evolution and synchronic variation will pose challenges to communication between doctors and patients and reduce the capacity for shared decision making. If clinicians are themselves not able to understand precisely what has changed between each update to a MAMLS system, or how, precisely, the system they are dealing with at this site, or in this patient, differs from other iterations, they may struggle to identify and explain the factors that are casually relevant to their ultimate decision about a diagnosis and/or treatment plan. In particular, they may find it difficult to provide the patient with counter factual information that might be relevant to shared decision-making about treatment. Importantly, this effect may occur even if the clinician is in fact justified—and can explain to the patient that they are justified—in relying on the MAMLS because of its superior accuracy relative to the alternatives.

It is sometimes argued that the use of AI and ML could allow clinicians more time to spend with their patients.[29,30] However, the various tasks associated with maintaining AI and ML systems could equally lead to increased administrative burdens for clinicians that could further interfere with the quality of care and empathy in the doctor–patient relationship.[31,32] This risk seems particularly acute in the case of MAMLS, because healthcare institutions will likely need to significantly expand the scope of their data collection policies and procedures to be able to provide the continuous stream of new data that training MAMLS will require.

The potential of MAMLS to evolve over time may also be expected to exacerbate the issues related to computers being "the third party in the room" in clinical consultations. As Christopher Pearce and others have noted, the introduction of computers into healthcare settings has transformed what was originally a dyadic relationship, between the doctor and patient into a triadic relationship between the doctor, the patient, and the doctor's computer.[33,34] Both doctor and patient now spend some—perhaps even much—of their time "together" looking at and relating to the computer: information provided by the computer shapes the course of the consultation. If the doctor's computer is—or accesses—a MAMLS then this will add an important temporal dimension to the relationship between the doctor and the computer and the patient and the computer. What the computer "says" may change over time. This alone may be sufficient to draw more of the doctor's and the patient's attention to the computer. However, the fact that the operations and the outputs of the MAMLS may change also opens up the possi-bility that doctors will become involved in trying to manage or shape those changes in order to meet their, and their patients, expectations. One might imagine clinicians trying to influence the evolution of the MAMLS by curating the data that they input into it, in the same way many of us now try to manage the recommendation engines of Spotify or Netflix. Clinicians' relationships with MAMLS will evolve along with the MAMLS and we should expect that at least some clinicians may want to be active in shaping the former evolution—and, thus, the latter.

### Challenges to informed consent

Insofar as it is not typically considered necessary for clinicians to inform patients about the technologies that they have used to inform their clinical recommendations, the use of MAMLS for decision support is unlikely to have implications for informed consent. However, where MAMLS assist in the delivery of medical *treatment* (eg, robotic surgery or, hypothetically, AI-guided radiation therapy) their nature may well be relevant to the process of securing informed consent to treatment.

The use of ML in treatments already involves new risks that may need to be disclosed to patients, such as the threat of cyberattack.[23,35] Adaptive learning will introduce additional risks, including the risk of catastrophic forgetting and of algorithmic biases developing postdeployment, which may need to be disclosed to patients in order to allow them to make an informed decision of the use of such systems. Moreover, the potential of MAMLS to evolve and to differ between sites and patients means that the provision of general or "standard" information about treatments guided by MAMLS may not be sufficient to secure informed consent to treatment. A patient who returns to a medical clinic for treatment involving a MAMLS after some time will undergo treatment that may differ subtly, or even significantly, from that they received in their previous visit. Similarly, a patient who moves from one hospital to another, which has implemented a version of the same MAMLS, may be subject to different levels of risk—indeed, different risks—in each location. Fully informed consent, then, may require that patients are made aware of the risks associated with treatment with the particular MAMLS that is involved in their treatment. However, diachronic evolution and synchronic variation, coupled with the characteristic opacity of ML systems, mean that it may not always be possible for manufacturers to provide information about the specific risks associated with a particular iteration of a MAMLS.

### Equity and obsolescence

One hopes that, with appropriate regulation, the continuous learning of MAMLS will lead to improved outcomes for patients over time. In-and-of-itself, then, diachronic evolution in the performance of MAMLS should not raise issues of equity.

The ability of MAMLS to adapt to specific patient cohorts and improve the performance of the system among these cohorts has the potential to promote health equity by better serving the needs of minority groups that are often under-represented in the training data used to train locked models. However, where synchronic variation is permitted, it is also possible that the difference in the performance of MAMLS at different sites or in different patients may become so pronounced as to generate serious issues of justice in relation to the quality of healthcare available to different cohorts. Particular instantiations of a MAMLS may have biases that are more pronounced, more numerous, or more consequential within the patient cohort that they serve, than other iterations of the product deployed at different sites. Moreover, it is possible that some iterations of a MAMLS product may become stuck in local minima during the learning process, such that their performance stagnates while others continue to improve. In some cases, these performance disparities may become so large that the MAMLS available to particular sites/patients are effectively obsolete.

### Costs to particular cohorts

The challenges that synchronic variation presents for equity may serve as another incentive for manufacturers and regulators to try to eliminate synchronic variation. However, although FL is likely to enhance the generalizability of MAMLS, as Futoma and coauthors have noted, "the demand for universal rules—generalisability—often results in [ML] systems that sacrifice strong performance at a single site for systems with mediocre or poor performance at many sites" [reference 36, see also reference 37]. Disease, symptoms, side-effects, and so on occur with differing probabilities across lines of race, sex, gender, ability, and so on, and the application of a one-size-fits-all model across different subpopulations will often result in a system having differing utility for members of different cohort. Indeed, it can result in a model that is suboptimal for all groups, or optimal only for the dominant subpopulation—a phenomenon known as "aggregation bias".[38] For this reason, the decision to prevent synchronic variation in MAMLS involves an ethical and political trade-off between prioritizing the health and well-being of dominant groups and the prioritization of the health and well-being of marginalized groups.

## CONCLUSION

We have argued that the implementation of MAMLS raises a number of challenging ethical issues that have thus far received little attention. We distinguished between 2 sorts of variance that such systems may exhibit—diachronic evolution (change over time) and synchronic variation (difference between cotemporaneous instantiations of the algorithm at different sites). Diachronic evolution complicates the hermeneutic task of clinicians and could interfere with downstream patient benefits. Maintaining the digital infrastructure and data collection requirements necessary to enable continual learning in MAMLS may generate greater administrative burdens for human physicians, resulting in compromised relations of care and empathy between doctors and patients. Synchronic variation has the potential to generate inequities between clinical sites using the same MAMLS. The choice between site-specific and FL approaches involves a trade-off between pursuing generalizability or local impact, and may be to the detriment of particular cohorts of patients. These ethical issues require sustained attention if we are to realize the benefits of continuous learning in medicine.

## FUNDING

This research was supported by the Australian Government through the Australian Research Council's Centres of Excellence funding scheme (project CE140100012). RS is also an Associate Investigator in the ARC Centre of Excellence for Automated Decision-Making and Society (CE200100005) and contributed to this article in that role. JH was supported by an Australian Government Research Training Program scholarship. The views expressed herein are those of the authors and are not necessarily those of the Australian Government or Australian Research Council.

## AUTHOR CONTRIBUTIONS

RS and JH were jointly responsible for the research design. JH completed the literature search and analysis. JH wrote the original draft, which was revised and edited by RS. Both authors then contributed to a further round of revisions and approved the final version of the manuscript for publication.

## ACKNOWLEDGMENTS

The authors thank Mark Howard for drawing their attention to relevant literature.

## CONFLICT OF INTEREST STATEMENT

None declared.

## DATA AVAILABILITY

No new data were generated or analyzed in support of this research.

## C. DISCUSSION AND FUTURE DIRECTIONS

In part B of this chapter, I have argued that clinicians' use of MAMLS presents a range of distinctive risks and challenges. In particular, MAMLS generate new risks to patient health and safety, new hermeneutic challenges for clinicians trying to interpret and evaluate the outputs of medical ML systems, expanded risks of increased administrative labour for clinicians, new threats to patients' informed consent, and new threats of discrimination and inequity in medicine.

Each of these risks and challenges are likely to negatively impact the quality of clinician-patient relationships. For instance, patients may be reluctant to place their trust in clinicians that use MAMLS to inform their judgements and recommendations insofar as these systems generate new and substantial threats to patients' health and safety, e.g. via the threat of catastrophic forgetting and the post-implementation emergence of algorithmic biases. Patients may also be reluctant to trust clinicians the use MAMLS insofar as these systems generate new threats of discrimination and inequity in medicine. Each of these additional threats to patient safety and health equity also provide further support for an ethical obligation that clinicians disclose their use of these systems to patients, discussed in chapter three. By expanding the risk of increased administrative labour for clinicians, the use of MAMLS is likely to further reduce the time that clinicians have to care for and empathise with their patients, as I discuss further in the following chapter.

The use of MAMLS also expands existing threats to clinicians' capacity to test the outputs of these systems against their own knowledge and expertise associated with the opacity of these systems, discussed in the previous chapter. This is because diachronic variation in these systems may confound clinicians' attempts to understand or anticipate the behaviour of these systems in clinical practice as they change over time. Synchronic variation may also increase these challenges for clinicians that work across multiple institutions insofar as the behaviour of each iteration of a MAMLS will differ at each institution. MAMLS thus expand existing threats to the quality of communication and understanding between clinicians and patients. These threats to communication and understanding also risk interfering with patients' informed consent insofar as they threaten to compromise patients' understanding of certain information that is material to their medical decisions.

MAMLS may also increase the risks that medical ML systems present to patient privacy, discussed in chapter one. Maintaining continual learning in MAMLS will require healthcare organisations to continuously collect patient health information to use as training data for these systems. This may expand existing incentives for healthcare organisation to collect and use patient data in ways that risk compromising their privacy. For instance, expanded data collection practices incentivised by the implementation of MAMLS appears likely to expand the risk of patient data being used in ways to which patients do not consent, or shared with persons or institutions that infringe patients' right to privacy. These privacy risks associated with the use of MAMLS provide further support for a broad ethical obligation that clinicians disclose their use of medical ML systems to protect patients' privacy and confidentiality, discussed in chapter three.

Over the past two chapters, I have been concerned with the impact of two specific types of medical ML systems – opaque and adaptive systems – on clinician-patient relationships. I have argued that both types of systems present distinctive risks to the quality of these relationships. However, I am yet to address what are arguably the most substantial threats to the quality of clinician-patient relationships, and Topol's vision for the coming age of AI in medicine. In particular, while Topol believes that medical ML systems are likely to enhance and promote the provision of care and empathy in the clinician-patient relationship, I argue in the following chapter that medical ML systems are more likely to greatly reduce the provision of empathetic caregiving in medicine.

## (6)   CARE, EMPATHY, AND THE PERILS OF 'DEEP' MEDICINE

### 1.   Introduction

Recently, several writers have argued that the use of ML systems could substantially improve the quality of care and empathy in medicine. Eric Topol, for instance, has argued that ongoing advances in the performance and capabilities of ML systems present an opportunity "to bring back real medicine: Presence. Empathy. Trust. Caring. Being Human" (Topol 2019: 309). This is because, according to Topol, medical ML systems are likely to reduce the administrative responsibilities of human clinicians, granting them more time to spend caring for their patients. Topol also argues that medical ML systems could also improve the quality of care and empathy in medicine by increasing the value of caregiving skills in medical education and employment, leading to the training and hiring of 'expert caregivers' (see also Graeber 2019; Susskind and Susskind 2015).

Topol presents a hopeful and optimistic vision for the future of ML-enabled medicine in Western medicine. However, I argue that clinicians' use of medical ML systems is likely to compromise the quality of care and empathy in Western medical contexts, rather than to improve it. This is because proponents of medical ML systems neglect a range of institutional, economic, and sociotechnical obstacles that medical ML systems would need to overcome to generate improvements in the quality of empathy and care in clinician-patient relationships. Moreover, proponents of medical ML systems neglect a series of risks associated with increased psychical and psychological distance between clinicians and patients that medical ML systems themselves present to the quality of empathy and care in medicine.

The remainder of this chapter proceeds as follows. In section two, I provide an overview of the nature and benefits of care and empathy in medicine. In section three, I provide an overview of the two arguments, noted above, that have recently been advanced in the literature to support the claim that clinicians' use of medical ML systems is likely to improve the quality of care and empathy in medicine. In section four, I argue that medical ML systems are more likely to decrease the time that clinicians can spend caring for patients due to a variety of institutional, sociotechnical, and economic factors. In section five, I argue that medical ML systems are more likely to recapitulate prevailing sociocultural values concerning the prioritisation of skills and labour due to the stability and inflexibility of sociocultural values over time. In section six, I argue that medical ML systems themselves are also likely to present a variety of threats to the quality of care and empathy in medicine, highlighting their potential to generate distance between clinicians and patients and dissociate clinicians from the first-person experiences of their patients. In section seven, I provide some concluding remarks.

## 2. The value of care and empathy

Before discussing the value of care and empathy in medicine, it is first necessary to specify what the terms 'care' and 'empathy' actually mean. Care and empathy are often highlighted by philosophers and bioethicists as essential skills and virtues of the practicing clinician due to their positive impact on clinician-patient relationships (Beauchamp and Childress 2019; Pellegrino and Thomasma 1993; Tong 1997). Care refers to "a species of activity that includes everything we do to maintain, contain, and repair our 'world' so that we can live in it as well as possible. That world includes our bodies, ourselves, and our environment" (Fisher and Tronto 1990: 35). The practice of caregiving involves three basic activities: *attentiveness*, which involves listening to patients' stated needs and anticipating unstated needs; *responsibility*, which involves actively taking responsibility for the needs of another; and *competence*, which refers to the skills and activities involved in practically meeting another's needs (Tronto 1998). In contrast, empathy refers to "a complex, imaginative process through which an observer simulates another person's situated psychological states while maintaining clear self-other differentiation" (Coplan 2011: 40). In medical contexts, demonstrating empathy requires that clinicians "(a) understand the patient's situation, perspective, and feelings (and their attached meanings), (b) communicate that understanding and check its accuracy and (c) act on that understanding with the patient in a helpful (therapeutic) way" (Mercer and Reynolds 2002: S10).

Care and empathy have a range of benefits for both patients and clinicians (Derksen, Bensing, and Lagro-Janssen 2013). For patients, being treated by a caring and empathetic clinician has been found to improve health outcomes (Kelley et al. 2014), e.g. by reducing the length and severity of the common cold, increasing the likelihood that patients will seek out medical care in the future, and increasing the likelihood that patients will adhere to recommended treatment regimens (Neumann et al. 2011; Rakel et al. 2011). Being treated by a caring and empathetic clinician has also been found to improve patients' psychological well-being by reducing their self-reported levels of psychological distress and anxiety, and improving their capacity to manage psychological problems (Buszewicz et al. 2006; Van Dulmen and Van Den Brink-Muinen 2004; Hojat et al. 2011). For clinicians, treating a patient in a caring and empathetic way can improve patient satisfaction with their treatment and care, and reduces the risk of malpractice litigation (Hickson et al. 2002; Hojat et al. 2011). Care and empathy have also been found to have a positive impact on communication between clinicians and patients. For instance, caring and empathetic clinicians have been found to communicate better with patients about psychosocial issues (Levinson and Roter 1995), and patients have been found to reveal more information to clinicians who demonstrate care and empathetic concern, particularly through being attuned to the non-verbal gestures and actions of their patients (Finset 2011; Finset and Mjaaland 2009).

### 3. The case for 'deep medicine'

Despite their many benefits, care and empathy have been overlooked and undervalued in medical education and practice for decades. In the 1960s, for instance, 'detached concern' was considered the ideal attitude for clinicians to demonstrate toward their patients (Erickson and Grove 2008; Halpern 2003; Lief 1963). In 1927, moreover, Francis Peabody claimed that:

> The most common criticism made at present by older practitioners is that young graduates have been taught a great deal about the mechanism of disease, but very little about the practice of medicine – or, to put it more bluntly, they are too 'scientific' and do not know how to take care of patients (Peabody 1927: 877).

Now, almost a century later, little has changed as medical students continue to experience 'empathy decline' throughout their education (Neumann et al. 2011), concerns about the neglect of caregiving in medicine are still expressed by practicing clinicians (Cassell 1997, 2004), and patients continue to perceive clinicians as rude, impatient, or interruptive (Coyle, Yen, and Elwyn 2022; Reader, Gillespie, and Roberts 2014; Rhoades et al. 2001).

According to Eric Topol (2019a), this current crisis of empathy and care is due to three core factors. First, clinicians and healthcare organisations currently face a range of powerful economic pressures and incentives that result in clinicians being "squeezed for maximal productivity and profits" (Topol 2019: 284-285). These economic pressures and incentives restrict the amount of time that clinicians can spend caring for patients. As Topol stresses, "[t]he average length of a clinic visit in the United States for an established patient is seven minutes; for a new patient, twelve minutes" (Topol 2019: 29). Second, patient consultations are increasingly being occupied by administrative tasks rather than caring for patients. In some cases, the time clinicians spend performing desk work and engaging with EHRs nearly doubles the amount of time they spend engaged in face-to-face patient care (Chaiyachati et al. 2019; Hill, Sears, and Melanson 2013; Sinsky et al. 2016). These statistics have prompted some practicing clinicians, including Abraham Verghese (2018), to argue that clinicians are rapidly being transformed from caring professionals into clerical workers. Third, there is a mental health crisis amongst clinicians insofar as clinicians are more likely to experience burnout or depression relative to the general population (Bovier and Perneger 2003; Friedberg et al. 2013; Oaklander 2015; Shanafelt et al. 2012; Shanafelt, Dyrbye, and West 2017). These ailments are likely to reduce the cognitive or emotional capacity that clinicians have to empathise with or care for their patients. As former paediatrician Victoria McEvoy has expressed, the great danger of burnout "is that you stop caring. The goal of each day and each night was simply to move everyone through, to clear the decks, rather than to deliberately and expertly care for those who need care and reassure those who did not" (Groopman 2007: 80).

However, Topol (2019) argues that medical ML systems could reduce the time pressures that clinicians currently face, reduce the time they spend engaging with EHRs, and ease the current crisis of burnout and depression in medicine. He anticipates that these improvements could enable clinicians to devote much more time and energy toward caring for and empathising with their patients. In particular, Topol argues that medical ML systems could increase clinicians' productivity by enabling them to perform clinical tasks more quickly and efficiently, as discussed in the introduction to this thesis. Indeed, the Institute for Public Policy Research has recently estimated that medical ML systems will generate time-savings of between 11%-57% for frontline healthcare workers (Darzi 2018). Time savings have also been found to be the most commonly identified anticipated benefit associated with medical ML systems amongst UK healthcare employees (Hardie et al. 2021). Thus, as Bertalan Meskó has expressed, by "taking away the repetitive parts of a physician's job, it might lead to being able to spend more precious time with their patients, improving the human touch" (Meskó 2017: 241).

Topol (2019) also argues that medical ML systems are likely to substantially reduce the overwhelming administrative burdens currently faced by clinicians. As noted in the introduction, ML systems are being developed to assist in a broad range of tasks associated with healthcare administration – including scheduling appointments, transcribing notes during patient consultations, updating patients' medical records, and processing drug prescriptions (Kocaballi et al. 2020; Wang et al. 2022). Topol argues that these types of ML systems could allow clinicians to delegate a substantial proportion of their administrative responsibilities to machines, a claim that is also supported by Shantanu Nundy and co-authors who claim that:

Through greater automation of low-value tasks, such as clinical documentation, it is possible that AI will free up physicians to identify patients' goals, barriers, and beliefs, and counsel them about their decisions and choices, thereby increasing trust (Nundy, Montgomery, and Wachter 2019: E1).

Moreover, Topol suggests that one of the greatest benefits of medical ML systems "will come from unshackling clinicians from electronic health records" (Topol 2019: 288). This is not only because EHR work takes up a large proportion of clinicians' time, but also because of the negative impact that Topol claims EHRs have had upon non-verbal empathetic communication between clinicians and patients (e.g. reduced eye contact).

Topol (2019) suggests that these improvements are likely to ease the mental health crisis currently facing practicing clinicians. This is because overwork, time pressures, and administrative burdens are identified as key contributing factors to the high rate of burnout and depression. Insofar as medical ML systems allow clinicians more time to spend caring for patients and reduce their administrative responsibilities, Topol anticipates that clinicians' mental health will broadly improve.

As if these benefits were not already enough, Topol (2019) also argues that medical ML systems are likely to increase the value of caregiving skills in medical education and employment. In particular, he argues that medical ML systems are likely to encroach on the cognitive niche of human clinicians by eventually surpassing their accuracy in the performance of clinical tasks, and that human clinicians will not be able to compete against the performance of medical ML systems in the long term. Topol argues that, in order to remain useful in medicine and competitive on the healthcare employment market, human clinicians will need to upskill in distinctively 'human' areas that cannot be outsourced to machines, e.g. caregiving, emotional intelligence, and empathy. He concludes that the "levelling of the medical knowledge landscape will ultimately lead to a new premium to find and train doctors who have the highest level of emotional intelligence" (Topol 2019: 18). Topol refers to this vision for the future of ML-enabled medicine as 'deep medicine'.

## 4. Sociotechnical obstacles and the costs of efficiency

Topol (2019) and others illustrate a compelling vision for the future of ML-enabled medicine in which the value of care and empathy in clinician-patient relationships is finally recognised. However, their arguments for these predictions either overlook or underestimate a broad range of sociotechnical, economic, and institutional obstacles that are likely to undermine the realisation of their vision for the future of ML-enabled medicine.

For instance, any time savings generated by the use of medical ML systems are likely to be used by healthcare administrators to ensure that clinicians see more patients each day, rather than spend more time with each patient. This is because care and empathy are unlikely to be prioritised over measurable indicators of productivity (e.g. the number of patients seen each day) in health systems that are increasingly concerned with productivity and efficiency. This is because care and empathy are abstract notions that are difficult to represent and measure quantitatively. Indeed, healthcare institutions are often restricted to evaluating the quality of care using proxy metrics such as consultation length and frequency of venous thromboembolisms. However, health systems are becoming increasingly influenced by the philosophy of 'new public management', in which economic and institutional objectives that prioritise the maximisation of measurable, principally economic, outcomes are promoted above all else (see Simonet 2011, 2015). Adopting medical ML systems to save time reinforces, rather than challenges, this prevailing system of values in healthcare organisations.

Topol acknowledges the potential for time savings to be "used by administrators as a means to rev up productivity, so doctors see more patients, read more scans or slides, and maximise throughput" (Topol 2019: 288). However, he argues that this risk can be overcome so long as clinicians "take on the role of activists" by advocating for improved working conditions and defending the value of caregiving, presence, and empathy in medicine against the economic incentives of healthcare administrators and managed care organisations (Topol 2019: 288). As Robert Sparrow and I have argued elsewhere, however, clinicians have a poor track record of success when it comes to activism, despite powerful professional institutions such as the AMA, that inspires little confidence in this prospect (Sparrow and Hatherley 2020).

It is also likely that the implementing, using, and maintaining medical ML systems will increase, rather than reduce, the administrative responsibilities of human clinicians. For instance, automating administrative tasks using ML systems may impose new demands on clinicians that could offset any time savings generated by their use. As Raja Parasuraman and co-authors express, "automation does not simply supplant human activity but rather changes it, often in ways unintended and unanticipated by the designers of automation, and as a result poses new coordination demands on the human operator" (Parasuraman, Sheridan, and Wickens 2000: 286-287; see Bradshaw et al. 2013). For instance, the use of administrative ML systems may require clinicians to review the notes of these systems, correct their mistakes, or adding information that was missed. New coordination demands of this sort may simply reproduce existing administrative responsibilities in a new guise. Indeed, in some cases, correcting the errors of administrative ML systems may prove more laborious for clinicians than simply performing the task themselves. As Thomas Maddox and co-authors highlight:

> history suggests that most forms of clinical decision support add to, rather than replace, the information clinicians need to process. As a result, there is a risk that integrating AI into clinical workflow could significantly increase the cognitive load facing clinical teams and lead to higher stress, lower efficiency, and poorer clinical care (Maddox, Rumsfeld, and Payne 2019: 32).

Medical ML systems are also likely to increase the administrative responsibilities of clinicians. This is because clinicians are likely to be responsible for ensuring that high-quality training data for medical ML systems are collected and maintained in EHRs. EHRs are one of the most common sources of training data for medical ML systems (see Cheng et al. 2016; Embi and Leonard 2012; Gianfrancesco et al. 2018; Miotto et al. 2016; Rajkomar et al. 2018; Shickel et al. 2018; Wachter and Cassel 2019) and clinicians are already responsible for entering patient data into these systems and maintaining patient health records. The information recorded in EHRs is typically restricted to that which is relevant to clinicians and, increasingly, healthcare administrators. However, widespread use of medical ML systems is likely to expand the range of information that clinicians must collect such that clinicians may soon be required to also collect information that is relevant to the designers of medical ML systems.

Moreover, widespread implementation of medical ML systems may warrant changes in the standard of information that clinicians record in EHRs that could generate additional administrative burdens. As noted in chapter one, developing and maintaining medical ML systems depends on the availability of largescale, comprehensive, and well-annotated datasets. However, as Maddox and co-authors observe, "most clinical data, whether from electronic health records (EHRs) or medical billing claims, remain ill-defined and largely insufficient for effective exploitation by AI techniques" (Maddox, Rumsfeld, and Payne 2019: 31). One reason for this is that, as Marzyeh Ghassemi and co-authors observe, "[c]linical data is almost exclusively documented without machine learning in mind" (Ghassemi et al. 2018: 2). However, it is also because data contained in EHR systems are notoriously patchy, repetitive and inconsistent. For instance, clinicians use different acronyms to refer to the same condition, patient data is often copied and pasted across visits instead of being recorded in detail, and important contextual features of clinical cases often go unrecorded. These shortcuts save time for clinicians under increasing time pressures, but generate substantial obstacles for AI developers turning to EHR systems for training data (Liu et al. 2022). Ensuring that EHR data meets certain minimum standards as training data for medical ML systems may increase the amount of data that clinicians must enter into these systems, and impose tighter restrictions on how these data are communicated and recorded. Rather than unshackling clinicians from EHRs, therefore, medical ML systems may simply reinforce clinicians' reliance upon them.

Regulatory requirements associated with the post-market surveillance of medical ML systems may also generate expanded administrative responsibilities for clinicians. In the US, the 21st Century Cures Act has recently encouraged the FDA to consider how to better incorporate 'real-world evidence' into the post-market surveillance of new medical devices. Real-world evidence "includes information generated through routine health care delivery, including electronic health records (EHRs), billing data, clinical registries, and other data sources" (Resnic and Matheny 2018: 596). However, EHRs are likely to become an increasingly influential source of real-world evidence (RWE). As Resnik and Matheny observe, as "EHRs are adopted ever more broadly, RWE will become a more accessible and lower-cost source of detailed clinical information that could help clinicians and regulators understand the performance of medical devices in real-world practice" (Resnic and Matheny 2018: 596). Ensuring sufficient post-market evaluation and monitoring of medical ML systems may thus further increase the range of information that clinicians are required to record in EHR systems.

In addition to increasing the administrative burdens of clinicians, medical ML systems may contribute to the current crisis of burnout and depression amongst practicing clinicians. Low self-reported measures of professional autonomy are a key factor in the historically high rate of burnout and depression (Friedberg et al. 2013). However, ML systems are likely to be used in ways that reduce clinicians' professional autonomy. For instance, the effects of increased surveillance on clinicians' sense of professional autonomy and professional satisfaction can be devastating (Gawande 2018; Klugman et al. 2018). Despite this, a broad range of ML systems have recently been developed to evaluate the competence and performance of human practitioners, their compliance with certain evidence-based standards, along with various other forms of generalised workplace surveillance (Dias et al. 2018; Haque, Milstein, and Fei-Fei 2020b; Martinez-Martin et al. 2021; Yilmaz et al. 2022).

Medical ML systems could also be designed to restrict the scope of professional autonomy by being designed as what Ursula Franklin (1990) refers to as 'prescriptive technologies'. Prescriptive technologies aim to restrict the capacity for users to exercise independent judgement and discretion, thereby promoting compliance with narrow and pre-defined objectives. Medical ML systems could be designed as prescriptive technologies by generating outputs designed to standardise the behaviours and professional judgments of human clinicians. Medical ML systems could also be designed to generate outputs that support the economic objectives of the healthcare institution for which they work, and clinicians may be asked to justify their decision to reject these outputs.

## 5. Sociocultural values and the future of healthcare work

As previously discussed, while medical ML systems are unlikely to save time for clinicians to spend caring for patients, Topol (2019) also argues that medical ML systems are likely to benefit clinician-patient relationships by increasing the value of skills associated with care and empathy in the training and employment of clinicians. In particular, Topol argues that once medical ML systems exceed human performance in clinical tasks, clinicians will be forced to adapt to the looming redundancy of their technical skillsets. To remain competitive on the healthcare employment market, Topol suggests that clinicians will need to reorient their skills toward caring for and empathising with their patients. In response, according to Topol, hospitals and teaching institutions will come to prioritise hiring clinicians and accepting medical students that demonstrate the highest levels of emotional intelligence.

Topol's argument is unconvincing, however, because clinicians and healthcare institutions are more likely to adapt to the looming redundancy of clinicians' technical skillsets in ways that simply recapitulate prevailing sociocultural values, and perhaps even further compromise the quality of care and empathy in medicine. In particular, clinicians are more likely to adapt to this state of affairs by developing their capabilities in the broad range of technical skills that will continue to be performed better by human clinicians. For instance, Topol himself elsewhere argues that radiologists and pathologists ought to adapt to the role of 'information specialists' to ensure that they remain clinically useful in light of increasing advances in the performance of medical ML systems. However, the core duties of information specialists involve the exercise of technical skills, rather than those associated with care and empathy. In particular, Topol anticipates that the role of the information specialist will be to "interpret the important data, advise on the added value of another diagnostic test, such as the need for additional imaging, anatomical pathology, or a laboratory test, and integrate information to guide clinicians" (Jha and Topol 2016: E2).

Indeed, as William Schwartz observed as early as 1970:

students and physicians exposed to prototype computer-based consulting programs commonly express anxiety and displeasure at the prospect of practicing medicine within a system that has as a major feature the surrender of many memory and analytical functions. It might be argued, of course, that the opportunity to deal more extensively with the emotional aspects of disease will compensate the physician for the expropriation of his diagnostic and therapeutic skills, but it is far from clear that most physicians are equipped by either temperament or training to accept change of this kind gracefully (Schwartz 1970: 1259-1260).

Even if clinicians do adapt to the looming redundancy of their technical skillsets by developing their capabilities in caring for and empathising with their patients, healthcare organisations are likely to adapt to the structural changes generated by medical ML systems by downsizing their human workforces, rather than hiring these 'expert caregivers' or 'empathy workers'. This is because, as I have argued previously, healthcare institutions are strongly incentivised to prioritise improvements to measurable indicators of economic efficiency over improvements associated with the quality of care and empathy. Since downsizing workforces where feasible is likely to improve economic efficiency, it seems more likely that healthcare institutions will take this approach over prioritising the hiring of expert caregivers.

Healthcare institutions may also be incentivised to downsize their human workforces due to the growing capabilities of robotic caregivers, which are becoming increasingly prevalent in healthcare settings, particularly those associated with aged care and disability support (Maalouf et al. 2018; Persson, Redmalm, and Iversen 2022; Pfadenhauer and Dukat 2015; Wang et al. 2017). For instance, healthcare institutions may elect to purchase and implement robotic caregivers rather than hire expert human caregivers if the former become more economically efficient. While Topol highlights that caregiving robots can only provide simulated care for patients, this is unlikely to deter healthcare organisations unless patients reject these systems in practice. However, caregiving robots may nevertheless be accepted by patients due to the tendency for humans to anthropomorphise machines and inanimate objects (Weizenbaum 1976). Indeed, some researchers have found that patients prefer to discuss sensitive information with computer systems rather than human clinicians (Pickard, Roster, and Chen 2016; Schuetzler et al. 2018). It is possible that some patients may even come to prefer being cared for by machines due to the consistency of care that these systems could provide, and the poor standard of care and empathy that is currently provided by human clinicians, as discussed above, whose caregiving can be negatively impacted by various contingencies including their mood and environment.

Sociocultural values associated with the prioritisation of skills and the distribution of labour in healthcare are deeply entangled with broader cultural attitudes concerning sex and gender. The skills associated with care and empathy, for instance, are culturally coded as 'feminine' and devalued in relation to technical skills, such as diagnostic competency, that are cultural coded as 'masculine'. As Rosemarie Tong has argued:

> although empathy can be taught as an epistemic skill and care can be taught as a moral virtue, the medical profession will not undertake these teaching projects wholeheartedly until society as a whole (1) comes to value culturally-associated 'female' or 'feminine' epistemic skills and moral virtues as much as culturally-associated 'male' or 'masculine' and (2) distributes its caregiving tasks and occupations to men and women equally (Tong 1997: 154).

However, it is extremely unlikely that the mere implementation of ML systems in medicine will achieve such monumental objectives. This is because, for the most part, ML systems do not challenge the prevailing system of gendered values in Western societies, but actively support and reinforce them (Collett and Dillon 2019; Costa and Ribas 2019; Leavy 2018; Lütz 2022; Nadeem, Marjanovic, and Abedin 2022; West, Whittaker, and Crawford 2019). Moreover, social values, as Michael Manfredo and co-authors observe, "are not stand-alone entities, readily vulnerable to change. Instead, they are deeply entangled in a web of material culture, collective behaviors, traditions, and social institutions" (Manfredo et al. 2017: 775). The implementation of a new medical technology is simply insufficient to disentangle these deep and complicated webs of co-determination and mutual influence. As Tristan Panch and co-authors express, merely "adding AI applications to a fragmented system will not create sustainable change" (Panch, Mattie, and Celi 2019: 1).

## 6. Increased distance, heightened dissociation

Thus far, I have argued that clinicians' use of medical ML systems is unlikely to improve clinician-patient relationships by improving the quality of care and empathy in medicine. In this section, however, I argue that the implementation of ML systems themselves generates a range of threats to the quality of care and empathy in medicine. In particular, I argue that clinicians' use of medical ML systems is likely to generate physical distance and psychological dissociation between clinicians and their patients.

Medical ML systems are likely to generate physical distance between clinicians and patients by expanding the range of clinical tasks that clinicians can perform remotely or in the patient's absence. For instance, clinicians may soon be able to diagnose patients, generate treatment plans, and check biometric signals without interacting with patients at all. Indeed, writers including Topol (2019a) and Robert Wachter (2015) argue that medical ML systems are likely to reduce clinicians and patients' reliance on brick-and-mortar hospitals for their medical treatment and care. Indeed, Topol (2019a) argues that medical ML systems are likely to facilitate substantial growth in the number and capacity of so-called 'virtual hospitals.' Virtual hospitals refer to healthcare institutions in which clinicians provide healthcare services to patients entirely remotely. For instance, Mercy Virtual Care Center is a hospital in St. Louis in which:

> doctors and nurses sit at carrels in front of monitors that include camera-eye views of the patients and their rooms, graphs of their blood chemicals and images of their lungs and limbs, and lists of problems that computer programs tell them to look out for. The nurses wear scrubs, but the scrubs are very, very clean. The patients are elsewhere (Allen 2017: 1).

Indeed, Richard Baldwin (2019) has argued that medical ML systems and other digital technologies are likely to incentivise healthcare organisations to capitalise on cheap labour offered by digital freelancers, or 'tele-migrants'. These tele-migrants are anticipated to have the capacity to perform clinical tasks remotely, from different cities, states, countries, and even continents.

By expanding the range of clinical tasks that clinicians can perform in the absence of patients, medical ML systems are likely to compromise the quality of empathy and care in medicine by reducing the frequency and duration of scheduled and incidental in-person contact between clinicians and patients. This is because physical distance is likely to decrease the accessibility of clinicians to their patients and reinforce current power imbalances between them. In particular, by expanding the range of clinical tasks that clinicians can perform in the absence of patients, medical ML systems are likely to reduce patients' power and control over when, where, and how they communicate or interact with their clinicians. This will enable clinicians to communicate and interact with patients on their terms, and their terms alone. Moreover, by reducing scheduled and incidental interactions between patients and clinicians, medical ML systems are likely to further compromise the quality of shared decision-making, as I discussed previously in chapter four. Medical ML systems are thus likely to grant clinicians fewer opportunities to interact with patients and gain a deeper understanding of their patients' values and preferences.

The use of medical ML systems is also likely to generate psychological distance between clinicians and patients. In particular, introducing medical ML systems into patient consultations is likely to prompt clinicians to devote more of their focus and attention toward operating technical systems rather than engaging in face-to-face patient care. This is because, as Stanley Joel Reiser (2009) has argued, the introduction of new technologies into clinical practice has created, and historically reinforced, the sociological phenomenon of 'modern technological distancing' in medicine. Modern technological distancing refers to the tendency for clinicians to become increasingly dissociated from the first-person knowledge and experiences of their patients. Reiser (2009) argues that modern technological distancing occurs because medical technologies expand the range of information and evidence that clinicians can gather about their patients' conditions without relying on the inconsistent and unreliable testimony of their patients. By enabling clinicians to gather more information and insight into their patients' condition without direct input from patients themselves, therefore, medical ML systems are likely to reinforce the current trend of modern technological distancing in clinical medicine.

Medical ML systems are also likely to reinforce the influence of 'disease-centred ontologies' in medicine. In philosophy, ontology refers to "the science of what is, of the kinds and structures of objects, properties, events, processes, and relations in every area of reality" (Smith 2003: 155). In the philosophy of medicine, the aim of ontology is to conceptualise the basic properties and entities under consideration in healthcare, particularly those of 'health', 'illness', and 'disease' (Simon 2011). Under disease-centered ontologies, health is conceptualised as the mere absence of disease, while disease is conceptualised as a foreign, harmful, and intrusive entity that has occupied a patient's body and must be removed (Cassell 1997, 2004). Disease-centered ontologies of this variety offer impoverished accounts of health, illness, and disease. This is because, as Havi Carel (2016) and Eric Cassell (1997, 2004) have argued, disease-centered ontologies neglect to account for the role of patients' first-person experiences in the constitution of health and illness.

Medical ML systems are likely to strengthen the influence of disease-centered ontologies in medicine by deepening clinicians' engagements with EHR systems. As I have argued above, medical ML systems are likely to increase the administrative responsibilities of clinicians by obligating them to collect and record data that is needed by the developers of medical ML systems to train and maintain these systems. Deepening clinicians' engagement with EHRs is likely to strengthen the influence of disease-centered ontologies because, as Abraham Verghese (2008) has argued, the introduction of EHR systems into clinical practice has prompted clinicians to neglect the first-person accounts of patients in favour of the composite of scans, test results, clinical notes, and medical images that make up a patients' clinical record. Medical ML systems are also likely to strengthen the influence of disease-centered ontologies by reinforcing the current fixation on data in medicine, and in society more generally (Bhageshpur 2019; Toonders 2014). Yuval Noah Harari (2015) refers to this cultural fixation as 'dataism'. Widespread use of medical ML systems is likely to reinforce the influence of dataism in medicine because big data are the core building blocks of these system. This is because ensuring the continued development, maintenance, and use of medical ML systems will demand a substantial increase in attention directed toward the collection, storage, and use of patient data that is likely to occupy an increasing amount of clinicians' time and attention.

Medical ML systems are also likely to strengthen the influence of disease-centered ontologies in medicine by virtue of their inability to incorporate patients' first-person perspectives into their training and input datasets. As Benjamin Chin-Yee and Ross Upshur (2019) have argued, medical ML systems invariably occupy a third-person epistemic vantage point. In particular, they argue that the use of ML systems is likely to strengthen existing commitments to positivistic clinical decision-making in medicine in which the subjective, narrative elements of a patients' illness and medical history overlooked. According to Chin-Yee and Upshur (2019), this limitation of medical ML systems is likely to diminish the significance of patient knowledge and experience, exacerbate epistemic injustice in medicine, distort understandings of the narrative and interpretive elements of clinical judgement, and interfere with the dialogic and interpretive elements of patient care.

By strengthening the influence of disease-centered ontologies, therefore, medical ML systems are likely to compromise the quality of care and empathy in medicine by reducing clinicians' focus and attention toward the first-person experiences of patients. In short, the outputs of medical ML systems are likely to be accepted by clinicians over the first-person experiences of their patients. This is because clinicians often perceive the outputs of medical ML systems as more impartial, objective, and reliable than human judgements (Aquino et al. 2023). When the datafied representations of patients generated through the use of medical ML systems conflict with their first-person experiences and testimony, clinicians may therefore be more likely to accept the ML system's output over the patients' first-person testimony. As Brent Mittelstadt and Luciano Floridi have expressed:

> greater reliance on data representations of patients brought about by adoption of Big Data practices may create new gaps in care or doctor-patient relationships […] Put another way, the patient's body and voice may increasingly be replaced or supplemented by data representations of state of being if Big Data practices are adopted in medicine (Mittelstadt and Floridi 2016: 328).

In other words, clinicians may be less likely to trust their patients over the outputs of medical ML systems. Medical ML systems could also compromise the quality of care and empathy in medicine by reducing patients' trust in their clinicians, providing additional support for my argument in chapter two. This is because the use of these systems is likely to negatively impact patients' perceptions of the competency of their clinicians and the quality of caregiving in medicine. In particular, and as previously noted in chapter four, patients often judge that clinicians who use diagnostic decision-support systems are less professional, less thorough and systematic, and less competent than clinicians who do not use clinical decision-support tools (Arkes, Shaffer, and Medow 2007; Shaffer et al. 2013). The more that clinicians use ML systems and rely on their outputs to inform their clinical judgements, the more that some patients may feel that their medical care is being provided by a machine rather than a fellow human being. Patients may therefore be less likely to adhere to their clinicians' recommendations or seek out medical treatment and care.

## 7. Conclusion

Despite high hopes that the implementation of ML systems will improve clinician-patient relationships, I have argued that clinicians' use of these systems in medicine is likely to compromise the quality of care and empathy in their relationships with patients. This is due to several sociotechnical, economic, and institutional obstacles that are likely to circumvent pundits' optimistic visions for the future of ML-enabled medicine. For instance, the value of care and empathy in medicine are routinely underrecognised by administrative and managerial bodies due to the strong focus on quantifiable measures and economic efficiency. The heavy data requirements of medical ML systems are also likely to enhance the data collection responsibilities of human clinicians, rather than to reduce them.

Clinicians' use of medical ML systems is also likely to negatively impact clinician-patient relationships because these systems themselves present serious risks to the quality of care and empathy in medicine. In particular, medical ML systems are likely to expand the range of clinical tasks that clinicians perform in the patient's absence, increasing the physical distance between clinicians and their patients. Medical ML systems are also likely to direct clinicians' focus and attention toward datafied representations of their patients' illness, rather than patients' own first-person experiences of suffering, thereby increasing the psychological distance between clinicians and their patients. Moreover, medical ML systems will likely be used by administrative and managerial bodies to impose stronger surveillance and monitoring mechanisms upon clinicians, augmenting their current degree of professional dissatisfaction, burnout, and depression, along with the downstream implications of such effects on the quality of patient caregiving.

In this chapter, I have addressed the negative impact that clinicians' use of ML systems is likely to have on clinician-patient relationships by interfering with the quality of care and empathy in medicine. This concludes my argument for the claim that medical ML systems are likely to impact negatively on the quality of clinician-patient relationships. In the concluding chapter of this thesis, I suggest that AI developers, healthcare organisations, and policy makers need to more carefully consider the effects of these systems on the relationship between clinicians and their patients to accurately assess the costs and benefits of these technologies prior to implementation.

# CONCLUSION

This thesis makes a significant contribution to the literature by tempering current expectations about the impact of ML systems in medicine. As discussed in the introduction to this thesis, a growing list of eminent researchers in AI and medicine, including Eric Topol, Geoffrey Hinton, and Abraham Verghese, anticipate that medical ML systems will revolutionise the practice of medicine and the delivery of healthcare services. In particular, these experts anticipate that medical ML systems will improve patient health and safety by greatly reducing the frequency of medical error, reduce health disparities by expanding the availability of medical expertise and treatment in under-serviced and under-resourced communities, and improve time- and cost-efficiency in medicine by enabling clinicians to delegate routine tasks to these systems and speed up the execution of more complex tasks.

In chapter one, I surveyed a series of concerns about the likely future impacts of ML systems in medicine. I argued that the implementation of medical ML systems into clinical workflows presents threats to patient health and safety due to a variety of persistent weaknesses and limitations in these systems, and human biases in using them. The use of medical ML systems also risks deepening current health disparities due to their susceptibility to adopting the biases of their designers and the societies in which they are embedded. The political, economic, and institutional incentives of the wealthy organisations driving medical research and technological innovation may also influence the kinds of problems that medical ML systems are designed to address, potentially causing an inequitable distribution of risks and benefits associated with the use of medical ML systems between socioeconomically advantaged and disadvantaged population groups. Intensified data collection practices generated by an acceleration in the development of medical ML systems also threaten to increase the risk of infringing patients' privacy and confidentiality, and to expand the scope of biased or intrusive surveillance by government agencies and private organisations. Finally, incorporating medical ML systems into clinician workflows threatens to compromise cultures of accountability in healthcare institutions, and the medical profession more broadly, by obfuscating the allocation of moral and legal responsibility for error, negligence, and harm that results from the use of these systems.

According to Eric Topol (2019a), however, the anticipated benefits of medical ML systems for patient health and safety, health equity, and efficiency are merely the "secondary gains" of the coming age of AI in medicine. This is because, according to Topol, this coming age of AI is "our chance, perhaps the ultimate one, to bring back real medicine: Presence. Empathy. Trust. Caring. Being Human" (Topol 2019: 309). In particular, Topol anticipates that medical ML systems will unburden clinicians of routine clinical and administrative labour, enabling them to spend more time caring for their patients face-to-face. In this thesis, however, I have argued that medical ML systems generate new and expanded threats to the quality of trust, care, empathy, and understanding in medicine that are likely to compromise the overall quality of clinician-patient relationships. Topol's vision for the future of clinician-patient relationships in the coming age of AI in medicine is thus fundamentally misguided.

In chapter two, I argued that medical ML systems are likely to interfere with relations of trust between clinicians and patients because, while medical ML systems themselves are the appropriate objects of reliance, they are not the appropriate objects of trust. Insofar as medical ML systems come to have an increasing role in medical decision-making, patients and clinicians will be required to rely on these systems. However, since medical ML systems cannot coherently be trusted, the use of these systems in medicine will compromise the quality and depth of relations of trust between clinicians and patients. Moreover, the current tendency to describe human relations with medical ML systems in terms of trust also risks compromising the quality of relations of trust between clinicians and patients by obfuscating accountability for error, negligence, and harm in medicine. Describing human relations with these systems in terms of trust risks implicitly attributing responsibility for error and harm to these systems rather than to the human agents involved in their development, maintenance, and use. Patients' trust in their clinician is likely to be compromised where clinicians are perceived as less-than-accountable for patient harm.

In chapter three, I argued that the use of medical ML systems for treatment recommendation also threatens to interfere with patient autonomy since ethical values often become embedded in these systems. These risks to patient autonomy serve to expand the scope of clinicians' ethical obligations with respect to communicating with patients about the tools they use in the formulation of their professional judgements and recommendations. In particular, the presence of these risks ethically obligate clinicians to disclose their use of medical ML systems to secure their patients' informed consent. However, clinicians are also ethically obligated to disclose their use of medical ML systems for several reasons beyond informed consent requirements. For instance, the use of medical ML systems generates a range of patient safety risks that warrant disclosure, and where clinicians fail to disclose their use of medical ML systems to patients, they also risk infringing their patients' right to privacy and their right to act on rational concerns about the future.

The use of two specific types of ML systems in medicine also generates specific threats to the quality of clinician-patient relationships.

The use of opaque medical ML systems is likely to interfere with the quality of these relationships by interfering with clinicians' capacity to test the outputs of medical ML systems against their own knowledge and experience. As I argued in chapter four, insofar as clinicians cannot test the outputs of these systems in this way, their ability to communicate appropriately with their patients about the reasons underlying their judgements and recommendations is likely to be restricted. While a range of critics have argued that clinicians are nevertheless justified in using opaque medical ML systems due to the overriding importance of accuracy over interpretability, these critics neglect a host of reasons that justify prioritising interpretable over opaque systems in medicine, including their benefits to patient safety and communication between clinicians and patients. By interfering with the capacity for clinicians to reliably appraise the veracity of the outputs of medical ML systems, opacity is likely to compromise the positive impacts that these systems are anticipated to have on patient health and well-being. Indeed, where medical ML systems are used to assist human clinicians in diagnosing or treating complex conditions, the use of ante-hoc interpretable systems is likely to generate more accurate and reliable judgements or recommendations, even where these systems may be less accurate than comparatively opaque systems.

A second sort of medical ML systems that are likely to generate distinctive threats to clinician-patient relationships are MAMLS, or medical adaptive ML systems, which continue learning from new data even after being implemented in a clinical setting. In chapter five, I argued that this is because MAMLS introduce two types of variation into the performance of medical ML systems, diachronic evolution and synchronic variation, each of which poses risks and challenges for clinicians and patients. For instance, diachronic evolution presents further obstacles to clinicians' ability to test the outputs of these systems against their own knowledge and expertise, and increase the hermeneutic labour that clinicians face when interpreting the outputs of these systems. By obfuscating the interpretation of MAMLS' outputs, diachronic evolution and synchronic variation also risk further compromising clinicians' ability to communicate basic information to patients about their medical treatment and care. In some cases, the negative impact of diachronic evolution on communication between clinicians and patients may compromise clinicians' capacity to secure their patients' informed consent. Synchronic variation also risks allowing the performance of MAMLS implemented at difference sites to diverge over time which may generate inequities in the quality of healthcare services between different settings or deepen existing health inequities. These additional risks may reduce patients' trust in the competency or fairness of their clinicians' judgements.

Finally, in chapter six, I argued that, contra Topol, medical ML systems are likely to compromise the quality of care and empathy in medicine. This is because medical ML systems are likely to recapitulate, or even intensify, existing pressures that reduce the time and energy that clinicians have available to spend caring for patients. For instance, any time savings generated by the use of medical ML systems are more likely to be used by healthcare institutions and administrators to improve more easily measurable indicators of productivity and economic efficiency than to allow clinicians to spend more time talking with patients. Meeting the increasing demand for data associated with the ongoing development, maintenance, and improvement of medical ML systems is also likely to add to the current administrative responsibilities of human clinicians. These expanding administrative duties are likely to have a negative impact on clinicians' capacity to provide empathetic caregiving to their patients. Moreover, healthcare organisations will be tempted to use ML systems to expand the scope and depth of workplace surveillance practices due to strong financial and practical incentives to standardise clinical processes, improve measurable indicators of productivity, and increase economic efficiency. The negative impact that such measures would have on clinicians' mental health and professional satisfaction threatens to curtail their cognitive and emotional capacity to provide empathetic care to their patients.

This thesis makes a significant contribution to the literature by drawing attention to the relational implications of medical ML systems. In particular, this thesis suggests that minimising the risks associated with these systems depends not only on improving their accuracy and performance, but also on attending to how these systems impact relationships between human beings, and how human beings relate to these systems themselves. AI developers, healthcare organisations, and policy makers need to more carefully consider the effects of these systems on the relationship between clinicians and their patients to accurately assess the costs and benefits of these technologies prior to implementation. It is critical that assessments of the anticipated benefits of medical ML systems remain realistic to avoid the costly implementation of medical ML systems without a clear understanding of their negative effects.

This thesis also makes a significant contribution to the literature by advocating for, and contributing to, a more clear-headed discussion of the use of ML systems in medicine to that treats that risks and benefits of these systems symmetrically. While the use of ML systems may deliver modest improvements in narrow domains of medicine, stakeholders must resist the temptation of becoming carried away by excitement over the anticipated benefits of these systems. In order to protect and preserve clinician-patient relationships in the coming age of AI in medicine, stakeholders must instead think substantially more about the costs they are likely to impose on these relationships. Failing to critically assess the likely impact of medical ML systems on clinician-patient relationships not only threatens these relationships themselves. It also threatens patient health, safety, and well-being insofar as each of these factors are critically dependent on the quality of clinician-patient relationships, as discussed in the introduction to this thesis.

Medical AI systems previously generated substantial hype in the 1970s in the form of expert systems. However, not only did these systems largely disappoint expectations; they also directly preceded the longest and deepest AI winter of the 20th century. Stakeholders must therefore aim to promote more realistic assessments of the benefits of these systems to avoid the risk of shepherding the discipline of AI into yet another period of disillusionment and despair.